\documentclass[fleqn,usenatbib]{mnras}

\usepackage{newtxtext,newtxmath}

\usepackage[T1]{fontenc}
\usepackage{CJKutf8}

\DeclareRobustCommand{\VAN}[3]{#2}
\let\VANthebibliography\thebibliography
\def\thebibliography{\DeclareRobustCommand{\VAN}[3]{##3}\VANthebibliography}

\usepackage{graphicx}	
\usepackage{amsmath}	
\usepackage{multirow}
\usepackage{enumitem}
\usepackage{float}
\usepackage{threeparttable}
\usepackage{enumitem}

\def\Oi{O\,{\sc i}}
\def\Oiii{O\,{\sc iii}}
\def\Hi{H\,{\sc i}}
\def\Mgii{Mg\,{\sc ii}}
\def\Civ{C\,{\sc iv}}
\def\Siiv{Si\,{\sc iv}}
\def\Nv{N\,{\sc v}}
\def\Niv{N\,{\sc iv}}
\def\Niii{N\,{\sc iii}}
\def\Heii{He\,{\sc ii}}
\def\Hii{H\,{\sc ii}}
\def\Cii{C\,{\sc ii}}

\def\Oii{[O\,{\sc ii}]}
\def\Neiii{[Ne\,{\sc iii}]}
\def\Ciii{C\,{\sc iii}}

\def\msun{{\rm M}_{\odot}}
\def\ysej{J1429$-$0104}
\def\yswl{J1450$-$0144}
\def\aa{\text{\AA}}

\title[Quasar Impostors at Cosmic Dawn]
{Quasar Impostors: Two Extremely UV-Bright ($M_{\rm UV}\approx-23.5$) Reionisation-Epoch Galaxies Powered by Very Massive Stars}
\author[D.~Yang et al.]{Daming Yang\begin{CJK}{UTF8}{gbsn} (羊达明)\end{CJK},$^{1}$\thanks{E-mail: dyang@strw.leidenuniv.nl}
Joseph F.~Hennawi,$^{1,2}$
Sarah E.~I.~Bosman,$^{3,4}$
Frederick B.~Davies,$^{4}$
\newauthor
Rychard Bouwens,$^{1}$
Timo Kist,$^{1}$
Eduardo Ba\~nados,$^{4}$
Jiamu Huang,$^{2}$
Alice E.~Shapley,$^{5}$
\newauthor
Marianne Vestergaard,$^{6,7}$
Hiddo S.~B.~Algera,$^{8}$
Silvia Belladitta,$^{4}$
Anna-Christina Eilers,$^{9,10}$
\newauthor
Xiaohui Fan,$^{7}$
Francesco Guarneri,$^{11}$
Xiangyu Jin,$^{12}$
Romain A.~Meyer,$^{13}$
Elia Pizzati,$^{1,14}$\thanks{NHFP Einstein Fellow}
\newauthor
Huub R\"ottgering,$^{1}$
Jan-Torge Schindler,$^{11}$
Mauro Stefanon,$^{15,16}$
Feige Wang$^{12}$
and Jinyi Yang$^{12}$
\\
$^{1}$Leiden Observatory, Leiden University, P.O. Box 9513, 2300 RA Leiden, The Netherlands\\
$^{2}$Department of Physics, University of California, Santa Barbara, CA 93106, USA\\
$^{3}$Institute for Theoretical Physics, Heidelberg University, Philosophenweg 12, D–69120, Heidelberg, Germany\\
$^{4}$Max-Planck-Institut f\"ur Astronomie, K\"onigstuhl 17, 69117 Heidelberg, Germany\\
$^{5}$Department of Physics and Astronomy, University of California, Los Angeles, 430 Portola Plaza, Los Angeles, CA 90095, USA\\
$^{6}$DARK, The Niels Bohr Institute, University of Copenhagen, Jagtvej 155, DK-2200 Copenhagen N, Denmark\\
$^{7}$Steward Observatory, University of Arizona, 933 North Cherry Avenue, Tucson, AZ 85718, USA\\
$^{8}$Institute of Astronomy and Astrophysics, Academia Sinica, 11F of Astronomy-Mathematics Building, No.\,1, Sec.\ 4, Roosevelt Rd, Taipei 106319, Taiwan, R.O.C.\\
$^{9}$MIT Kavli Institute for Astrophysics and Space Research, 77 Massachusetts Avenue, Cambridge, MA 02139, USA\\
$^{10}$Department of Physics, Massachusetts Institute of Technology, 77 Massachusetts Avenue, Cambridge, MA 02139, USA\\
$^{11}$Hamburg Observatory, University of Hamburg, Gojenbergsweg 112, D-21029 Hamburg, Germany\\
$^{12}$Department of Astronomy, University of Michigan, 1085 South University Avenue, Ann Arbor, MI 48109, USA\\
$^{13}$Department of Astronomy, University of Geneva, Chemin Pegasi 51, 1290 Versoix, Switzerland\\
$^{14}$Center for Astrophysics $\vert$ Harvard \& Smithsonian, Cambridge, MA 02138, USA\\
$^{15}$Departament d'Astronomia i Astrof\'isica, Universitat de Val\`encia, C. Dr. Moliner 50, E-46100 Burjassot, Val\`encia, Spain\\
$^{16}$Unidad Asociada CSIC ``Grupo de Astrof\'isica Extragal\'actica y Cosmolog\'ia'' (Instituto de F\'isica de Cantabria -- Universitat de Val\`encia), Spain
}

\date{Accepted XXX. Received YYY; in original form ZZZ}

\pubyear{\the\year{}}

\begin{document}
\label{firstpage}
\pagerange{\pageref{firstpage}--\pageref{lastpage}}
\maketitle

\begin{abstract}
The extreme bright end of the galaxy UV luminosity function during reionisation remains poorly constrained, particularly where the galaxy and quasar luminosity functions overlap and source classification becomes ambiguous. We present \textit{JWST}/NIRSpec and ALMA Band-6 observations of J1450$-$0144 ($z=6.627$) and J1429$-$0104 ($z=6.796$), two $M_{\rm UV}\simeq-23.5$ sources originally classified as faint quasars by SHELLQs. NIRSpec reveals blue UV continua, strong P~Cygni profiles in \ion{N}{v}, \ion{Si}{iv}, and \ion{C}{iv}, broad \ion{He}{ii}\,$\lambda1640$ emission with rest-frame equivalent widths of $8.8\pm1.2$ and $3.7\pm1.1$\,\AA, respectively, and narrow nebular lines, reclassifying both as extremely UV-luminous galaxies. Standard population-synthesis models cannot simultaneously reproduce the strong \ion{He}{ii} and wind features, whereas models incorporating very massive stars (VMS; $M\gtrsim100\,M_\odot$) with dedicated wind prescriptions can. These models favor a star-formation duration of $2$--$4$\,Myr for J1450$-$0144, with a broader allowed range for J1429$-$0104, stellar masses of $\log(M_\star/M_\odot)\approx9.2$--$9.9$, and star-formation rates of $\simeq300$--$540\,M_\odot\,{\rm yr}^{-1}$. Under the same VMS wind models, equivalent-width diagnostics imply $M_{\rm up}\gtrsim225\,M_\odot$ for J1429$-$0104, while J1450$-$0144 lies beyond even the $M_{\rm up}=475\,M_\odot$ grid. ALMA detects luminous [\ion{C}{ii}]\,$158\,\mu$m emission in both systems, with $L_{\rm [CII]}\approx0.8$ and $4.1\times10^{9}\,L_\odot$, respectively. J1429$-$0104 additionally shows bright dust continuum, with both [\ion{C}{ii}] and dust offset by $\sim5.4$\,kpc from its UV emission. These sources demonstrate that VMS can power some of the most UV-luminous galaxies at cosmic dawn and show that source classifications, and hence the inferred demographics of both galaxies and quasars in the crossover regime, need revisiting.
\end{abstract}




\section{Introduction}\label{sec:intro}

The bright end of the galaxy UV luminosity function (UVLF) during the epoch of reionisation ($z \gtrsim 6$) remains poorly constrained. At $z \sim 7$, the bright end of the UVLF in \citet{Harikane2025} is constrained by only four galaxies with $M_{\rm UV} < -23$, and in their brightest bin, $-24.5 < M_{\rm UV} < -23.5$, the number density is only a lower limit, based on a single galaxy. With so few sources contributing, the two LF forms considered in \citet{Harikane2025} give rather different number densities in these luminosity regimes: at $M_{\rm UV} = -23$ the double power-law (DPL) gives $154~\mathrm{Gpc}^{-3}\,\mathrm{mag}^{-1}$ against $40~\mathrm{Gpc}^{-3}\,\mathrm{mag}^{-1}$ for the Schechter function, and at $M_{\rm UV} = -24$, the extrapolation of the two diverge by four orders of magnitude, with the DPL giving $\sim5~\mathrm{Gpc}^{-3}\,\mathrm{mag}^{-1}$ and the Schechter function $\sim4 \times 10^{-4}~\mathrm{Gpc}^{-3}\,\mathrm{mag}^{-1}$. Even finding a modest sample of ten sources at $-24.5 < M_{\rm UV} < -23.5$ would require surveying $\sim160~\mathrm{deg}^2$ if the bright end follows the DPL, or close to the entire sky if it follows the Schechter function, far beyond the pencil-beam fields of \textit{JWST} deep imaging programmes.
However, the galaxies that do reach this regime are among the most extreme actively star-forming systems known at any epoch, with rest-frame UV luminosities ($\nu L_\nu\gtrsim10^{45}\,{\rm erg\ s^{-1}}$ at 1500\,\AA) corresponding to star-formation rates of $\sim 100$--$1000\,\msun\,{\rm yr}^{-1}$. 
Whether such rates can be accommodated by standard star-formation prescriptions, or whether they instead require new ingredients, is unresolved. Proposed routes to sustain such intense UV emission include an elevated baryon-to-star conversion efficiency, in which a larger fraction of a halo's baryons are converted into stars \citep{Dekel2023, Li2024}; a non-standard IMF, either top-heavy or hosting substantial populations of very massive stars, raising the UV output per unit stellar mass formed \citep{Trinca2024, Cueto2024, Schaerer2025}, stochastic, bursty star-formation histories, under which these galaxies are captured at the peak of their UV fluctuations \citep{Mason2023, Sun2023, Pallottini2023, Shen2023}, and radiation-driven outflows that decouple dust from the UV-emitting stars and reduce attenuation \citep{Ferrara2023, Ziparo2023, Ferrara2025}. Many of these mechanisms have also been invoked to explain the unexpected abundance of UV-luminous galaxies at $z > 10$ uncovered by \textit{JWST} \citep[e.g.,][]{Naidu2022, Castellano2022, Harikane2023a, Finkelstein2023, Robertson2024}. Those galaxies are fainter ($M_{\rm UV} \approx -19$ to $-22$) than the $M_{1450} \lesssim -23$ regime considered here, and at those redshifts individual sources rarely yield the spectroscopic and far-infrared signal-to-noise ratio needed to discriminate between the proposed mechanisms. Sources at the bright end of the reionisation-epoch UVLF are therefore the natural testbeds for these theories, where detailed multi-wavelength follow-up of individual systems becomes feasible.

Observational constraints on these mechanisms have so far been limited due to the small existing spectroscopic sample of bright galaxies during cosmic dawn. Identifying galaxies at $M_{1450} \lesssim -23$ at $z > 6$ in meaningful numbers requires hundreds to thousands of square degrees of imaging, accessible only to wide-area ground-based or space-based surveys. ALMA programmes such as REBELS \citep{Bouwens2022} have followed up UV-bright candidates at $z \sim 7$ 
across $\sim 7$\,deg$^2$ of the COSMOS, UDS, and XMM-LSS fields, yielding detailed [\Cii]\,$158\,\mu$m and dust-continuum constraints, but reach only $M_{\rm UV} \approx -22.9$ at their brightest 
and do not extend to $M_{1450} \lesssim -23$. \citet{Harikane2025} recently compiled a broader sample of spectroscopically confirmed bright galaxies at $z \sim 7$--$14$ drawn from a wide range of programmes, covering $\gtrsim 100$\,deg$^2$ in total. While they showed that the bright end of the $z \sim 7$ UV luminosity function is in excess of a Schechter extrapolation and better described by a DPL, the compilation contains only a handful of sources at $M_{1450} \lesssim -23$ and $z > 6$: HSC J023536$-$031737 \citep[$z = 6.913$, $M_{\rm UV} = -23.6$;][]{Ono2018, Harikane2022a}, UVISTA-304416 \citep[$z = 7.060$, $M_{\rm UV} = -23.1$;][]{Bowler2017, Schouws2023}, HSC J084916+005311 \citep[$z = 6.606$, $M_{\rm UV} = -23.1$;][]{Harikane2022a, Harikane2025}, and HSC J091519$-$012630 \citep[$z = 7.0$, $M_{\rm UV} = -23.9$;][]{Harikane2022a, Harikane2025}. Several of these sources, however, still lack rest-frame UV spectroscopy of sufficient depth and wavelength coverage to characterise their stellar population, leaving both the AGN contamination and the physical drivers of the UV output open questions.

At the same time, the Subaru High-$z$ Exploration of Low-Luminosity Quasars survey \citep[SHELLQs;][]{Matsuoka2016, Matsuoka2018a,Matsuoka2018b,Matsuoka2019,Matsuoka2022,Matsuoka2025b} has also yielded a substantial sample of UV-bright galaxy candidates with $-22>M_{\rm 1450}>-25$ at $z\sim6$ during 
their quasar searches. \citet{Marques-Chaves2025} further stacked a subset of these galaxies with strong but narrow Ly$\alpha$ lines and showed that these sources are consistent with very young ($\sim 6\,{\rm Myr}$) starbursts. However, the luminosity regime of $M_{1450} \approx -23$ to $-25$ is exactly where the faint end of the quasar luminosity function and the extreme bright end of the galaxy UVLF intersect, and a blue UV continuum paired with a prominent, moderately broad Ly$\alpha$ line 
is consistent with either a faint type-I quasar or an intensely star-forming galaxy. Distinguishing the two populations on the basis of discovery spectroscopy with low sensitivity alone is inherently ambiguous \citep[e.g.,][]{Matsuoka2018a, Harikane2025}. As a result, some fraction of sources classified as faint quasars may in fact be galaxies, and vice versa, with direct consequences for both the faint-end slope of the quasar luminosity function and the bright-end slope of the galaxy luminosity function.
 
Despite the observational difficulties at $z > 6$, similarly bright galaxies at $z \sim 2$--$4$ have been uncovered and studied in detail, dubbed the Extreme UV Luminous Galaxies \citep[EUVLGs;][]{Marques-Chaves2020, Marques-Chaves2021, Marques-Chaves2022, Upadhyaya2024, Dessauges-Zavadsky2025}. These galaxies 
have $M_{\rm UV}$ between $-23$ and $-24.5$ and SFRs of several hundred $\msun\,{\rm yr}^{-1}$. 
Their rest-frame UV spectra display blue continua, prominent P~Cygni profiles in the high-ionisation wind lines 
\Nv\,$\lambda1240$, \Siiv\,$\lambda1400$, and \Civ\,$\lambda1550$, and narrow nebular emission consistent with 
\Hii-region rather than AGN excitation. SED modelling 
returns stellar ages of $\lesssim 10\,$Myr and specific star-formation rates of $\sim 100\,{\rm Gyr}^{-1}$, identifying EUVLGs as short-lived but extreme starburst phases. Cosmic-noon EUVLGs thus provide a template for what a bright, AGN-free star-forming galaxy looks like in this regime.

One of the most striking findings from these EUVLGs is the frequent detection of strong, broad \Heii\,$\lambda1640$ emission, with ${\rm EW_0}\simeq1.4$--$4.6$\,\AA\ \citep{Upadhyaya2024}. Conventional 
population-synthesis models with IMF upper-mass cutoffs of $\sim 100\,\msun$ cannot reproduce these signatures, which are instead characteristic of the dense, radiatively driven winds of very massive stars \citep[VMS; $M \gtrsim 100\,\msun$; see][for a review]{Vink2015}. Comparable broad and strong \Heii\ features are seen most directly in the R136 cluster in the Large Magellanic Cloud, where individual VMS have been resolved and identified \citep{Crowther2010, Crowther2016}, and in other local VMS-dominated star clusters or star-forming regions, such as Mrk\,71-A, II\,Zw\,40-A, and SB\,126 \citep{Leitherer2018, Senchyna2021, Smith2023, Martins2023}. 
Most recently, \citet{Marques-Chaves2026} reported direct evidence for VMS in the reionisation epoch using data from the JWST program SPectroscopic Ultra-deep Reionisation-era Survey (SPURS, GO~9214; PIs: C. Mason \& D. Stark). They reported broad \Heii\ emission and N\,{\sc v}/\Civ\ P~Cygni profiles in two $z \simeq 8.7$ galaxies, CEERS-1019 ($M_{\rm UV} = -22.4$, \Heii\ EW $= 4.53 \pm 0.36$~\AA) and CEERS-1025 ($M_{\rm UV} = -21.2$, \Heii\ EW $= 4.09 \pm 0.71$~\AA), that are reproduced only by models including VMS. From the same programme, ultra-deep spectroscopy of GN-z11 ($M_{\rm UV} = -21.6$, $z = 10.6$) has revealed P~Cygni wind features and broad \Heii\ emission that are jointly reproduced by models incorporating VMS at low metallicity and young ages ($\lesssim 3$\,Myr), extending the evidence for VMS to $z > 10$ \citep{Chen2026}. These discoveries have hinted at a potential link between extreme UV-brightness and the presence of VMS and/or atypical IMF emergence.


In this work, we address these questions using two exceptionally UV-luminous sources in the epoch of reionisation. We present \textit{JWST}/NIRSpec and ALMA Band-6 observations of \yswl\ ($z = 6.627$) and \ysej\ ($z = 6.796$), originally classified as faint quasars by the SHELLQs survey \citep{Matsuoka2018a, Matsuoka2022}. The new NIRSpec spectra show that both objects are better interpreted as extreme UV-luminous galaxies, rather than unobscured
quasars, with rest-frame UV emission dominated by very young, intensely star-forming stellar populations. With $M_{1450} \approx -23.5$, these sources lie among 
the UV-brightest spectroscopically confirmed objects known at $z \gtrsim 6.5$. We use the combination of rest-UV spectroscopy, ALMA [\Cii]\,$158\,\mu$m observations, and dust-continuum constraints to investigate the physical nature of these sources and their implications for the abundance of EUVLGs at $z\sim7$.

The paper is organised as follows. Sect.~\ref{sec:observation} describes the discovery of \yswl\ and \ysej\ and the follow-up JWST/NIRSpec and ALMA observations. In Sect.~\ref{sec:results} we establish the EUVLG nature of both sources, model their stellar populations which could contain a substantial fraction of very massive stars, present the ALMA gas and dust measurements, and derive a lower limit on the EUVLG number density at $z\sim7$. Sect.~\ref{sec:discussion} discusses the combined UV--FIR view of the two systems and the implications of their VMS content, and we summarise our findings in Sect.~\ref{sec:conclusions}. Throughout this work we adopt a flat $\Lambda$CDM cosmology with $H_0 = 70\,{\rm km\,s}^{-1}\,{\rm Mpc}^{-1}$, $\Omega_{\rm m} = 0.3$, and $\Omega_\Lambda = 0.7$. All magnitudes are given in the AB system.

\section{Discovery and follow-up observations of \ysej\ \& \yswl}\label{sec:observation}

\subsection{Initial discovery}\label{sec:obs_discovery}
\ysej\ and \yswl\ were first discovered in the Subaru High-$z$ Exploration of Low-Luminosity Quasars \citep[SHELLQs; e.g.,][]{Matsuoka2016} program.
\ysej\ was identified as a quasar at $z=6.8$ using the Optical System for Imaging and low-Intermediate-Resolution Integrated Spectroscopy \citep[OSIRIS;][]{Cepa2000osiris} on the Gran Telescopio Canarias (GTC). Its UV magnitude was estimated to be $M_{\rm 1450}\approx -23$, and its Ly$\alpha$ line was measured to have ${\rm FWHM}=1400\pm100\ {\rm km\,s^{-1}}$ and $L_{\rm Ly\alpha}\approx10^{43.95}\ {\rm erg\,s^{-1}}$ \citep{Matsuoka2018a}. \yswl\ was identified as a quasar at $z=6.63$ with $M_{\rm 1450}\approx -23.73$ using the Faint Object Camera and Spectrograph \citep[FOCAS;][]{Kashikawa2002FOCAS} on Subaru. Its Ly$\alpha$ line was measured to have ${\rm FWHM}=900\pm300\ {\rm km\,s^{-1}}$ and $L_{\rm Ly\alpha}\approx10^{43.48}\ {\rm erg\,s^{-1}}$ \citep{Matsuoka2022}. The classification of both sources was based on the onset of a prominent Ly$\alpha$ emission line and their blue continua. In the discovery spectrum of \ysej, a dip between the Ly$\alpha$ and \Nv\ lines was reported and interpreted as absorption associated with a broad absorption line (BAL) system \citep{Matsuoka2018a}.

A summary of the basic properties of \ysej\ and \yswl\ is presented in Table~\ref{tab:source_properties}. Note that we have updated the redshifts using the [\Cii] line from our ALMA observations (Sect.~\ref{sec:alma}) and $M_{1450}$ from our JWST/NIRSpec data (Sect.~\ref{sec:specmod}). In Fig.~\ref{fig:z_muv} we show the two galaxies in the redshift versus $M_{\rm UV}$ plane with these updated values.

\subsection{JWST/NIRSpec observation}\label{sec:obs_jwst}

\subsubsection{Observations and data reduction}

Both sources were later observed with JWST as part of the General Observers (GO) program 9180 (PI: J.~Hennawi) using JWST/NIRSpec fixed-slit (FS) mode with the G140H/F070LP and G235H/F170LP configurations. 
The aim of the program include measuring the cosmic reionisation history from the damping wing signatures on the spectra of 27 faint quasars at $6.5 < z < 7.7$. Twelve of the 27 targets are discovered by \textit{Euclid} \citep{Yang2026}, while the remainder are drawn primarily from SHELLQs, with preference given to sources with existing ALMA coverage (see Sect.~\ref{sec:obs_alma}). These setups provide a spectral resolution of $R\sim2700$ and a continuous wavelength coverage of $0.81$--$3.05\,\mu{\rm m}${\footnote{For typical targets, the red end of the G140H/F070LP configuration is truncated at $1.27\,\mu{\rm m}$ because of second-order contamination. The targets in this work lie at $z>6.5$ and therefore possess a strong IGM absorption trough blueward of Ly$\alpha$. As a result, second-order contamination does not impact our observations, enabling continuous spectral coverage.}}, with both S200A1 and S200A2 used to bridge the inter-detector gap. For the two sources presented in this work, these observations enable rest-frame coverage from Ly$\alpha$ to approximately 4000\,\AA. The total exposure times for \ysej\ are 9103\,s and 5194\,s for the G140H and G235H setups, respectively, while those for \yswl\ are 3268\,s and 2480\,s.

We reduced the JWST/NIRSpec fixed-slit data starting from the \texttt{\_uncal} exposures retrieved from MAST, using the JWST calibration pipeline (v1.20.2; CRDS context \texttt{jwst\_1464.pmap}). Stage~1 (\texttt{calwebb\_detector1}) performed the standard detector-level calibration including the \texttt{clean\_flicker\_noise} correction to produce \texttt{\_rate} files. Stage~2 (\texttt{calwebb\_spec2}) carried out the 2D spectroscopic calibration, including the wavelength calibration, flat-fielding, pathloss corrections, skipping the default background subtraction, 1D extraction, and resampling steps. Instead, sky subtraction and spectral extraction were performed with a custom interface between the Stage~2 products and \texttt{PypeIt} \citep{Prochaska2020}. 
For each slit, 2D science images, inverse-variance maps, and bad-pixel masks were constructed from the pipeline calibration arrays, with a custom variance model propagated from the JWST Poisson and read-noise estimates and an empirical rescaling factor of 1.5. The sky was modelled directly from the slit using \texttt{PypeIt}'s global B-spline algorithm. Sources were identified on the sky-subtracted 2D images and extracted using optimal extraction \citep{Horne1986}. The individual 1D spectra from separate exposures and disperser settings were coadded with \texttt{pypeit\_coadd\_1dspec} onto a common wavelength grid. The final 1D spectra are displayed in Fig.~\ref{fig:spec_compare}.

\begin{table*}
\centering
\caption{Source properties. Coordinates are J2000. The systemic redshifts $z_{\mathrm{[C\textsc{ii}]}}$ are measured from the ALMA [\Cii]\,$158\,\mu$m line (Sect.~\ref{sec:alma}). The $y$-band magnitudes $y_{\rm AB}$ are taken from the SHELLQs photometry \citep{Matsuoka2018a, Matsuoka2022} and set the absolute flux scale of the NIRSpec spectra (Sect.~\ref{sec:specmod}). $M_{1450}^{\rm raw}$ is the absolute UV magnitude measured directly from the NIRSpec spectrum as observed through the slit, whereas $M_{1450}$ is measured after rescaling the spectrum to the $y$-band photometry and is the value adopted throughout this work. The difference for \ysej\ is substantial, consistent with its extended morphology (Sect.~\ref{sec:feature_morph}). The spectral slope $\beta$ ($f_\lambda \propto \lambda^\beta$) is measured from the rescaled spectra as described in Sect.~\ref{sec:feature_continua}.}
\label{tab:source_properties}
\begin{tabular}{lccccccc}
\hline
Source & R.A. & Dec. & $z_{\mathrm{[C\textsc{ii}]}}$ & $y_{\rm AB}$ & $M_{1450}^{\rm raw}$ & $M_{1450}$ & $\beta$ \\
\hline
\yswl & 14:50:05.39 & $-$01:44:38.9 & $6.6270\pm0.0018$ & $23.45\pm0.08$ & $-23.23\pm 0.02$ & $-23.45\pm0.08$ & $-2.20\pm 0.09$ \\
\ysej & 14:29:03.08 & $-$01:04:43.4 & $6.7964\pm0.0002$ & $23.73\pm0.09$ & $-22.68\pm0.01$ & $-23.45\pm0.09$ & $-2.06\pm 0.07$ \\
\hline
\end{tabular}
\end{table*}

\subsubsection{Spectral modelling}\label{sec:specmod}

Before performing any measurements from the NIRSpec spectra we apply two preparatory steps: an absolute flux correction with Subaru photometry information, and the construction of a continuum model that is subsequently used for all line and equivalent-width measurements.

The apparent UV magnitudes measured directly from the NIRSpec spectra are $M_{1450}^{\rm raw}=-23.23$ for \yswl\ and $-22.68$ for \ysej\ (Table~\ref{tab:source_properties}). Due to the potential slit-losses, we rescale each spectrum to match the Subaru/HSC $y$-band magnitudes from \citet{Matsuoka2018a,Matsuoka2022}, corresponding to multiplicative flux corrections of $\simeq1.2$ for \yswl\ and $\simeq2.0$ for \ysej. The large correction factor for \ysej\ is consistent with its extended morphology (see Sect.~\ref{sec:feature_morph}). The rescaled spectra, which give the adopted $M_{1450}=-23.45$ for both sources, are used for all subsequent measurements. We note that the HSC imaging was taken several years ago, so this rescaling attributes the entire offset to aperture losses, ignoring any intrinsic variability between the two epochs. Future JWST/NIRCam imaging would help clarify this issue.

We model the rest-frame UV/optical continuum as a smoothing cubic spline (\texttt{scipy.interpolate.UnivariateSpline}), fit directly to the inverse-variance-weighted JWST spectrum after masking out possible line features. The smoothing parameter is chosen empirically so that the reduced $\chi^2$ over the line-free pixels is close to unity while keeping the model free of structure on the scale of individual emission features. Continuum uncertainties are propagated by repeating the fit on 500 Monte Carlo realisations of the spectrum, each generated by perturbing every pixel by a Gaussian draw from its noise vector. The best-fit continuum is taken as the median across realisations. We show the resulting continuum models, together with the per-pixel residuals, in Appendix~\ref{app:continuum}.

We fit the following emission lines: \Niv]\,$\lambda1486$, \Heii\,$\lambda1640$, \Oiii]\,$\lambda1666$, \Ciii]\,$\lambda1907,1909$, \Mgii\,$\lambda2796,2804$, \Oii\,$\lambda\lambda3727,3730$, and \Neiii\,$\lambda\lambda3870,3969$ with the continuum-subtracted spectrum using \texttt{lmfit} \citep{Newville2025}. 
We fit with Gaussian line models with the line redshift, amplitude, and width as free parameters. The redshift is allowed to vary within $\Delta z = \pm 0.02$ of the systemic value ($z_{\rm [CII]}$). For known doublets (\Niv], \Oiii], \Ciii], \Mgii, \Oii), we fit with two Gaussians with a shared $\sigma$ and redshift but independent amplitudes. 
All derived FWHMs are deconvolved from the line-spread function assuming a Gaussian instrumental profile with $R = 2700$. Uncertainties on all fitted quantities are estimated from 500 Monte Carlo realisations in which the science spectrum is perturbed by its per-pixel error and refit with the continuum held fixed. We quote the 16th and 84th percentiles of the resulting distributions as the asymmetric $1\sigma$ errors. The best-fit models are displayed in Fig.~\ref{fig:all_nebular_lines} and the measurements are shown in Table~\ref{tab:nebular_lines}.

\subsection{ALMA observations}\label{sec:obs_alma}

These two sources were also observed by ALMA in the cycle-10 program CISTERN ([C\,{\sc ii}]-SelecTed Early universe RefereNce, PI: Bouwens; Bouwens et~al.~2026, in prep), targeting the [C\,{\sc ii}]$_{\rm 158\mu m}$ line from 92 $UV$-bright sources with spectroscopic redshifts.  The integration time for \ysej\ and \yswl\ were 5.0 minutes and 5.5 minutes, respectively, yielding a line sensitivity of 0.5 mJy beam$^{-1}$
per 66 km s$^{-1}$ channel and a continuum sensitivity of 73 $\mu$Jy beam$^{-1}$.
This enables the detection of [C\,{\sc ii}] ISM cooling lines at 5$\sigma$ down to luminosities of $\sim$4$\times$10$^{8}$ $L_{\odot}$ (corresponding to star-formation rates of 40 $M_{\odot}$ yr$^{-1}$ based on the \citet{deLooze2014} SFR--L$_{\rm [CII]}$ relation). 
A synthesized beam size larger than 0.9$^{\prime\prime}$ was requested to avoid resolving out the typically extended [CII] emission observed in luminous star-forming galaxies and quasars at $z\gtrsim6$. Additional details on the observations and source-by-source results will be presented in the forthcoming CISTERN survey paper (R.J. Bouwens et al. 2026, in prep.).

The ALMA data for each target were processed as follows. Measurement sets were generated from the QA2 delivery products using the ScriptForPI.py script within CASA \citep{CASA}. The visibilities were averaged in 30-second intervals and subsequently reimaged with the \textsc{tclean} task using natural weighting. This weighting scheme maximizes the signal-to-noise ratio achievable for both line and continuum detections in CISTERN, providing a modest improvement relative to the standard JAO reductions, which adopt a Briggs weighting with a \textsc{robust} parameter of 0.5. For continuum measurements, frequency ranges within $\pm$0.4 GHz of any candidate ($>$5$\sigma$) [CII] line detections were excluded when generating continuum images of the targeted quasars.

We determine the systemic redshifts of \yswl\ and \ysej\ from the [\Cii]\,$158\,\mu$m line detected in our ALMA observations, with values reported in Table~\ref{tab:source_properties}. Details of the [\Cii] line measurement are given in Sect.~\ref{sec:alma} and Appendix~\ref{app:alma_measurements}.

\section{Results and Analysis}\label{sec:results}


\subsection{Nature of these two sources: Extreme UV Luminous Galaxies (EUVLGs) at cosmic dawn\label{sec:euvlg}}

With the JWST/NIRSpec spectra in hand, we are now in a position to revisit the nature of \yswl\ and \ysej. Fig.~\ref{fig:spec_compare} shows the G140H$+$G235H spectra of both sources, covering rest-frame wavelengths from Ly$\alpha$ to $\sim 4000$\,\AA. Despite their quasar-like UV luminosities ($M_{1450} \approx -23.5$), neither spectrum resembles that of a typical type-I quasar. Instead, both display the hallmark features of EUVLG population, including blue UV continua, prominent P~Cygni profiles in high-ionisation wind lines (N\,{\sc v}\,$\lambda 1240$, Si\,{\sc iv}\,$\lambda 1400$, and C\,{\sc iv}\,$\lambda 1550$, N\,{\sc iv}\,$\lambda 1719$), broad \Heii\,$\lambda 1640$ emission, and strong narrow nebular lines (e.g., \Oii\,$\lambda\lambda 3727, 3730$, \Neiii\,$\lambda\lambda 3870, 3969$). To make this comparison straightforward, Fig.~\ref{fig:spec_compare_euvlg_r136} shows the Ly$\alpha$ to \Heii\ portion of each spectrum normalised by the local continuum, overlaid with the cosmic-noon EUVLG composite of \citet{Upadhyaya2024} (top) and the stacked spectrum of the seven most massive stars ($M_{\rm init}\sim100-300\,M_\odot$) in R136 from \citet{Crowther2016} (bottom). Both \yswl\ and \ysej\ closely match the EUVLG composite and resemble the composite spectrum of the seven very massive stars in the R136. The P~Cygni wind profiles in the high-ionisation resonance lines recall classical Wolf--Rayet (WR) stars. They are evolved massive stars whose dense, radiatively driven winds have stripped the hydrogen envelope to expose core-helium-burning layers. The features here, however, are not those of classical WR: as in R136, the strong broad \Heii, together with the P~Cygni lines, point instead to the H-rich WR stars, usually denoted as the WNh subclass. These stars are luminous hydrogen-rich stars close to the Eddington limit whose dense winds produce a WR-like emission spectrum while they still burn hydrogen in their cores, i.e. they are still in the main sequence. Both the P~Cygni profiles and the broad \Heii\ emission arise in these winds, but only the resonance lines (\Nv, \Siiv, \Civ, \Niv) develop the blueshifted P~Cygni absorption, whereas the non-resonant \Heii\, appears predominantly in emission and only occasionally as a P~Cygni profile. As suggested by many works \citep[e.g.,][]{Vink2015,Crowther2016}, very massive stars on the main sequence appear spectroscopically as WNh stars, so these features are hinting that VMS contribute substantially to the UV spectra, which we develop further in Sect.~\ref{sec:sps_vms} and \ref{sec:vms_discussion}.

\begin{figure}
	\includegraphics[width=\linewidth]{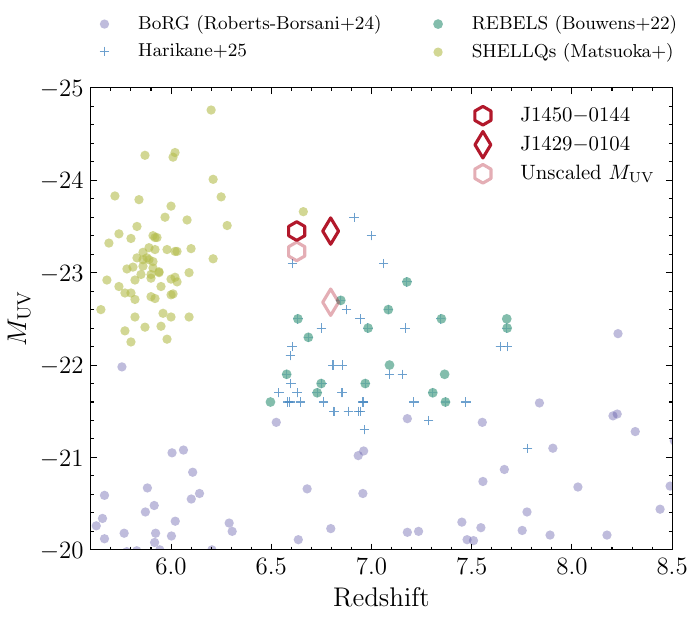}
    \caption{The two galaxies presented in this work (red) in the $z$--$M_{\rm UV}$ plane, compared with literature samples at $z \sim 6$--8.5: SHELLQs galaxies \citep{Matsuoka2016,Matsuoka2018a,Matsuoka2018b,Matsuoka2019,Matsuoka2022}, REBELS galaxies \citep{Bouwens2022}, compilation samples in \citet{Harikane2025}, and BoRG-JWST galaxies \citep{Roberts-Borsani2024,Roberts-Borsani2025}. Our sources lie among the UV-brightest spectroscopically confirmed galaxies known at $z > 6.5$.}
    \label{fig:z_muv}
\end{figure}

\begin{figure*}
	\includegraphics[width=\linewidth]{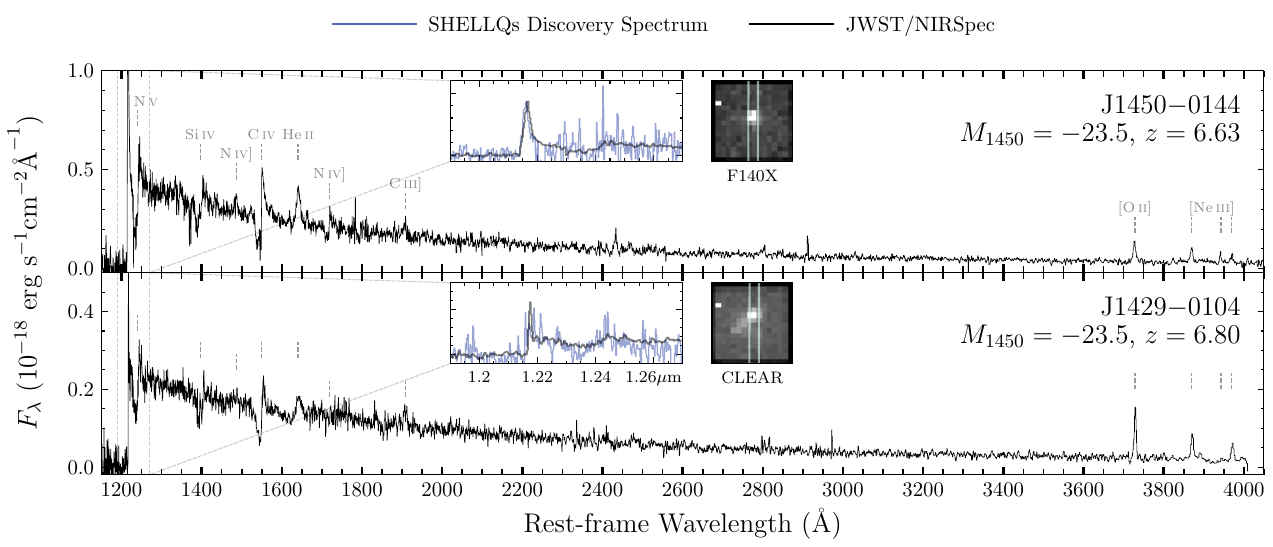}
    \caption{JWST/NIRSpec of \yswl\ (top, $z = 6.63$) and \ysej\ (bottom, $z = 6.80$), covering $\sim$1200--4000\,\AA\ in rest-frame. Black curves show the JWST/NIRSpec G140H$+$G235H spectra, blue curves in the inset show the original SHELLQs discovery spectra from GTC/OSIRIS and Subaru/FOCAS, respectively \citep{Matsuoka2018a, Matsuoka2022}. Both sources exhibit the characteristic features of extreme UV-luminous star-forming galaxies: blue UV continua, prominent P~Cygni profiles in the high-ionisation wind lines \Nv\,$\lambda1240$, \Siiv\,$\lambda1400$, and \Civ\,$\lambda1550$, broad \Heii\,$\lambda1640$ emission, absence of broad Ly$\alpha$ and \Mgii, and narrow nebular lines including \Oii\,$\lambda\lambda3727,3730$ and \Neiii\,$\lambda\lambda3870,3969$. The inset images on the right show the NIRSpec WATA stamps ($1.6'' \times 1.6''$) obtained with the F140X filter (\yswl) and CLEAR filter (\ysej). \ysej\ shows extended morphology. }
    \label{fig:spec_compare}
\end{figure*}

\begin{figure*}
	\includegraphics[width=\linewidth]{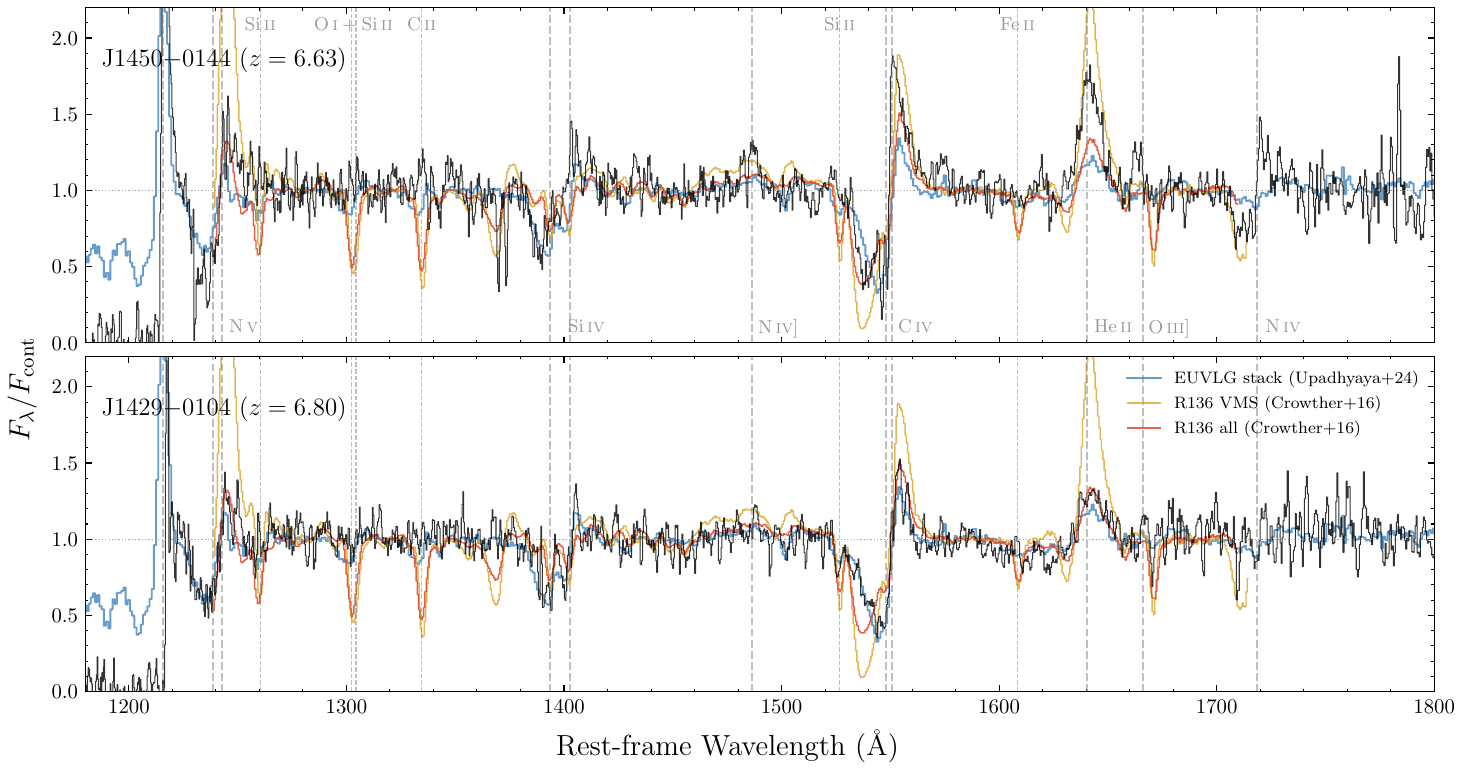}
    \caption{Rest-frame UV spectra of \yswl\ (top; $z=6.63$) and \ysej\ (bottom; $z=6.80$), each overlaid with the cosmic-noon EUVLG composite of \citet{Upadhyaya2024} (blue), the stacked spectrum of the seven most massive stars ($M_{\rm init}\sim100$--$300\,M_\odot$) in the R136 cluster (yellow) and the entire R136 (red) from \citet{Crowther2016}. Black curves show the JWST/NIRSpec spectra for the two galaxies, smoothed with an inverse-variance-weighted 3-pixel window. All spectra are normalised by the local UV continuum. Key high-ionisation wind and nebular features are labelled. Both reionisation-epoch galaxies closely resemble the cosmic-noon EUVLG composite and the R136 VMS spectrum across the high-ionisation P~Cygni lines. See Sect.~\ref{sec:euvlg} for further discussion.}
    \label{fig:spec_compare_euvlg_r136}
\end{figure*}

Finally, with the updated redshifts (Sect.~\ref{sec:alma}) and rest-frame UV magnitudes ($M_{1450} \approx -23.5$), they sit among the UV-brightest spectroscopically confirmed galaxies known at $z>6.5$, as shown in Fig.~\ref{fig:z_muv}. These UV luminosities correspond to unobscured star-formation rates of $\mathrm{SFR_{UV}} \simeq 135\pm10\,M_\odot\,{\rm yr}^{-1}$ for both sources, adopting $\kappa_{\rm UV}=1.3\times10^{-28}\,{\rm M_\odot\,yr^{-1}/(erg\,s^{-1}\,Hz^{-1})}$ from \citet{Upadhyaya2024}, derived for the young stellar populations of cosmic-noon EUVLGs. However, as shown by \citet{Schaerer2025}, for such young populations, and especially when VMS are present, $\kappa_{\rm UV}$ is a strong function of stellar age, declining by roughly an order of magnitude from $\simeq10\times10^{-28}$ at $\sim1$\,Myr to $\simeq1\times10^{-28}$ after $\sim10$\,Myr of constant star formation. The $\mathrm{SFR_{UV}}$ reported here should therefore be regarded as a first-order estimate.


\begin{table}
\centering
\caption{Emission-line properties measured from Gaussian fits to the continuum-subtracted JWST/NIRSpec spectra (Sect.~\ref{sec:feature_nebularline}). Lines marked with $^{\dagger}$ are fitted as doublets (see Sect.~\ref{sec:specmod}). Their $\log L$ and EW$_{0}$ are summed over both components, while FWHM and $\Delta v$ are shared. The \Neiii\ $\lambda3969$ line (marked with $^{\ddagger}$) is blended with H$\epsilon$ $\lambda3971$ and fitted as a single Gaussian, the fitted values therefore absorb the H$\epsilon$ contribution.}
\label{tab:nebular_lines}
\renewcommand{\arraystretch}{1.4}
\begin{tabular}{lcccc}
\hline
Line & FWHM & $\log L$ & EW$_{0}$ & $\Delta v$ \\
     & (km\,s$^{-1}$) & (erg\,s$^{-1}$) & (\AA) & (km\,s$^{-1}$) \\
\hline
\multicolumn{5}{c}{{\yswl}} \\
\hline
N\,{\sc iv}] $\lambda1486^{\dagger}$ & $1477^{+209}_{-825}$ & $42.42^{+0.06}_{-0.13}$ & $1.9^{+0.3}_{-0.5}$ & $+653^{+0}_{-602}$ \\
\Heii\ $\lambda1640$ & $1906^{+119}_{-135}$ & $42.96^{+0.02}_{-0.03}$ & $8.1^{+0.5}_{-0.5}$ & $+227^{+51}_{-53}$ \\
O\,{\sc iii}] $\lambda1666^{\dagger}$ & $813^{+91}_{-218}$ & $42.16^{+0.09}_{-0.08}$ & $1.3^{+0.3}_{-0.2}$ & $-509^{+846}_{-18}$ \\
\Ciii]\ $\lambda1909^{\dagger}$ & $1270^{+485}_{-552}$ & $42.41^{+0.10}_{-0.12}$ & $3.2^{+0.8}_{-0.8}$ & $-294^{+243}_{-144}$ \\
Mg\,{\sc ii}\ $\lambda2800^{\dagger}$ & $527^{+44}_{-66}$ & $42.21^{+0.05}_{-0.05}$ & $4.8^{+0.5}_{-0.5}$ & $-10^{+48}_{-56}$ \\
\Oii\ $\lambda3728^{\dagger}$ & $666^{+67}_{-96}$ & $42.63^{+0.03}_{-0.03}$ & $25.7^{+1.8}_{-2.0}$ & $-67^{+37}_{-13}$ \\
\Neiii\ $\lambda3870$ & $508^{+68}_{-70}$ & $42.40^{+0.04}_{-0.04}$ & $13.9^{+1.5}_{-1.3}$ & $+21^{+25}_{-30}$ \\
\Neiii\ $\lambda3969^{\ddagger}$ & $657^{+113}_{-134}$ & $42.20^{+0.07}_{-0.08}$ & $11.1^{+1.9}_{-1.8}$ & $+151^{+58}_{-68}$ \\
\hline
\multicolumn{5}{c}{{\ysej}} \\
\hline
N\,{\sc iv}] $\lambda1486^{\dagger}$ & $2126^{+863}_{-1745}$ & $42.17^{+0.17}_{-0.18}$ & $1.0^{+0.5}_{-0.3}$ & $-19^{+420}_{-393}$ \\
\Heii\ $\lambda1640$ & $2193^{+149}_{-194}$ & $42.69^{+0.04}_{-0.04}$ & $4.0^{+0.4}_{-0.4}$ & $+520^{+106}_{-108}$ \\
O\,{\sc iii}] $\lambda1666^{\dagger}$ & $456^{+32}_{-327}$ & $41.76^{+0.12}_{-0.13}$ & $0.5^{+0.2}_{-0.1}$ & $+105^{+71}_{-102}$ \\
\Ciii]\ $\lambda1909^{\dagger}$ & $1322^{+204}_{-149}$ & $42.57^{+0.05}_{-0.06}$ & $4.0^{+0.5}_{-0.5}$ & $+92^{+107}_{-113}$ \\
Mg\,{\sc ii}\ $\lambda2800^{\dagger}$ & $192^{+28}_{-23}$ & $42.17^{+0.07}_{-0.05}$ & $3.8^{+0.6}_{-0.4}$ & $+102^{+17}_{-16}$ \\
\Oii\ $\lambda3728^{\dagger}$ & $364^{+19}_{-21}$ & $42.85^{+0.01}_{-0.02}$ & $34.1^{+1.2}_{-1.3}$ & $+36^{+10}_{-10}$ \\
\Neiii\ $\lambda3870$ & $328^{+22}_{-21}$ & $42.68^{+0.02}_{-0.02}$ & $27.5^{+1.5}_{-1.2}$ & $+118^{+8}_{-8}$ \\
\Neiii\ $\lambda3969^{\ddagger}$ & $491^{+41}_{-44}$ & $42.53^{+0.04}_{-0.04}$ & $22.8^{+2.0}_{-1.9}$ & $+213^{+23}_{-23}$ \\
\hline
\end{tabular}
\end{table}

\subsubsection{Blue UV continua}\label{sec:feature_continua}
The rest-frame UV continua of both \yswl\ and \ysej\ are blue and approximately power-law. We characterise the continuum slope by fitting $f_\lambda \propto \lambda^\beta$ to four line-free windows centred at rest-frame 1300, 1450, 1600, 1860\,\AA\ (each 20\,\AA\ wide), obtaining $\beta = -2.20 \pm 0.09$ for \yswl\ and $\beta = -2.06 \pm 0.07$ for \ysej\ (Table~\ref{tab:source_properties}). Both slopes are consistent with young, star-forming galaxies. We note, however, that the observed UV slope of such a population is not set by the stellar continuum alone: nebular continuum emission (primarily two-photon and bound-bound) can redden the intrinsic slope by $\Delta\beta \sim 0.3$--$0.6$ at the youngest ages \citep{Schaerer2025}, an effect that becomes important precisely in the young-burst regime relevant here. The SED fitting we present in Sect.~\ref{sec:sps} does not include a separate nebular-continuum component. Dust attenuation and nebular reddening are therefore partially degenerate in the inferred $E(B-V)$, and the values quoted there should be regarded as upper limits on the true stellar dust attenuation. For context, the cosmic-noon EUVLG sample of \citet{Upadhyaya2024} has $\langle\beta\rangle = -2.33$. \yswl\ and \ysej\ sit at the redder end of this distribution but well within its spread.

\subsubsection{P~Cygni wind profiles}\label{sec:feature_pcygni}

Both sources show prominent P~Cygni profiles in the high-ionisation lines N\,{\sc v}\,$\lambda 1240$, Si\,{\sc iv}\,$\lambda 1400$, and C\,{\sc iv}\,$\lambda 1550$, each consisting of a blueshifted absorption trough and a redshifted emission peak at the systemic redshift (Fig.~\ref{fig:spec_compare_euvlg_r136}). P~Cygni profiles in these resonance lines are the canonical signature of radiatively driven winds from massive OB stars \citep[e.g.,][]{Vink2001}, with their strength enhanced by Wolf--Rayet-like outflows near the Eddington limit \citep[e.g.,][]{Crowther2007, Crowther2016}.

The C\,{\sc iv} and N\,{\sc v} profiles are particularly strong, with absorption components reaching $F_\lambda/F_{\rm cont}\sim0.2$--$0.5$ and comparably prominent emission peaks. In \yswl\ we also identify a clear N\,{\sc iv}\,$\lambda1719$ P~Cygni profile. Compared with the cosmic-noon EUVLG composite of \citet{Upadhyaya2024} (blue in Fig.~\ref{fig:spec_compare_euvlg_r136}), \ysej\ matches the composite closely in all three P~Cygni profiles, while \yswl\ shows systematically stronger N\,{\sc v}, Si\,{\sc iv}, and C\,{\sc iv} features; the N\,{\sc iv} P~Cygni profile is not seen in the cosmic-noon composite. Compared with the R136 VMS stack of \citet{Crowther2016} (red), the C\,{\sc iv} P~Cygni profiles in both sources match R136 closely in depth and shape, but R136 shows a substantially stronger N\,{\sc v} emission component than either galaxy. Conversely, P~Cygni Si\,{\sc iv} is absent in the R136 stack yet present in both of our sources. \citet{Crowther2016} attribute the absence of Si\,{\sc iv}\,$\lambda1400$ P~Cygni emission to a lack of mid- to late-type O supergiants.

The strength and consistency of the profiles observed here imply a young stellar population in which radiatively driven winds dominate the rest-UV resonance-line spectrum, and the systematically stronger P~Cygni profiles in \yswl\ relative to \ysej\ are consistent with a younger age, since wind-line strengths peak in the first $\sim 2$\,Myr of a starburst. We model both populations quantitatively in Sect.~\ref{sec:sps}.

The P~Cygni morphology of \yswl\ and \ysej\ further rules out a broad absorption line (BAL) quasar origin, a class of object that can sometimes superficially mimic a starburst spectrum at low spectral resolution \citep{Appenzeller2005}. The crucial distinction is that each of the three high-ionisation absorption features in our spectra is accompanied by a redshifted emission component anchored at the systemic redshift, producing the symmetric P~Cygni morphology characteristic of stellar winds. BAL outflows, by contrast, produce only the blueshifted absorption component, typically offset by thousands of km\,s$^{-1}$ from systemic and without a compensating emission peak. The bona fide P~Cygni nature of the profiles in \yswl\ and \ysej\ is further confirmed by the population-synthesis fits in Sect.~\ref{sec:sps} (also see Fig.~\ref{fig:J1450_fit} and \ref{fig:J1429_fit}), which reproduce both the blueshifted absorption and the redshifted emission of each line consistently.

\subsubsection{Broad \Heii\ and other wind-line features}\label{sec:feature_heii}

In addition to the P~Cygni profiles described above, several emission features in \yswl\ and \ysej\ are significantly broader than the narrow nebular lines (FWHM $\lesssim 1200$\,km\,s$^{-1}$) and likely originate from the optically-thick stellar winds \citep[e.g.,][]{Martins2022,Martins2025}, including \Niv]\,$\lambda 1486$, \Heii\,$\lambda 1640$, and \Ciii]\,$\lambda 1909$. They are shown in Fig.~\ref{fig:all_nebular_lines}, and the Gaussian-fit results are listed in Table~\ref{tab:nebular_lines}. 
The \Heii\ emission is remarkably strong, with FWHM $\sim 1900$--$2200$\,km\,s$^{-1}$ and rest-frame equivalent widths ${\rm EW_0} = 8.1 \pm 0.5$\,\AA\ in \yswl\ and $4.0 \pm 0.4$\,\AA\ in \ysej. The strength of the \Heii\ in \yswl\ is comparable with the seven most massive stars in R136 and the one for \ysej\ is comparable to the cosmic-noon EUVLG stack, as can be seen from Fig.~\ref{fig:spec_compare_euvlg_r136}, placing both galaxies among the strongest broad-\Heii\ emitters known in star-forming galaxies and pointing to a substantial population of very massive stars. In Sect.~\ref{sec:vms_discussion} we inspect the VMS content based on the \Heii\ strength. To compare directly with cosmic-noon EUVLG measurements in \citet{Upadhyaya2024}, we there adopt EW$_{0}$ measured by direct integration of the continuum-subtracted flux rather than the Gaussian-fit values. The two measurements are consistent within the uncertainties, with the non-parametric ${\rm EW_0} = 8.8 \pm 1.2$ and $3.7 \pm 1.1$\,\AA\ for \Heii\ in \yswl\ and \ysej, respectively.
For \Niv]\,$\lambda 1486$, \yswl\ has a strong detection with ${\rm EW_0} = 1.9^{+0.3}_{-0.5}$\,\AA\ and FWHM $= 1477^{+209}_{-825}$\,km\,s$^{-1}$, while for \ysej\ it is only marginal (${\rm EW_0} = 1.0^{+0.5}_{-0.3}$\,\AA), with a poorly constrained width (FWHM $= 2126^{+863}_{-1745}$\,km\,s$^{-1}$).
The semi-forbidden line \Ciii]\,$\lambda 1909$ is usually taken to be of nebular origin, but here it also exceeds the typical nebular limit, with FWHM $\sim 1300$\,km\,s$^{-1}$ and ${\rm EW_0} \approx 3$--$4$\,\AA\ in both sources, suggesting a partial stellar-wind contribution to the total emission.

The widths of these lines rules out a purely nebular origin, and a quasar broad-line region (BLR) is also disfavored for several reasons. First, if these lines were produced in a BLR, the resonant Ly$\alpha$ line should appear as a strong, broad feature at least as broad as \Heii. Instead, no broad Ly$\alpha$ is seen in either source (Figs.~\ref{fig:spec_compare} and \ref{fig:spec_compare_euvlg_r136}). Second, the high-ionisation resonance lines \Civ\,$\lambda1549$ and \Nv\,$\lambda1240$ show P~Cygni profiles (Sect.~\ref{sec:feature_pcygni}) rather than the symmetric broad emission expected from a BLR, or a broad absorption trough as in BAL. Their blueshifted absorption troughs are a characteristic feature of expanding stellar winds, not photoionised BLR gas. Third, the \Mgii\,$\lambda\lambda2796,2804$ doublet is detected with a narrow width (FWHM $\sim 200$--$500$\,km\,s$^{-1}$; see Table~\ref{tab:nebular_lines} and Sect.~\ref{sec:feature_nebularline}), consistent with nebular rather than BLR gas. The broad lines are therefore most naturally produced in the dense, radiatively driven winds of young massive stars, whose terminal velocities set the line widths and whose mass-loss rates and wind densities set the equivalent widths.

\subsubsection{Narrow nebular emission lines}\label{sec:feature_nebularline}

\begin{figure*}
	\includegraphics[width=0.95\linewidth]{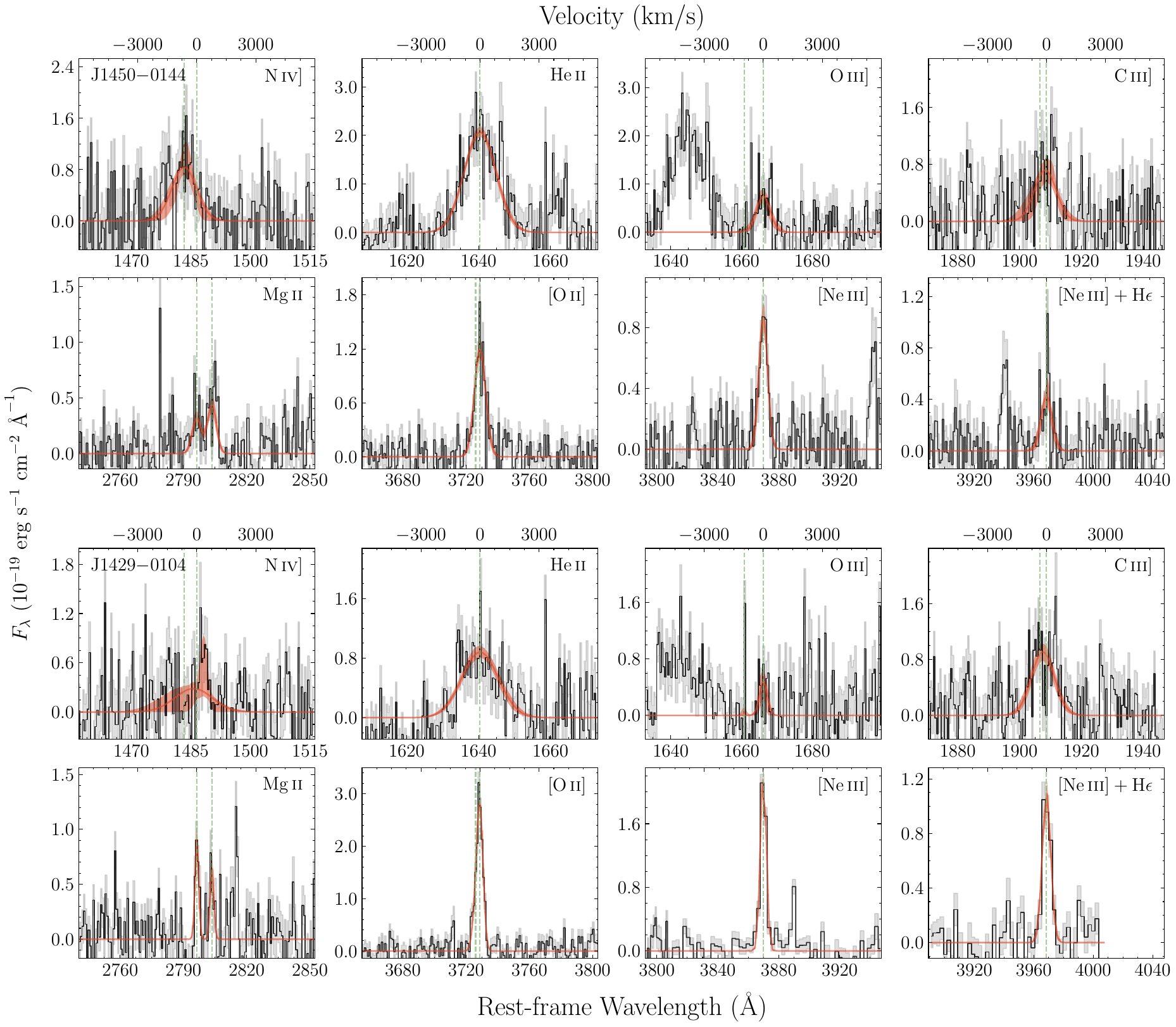}
    \caption{Continuum-subtracted data of the emission lines and their best-fit Gaussian (or double-Gaussian, for doublets) models in \yswl\ (top two rows) and \ysej\ (bottom two rows). From left to right and top to bottom in each block: \Niv]\,$\lambda\lambda1483,1486$, \Heii\,$\lambda1640$, \Oiii]\,$\lambda\lambda1661,1666$, \Ciii]\,$\lambda\lambda1907,1909$, \Mgii\,$\lambda\lambda2796,2804$, \Oii\,$\lambda\lambda3727,3730$, \Neiii\,$\lambda3870$, and \Neiii\,$\lambda3969$. See Sect.~\ref{sec:specmod} for details of the continuum modelling and line-fitting procedure. The resulting line measurements are reported in Table~\ref{tab:nebular_lines}.}
    \label{fig:all_nebular_lines}
\end{figure*}

In contrast to the broad wind-driven features discussed above, both spectra also show narrow emission lines characteristic of photoionised gas. Fig.~\ref{fig:all_nebular_lines} shows close-up views of the individual lines, and Table~\ref{tab:nebular_lines} summarises the measured properties. Details of the continuum modelling and Gaussian fitting are given in Appendix~\ref{app:continuum}. The forbidden lines \Oii\,$\lambda\lambda 3727, 3730$ and \Neiii\,$\lambda\lambda 3870, 3969$ are detected at high significance ($\gtrsim 5\sigma$) in both sources, with FWHM $\sim 330$--$670$\,km\,s$^{-1}$. The cosmic-noon EUVLG sample shows similarly narrow nebular emission whose line ratios favour photoionisation by young stars over AGN activity \citep[e.g.,][]{Marques-Chaves2022}. Our current wavelength coverage does not extend to the rest-frame optical diagnostic lines (H$\alpha$, [O\,{\sc iii}]\,$\lambda\lambda 4959, 5007$, [N\,{\sc ii}]\,$\lambda\lambda 6548, 6584$) needed to confirm this directly, but the close spectral similarity in the rest-frame UV suggests analogous excitation conditions.

Using the \Neiii/\Oii\ ratio (Ne3O2) as a metallicity indicator, we obtain $\log(\mathrm{Ne3O2}) \approx -0.17$ for both sources. Applying the high-redshift empirical calibration of \citet{Sanders2025} yields $12 + \log(\mathrm{O/H}) \approx 8.0$, corresponding to $Z \approx 0.2\,Z_\odot$ for a solar oxygen abundance of $12 + \log(\mathrm{O/H})_\odot = 8.69$ \citep{Asplund2021}. We adopt this metallicity when fixing the template grid 
in the SED fitting (Sect.~\ref{sec:sps}). One caveat is that if EUVLGs are Lyman-continuum leakers, as suggested by \citet{Marques-Chaves2022}, density-bounded H\,{\sc ii} regions could preferentially suppress low-ionisation lines such as \Oii\ and bias the Ne3O2-derived metallicity low. However, the Ne3O2 values we measure are not unusually elevated and fall well within the calibrated range, so we retain them as working estimates pending rest-frame optical follow-up.

\subsubsection{Weak low-ionisation interstellar absorption}\label{sec:feature_lis}

Equally notable is the lack of low-ionisation interstellar (LIS) absorption lines tracing the neutral ISM, such as Si\,{\sc ii}\,$\lambda1260$, \Oi\,$\lambda1302$\,$+$\,Si\,{\sc ii}\,$\lambda1304$, \Cii\,$\lambda1334$, and Si\,{\sc ii}\,$\lambda1526$. They are not apparent in either galaxy at the current depth (Fig.~\ref{fig:spec_compare_euvlg_r136}). Unlike the high-ionisation wind features of Sect.~\ref{sec:feature_pcygni}, these transitions arise in neutral gas in the ISM along the line of sight, and in typical star-forming galaxies at $z\sim2$--$3$ they appear with $\mathrm{EW_0}\sim1$--$2$\,\AA\ \citep[e.g.,][]{Shapley2003}. Cosmic-noon EUVLGs span a wide range of LIS strengths, from clearly detected absorption in BOSS-EUVLG1 and J0121$+$0025 \citep{Marques-Chaves2020, Marques-Chaves2021} to undetected LIS lines in J1316$+$2614 \citep{Marques-Chaves2022}. Our two sources resemble the latter. Similarly weak interstellar absorption is also seen in the young, low-metallicity galaxy Q2343-BX418 at $z=2.3$ \citep{Erb2010}, and the ultra-deep SPURS spectrum of GN-z11 at $z = 10.6$ likewise shows no detectable LIS absorption, interpreted as a low covering fraction of enriched neutral gas in front of its UV-emitting regions \citep{Chen2026}.

Because the LIS transitions are typically saturated, their depth measures the covering fraction of neutral gas rather than its column density, so their weakness signals a low \Hi\ covering fraction and favourable conditions for the escape of Lyman-continuum (LyC) photons \citep[e.g.,][]{Reddy2016, Gazagnes2018, Saldana-Lopez2022}. Among the cosmic-noon EUVLGs, the LIS strength indeed tracks the measured LyC leakage: J0121$+$0025, with detected LIS absorption, is a leaker with $f_{\rm esc}({\rm LyC})\approx40$ per cent, whereas J1316$+$2614, with no detectable LIS lines, leaks at $f_{\rm esc}({\rm LyC})\approx90$ per cent \citep{Marques-Chaves2021, Marques-Chaves2022}. The strong, narrow Ly$\alpha$ emission that drove the original quasar classification of both sources (Sect.~\ref{sec:obs_discovery}) fits naturally into the same picture, since Ly$\alpha$ escape is likewise regulated by the \Hi\ covering fraction, and a low covering fraction would also act in the same direction as the density-bounded scenario flagged in Sect.~\ref{sec:feature_nebularline}. We caution that this interpretation remains qualitative: the continuum signal-to-noise ratio limits the constraining power of the non-detections, the LIS lines may not be fully saturated at the relatively low metallicity, and for \ysej\ any escape must be strongly direction dependent given the $\sim5.4$\,kpc offset between its UV and dust/[\Cii] emission (Sect.~\ref{sec:alma}). Quantitative constraints on the covering fraction and the escape of ionising radiation are deferred to future work.

\subsubsection{Morphology}\label{sec:feature_morph}

Finally, Fig.~\ref{fig:spec_compare} also includes NIRSpec Wide Aperture Target Acquisition (WATA) images ($1.6^{\prime\prime}\times1.6^{\prime\prime}$) of the two objects, taken with the F140X filter for \yswl\ and the CLEAR filter for \ysej. \yswl\ is consistent with a point source, unresolved at the $\sim0.1^{\prime\prime}$ FWHM of the WATA PSF, which corresponds to $\sim 500$\,pc at $z=6.63$ and sets an upper limit on the size of the rest-frame UV emitting region. 
\ysej\ also shows a bright, compact core on which the NIRSpec slit is centred, but additionally exhibits clearly extended structure to the south--east (see also Fig.~\ref{fig:alma}). The compact component dominates the rest-frame UV light entering the slit and therefore drives the spectrum of \ysej. Whether the extended component traces the rest-frame UV or optical light is unclear, since the CLEAR configuration transmits the full NIRSpec range ($\sim0.6$--$5.3\,\mu{\rm m}$) and so the image is sensitive to Ly$\alpha$ and to strong rest-frame optical lines such as [\Oiii]\,$\lambda5007$. Higher-resolution imaging with JWST/NIRCam with multiple filters, or deep JWST/NIRSpec IFU observations, are needed to characterise the detailed morphologies and individual components of both galaxies.


\subsection{Young, star-forming stellar populations\label{sec:sps}}


The same UV wind signatures, blue continua, and nebular-line-dominated emission seen in cosmic-noon EUVLGs support the interpretation that the extreme UV luminosities of our two sources are powered primarily by young stars rather than by AGN activity. We therefore infer their stellar populations by fitting the rest-frame UV P-Cygni wind features with stellar population synthesis (SPS) models. These features arise in the radiatively driven winds of short-lived massive stars ($\lesssim 10$\,Myr for $M \gtrsim 15\,\msun$). Both their presence and their relative strength evolve rapidly over the first few Myr of star formation, which makes them sensitive diagnostics of the recent star-formation history.

However, as shown in Sect.~\ref{sec:feature_heii}, both galaxies exhibit strong broad \Heii\,$\lambda1640$ emission, with equivalent widths well above those of typical star-forming galaxies. Cosmic-noon EUVLGs show similarly strong \Heii, which has been attributed to the presence of very massive stars \citep{Upadhyaya2024}. Canonical SPS models typically truncate the IMF at $M_{\rm up} = 100$--$120\,\msun$ and lack dedicated VMS atmosphere and wind models, and may therefore be unable to reproduce the observed spectra. In Sect.~\ref{sec:sps_models} we describe the two sets of SPS models we use: standard \texttt{BPASS} models and their extension with dedicated VMS treatment from \citet{Upadhyaya2024}. We present the fitting results with the standard models in Sect.~\ref{sec:sps_bpass} and with the VMS extension in Sect.~\ref{sec:sps_vms}.

\subsubsection{Population synthesis models and fitting method\label{sec:sps_models}}

The first family is the standard \texttt{BPASS} v2.2.1 binary grid \citep{Eldridge2017,Stanway2018}, which adopts a broken power-law IMF ($\alpha=-1.30$ below $0.5\,\msun$ and $\alpha=-2.35$ above) with upper-mass limits of $100$ and $300\,\msun$. Following \citet{Upadhyaya2024}, we convert the native $10^6\,\msun$ instantaneous-burst spectra to a constant star-formation rate (CSFR) of $1\,\msun\,{\rm yr}^{-1}$ by summing the burst spectra over the native BPASS age bins. The grid is sampled at nominal $\log(t/{\rm yr})=6.0$--$8.0$ in steps of $0.1$. We fix the metallicity at $Z=0.006$, matching the metallicity of the \texttt{BPASS+VMS} models introduced next. This is higher than the $Z=0.004$ suggested by the Ne3O2 ratio in Sect.~\ref{sec:feature_nebularline}, but the difference has small effect on the fitted parameters. We therefore adopt $Z=0.006$ for consistency with the \texttt{BPASS+VMS} models, which are available only at this metallicity. Hereafter, we refer to this model family as \texttt{BPASS-only}.

The second family is the \texttt{BPASS+VMS} grid from \citet{Upadhyaya2024}, which augments \texttt{BPASS} v2.2.1 with the VMS models of \citet{Martins2022} and extends the IMF to five upper-mass limits between $M_{\rm up} = 175$ and $475\,\msun$. The VMS models adopt the \citet{Grafener2021} mass-loss prescription for stars near the Eddington limit, which \citet{Martins2022} showed is required to reproduce the strong broad \Heii\, emission of the VMS in R136 \citep{Crowther2016}. We return to this in Sect.~\ref{sec:discussion_imf}. The grid adopts a Salpeter IMF ($\alpha = -2.35$), identical to the standard grid above $0.5\,\msun$. The difference below this mass is irrelevant for the rest-UV spectrum. We also use models with CSFR, sampled at $\log(t/{\rm yr}) = 6.0$--$8.0$ in steps of $0.1$. This choice is mainly because of the short lifetimes of VMS, i.e. in a burst model, all VMS die within a few Myr, and the models lose their constraining power on $M_{\rm up}$. 

We fit both families with the same procedure. Rather than fitting the absolute-flux spectrum, we compare continuum-normalised line profiles in three rest-frame windows: 1230--1245\,\aa\ (\Nv), 1530--1560\,\aa\ (\Civ), and 1630--1655\,\aa\ (\Heii). The observed spectrum is normalised by the spline continuum of Sect.~\ref{sec:specmod}, with the continuum uncertainty propagated into the normalised errors, and each model is normalised by the median flux in flanking line-free sidebands. Both dust attenuation and the absolute flux calibration cancel in normalised space, so the fit has no free scale or reddening parameter: we compute $\chi^2$ over all pixels in the three windows, constraining the burst age for the standard grid and the duration of star formation and $M_{\rm up}$ for the \texttt{BPASS+VMS} grid. We subsequently estimate $E(B\!-\!V)$ by reddening the best-fit template with the \citet{Calzetti2000} law ($R_V = 4.05$) until its UV slope matches the observed $\beta$. Parameter uncertainties are derived from the $\chi^2$ grid by weighting each template by $\exp(-\Delta\chi^2/2)$, rescaling $\Delta\chi^2$ by the minimum reduced $\chi^2$ where it exceeds unity, and quoting the median and 16th--84th percentile interval.

\subsubsection{\texttt{BPASS-only} models fail to produce the broad \Heii\ emission in these two galaxies\label{sec:sps_bpass}}

\begin{table*}
\centering
\caption{Summary of stellar population parameters inferred from template fitting (Sect.~\ref{sec:sps_models}). Metallicities are fixed at $Z=0.006$ ($Z/Z_\odot=0.3$ with $Z_\odot=0.02$). All models assume a constant star-formation history, and the age column denotes the duration of star formation. The SFR is obtained by scaling each best-fit template, reddened with its $E(B-V)$, to the observed flux level, and the stellar mass follows as $M_\star = {\rm SFR}\times{\rm age}$.}
\label{tab:template_fits}
\begin{tabular}{l c c c c c}
\hline
Template & Age (Myr) & $E(B\!-\!V)$ & $\log(M_\star/M_\odot)$ & $M_{\rm up}$ ($M_\odot$) & SFR ($M_\odot$/yr)\\
\hline
\multicolumn{6}{c}{\yswl} \\
\hline
\texttt{BPASS-only} & 2.00 & 0.10 & 9.29 & 300 & 986.42 \\
\texttt{BPASS+VMS}  & 3.16 & 0.09 & 9.24 & 275 & 543.71 \\
\hline
\multicolumn{6}{c}{\ysej} \\
\hline
\texttt{BPASS-only} & 3.98 & 0.12 & 9.37 & 300 & 594.91 \\
\texttt{BPASS+VMS} & 25.12 & 0.11 & 9.88 & 325 & 299.50 \\
\hline
\end{tabular}
\end{table*}

Figs.~\ref{fig:J1450_fit} and \ref{fig:J1429_fit} show the best-fit \texttt{BPASS-only} templates for \yswl\ and \ysej\ (blue curves), with the corresponding parameters summarised in Table~\ref{tab:template_fits}. The \texttt{BPASS-only} models favour very young stellar populations in both galaxies, with best-fit ages of $2.0$\,Myr for \yswl\ and $4.0$\,Myr for \ysej, and mild reddening of $E(B-V)\simeq0.1$. Scaling each reddened best-fit template to the observed flux yields ${\rm SFR}\simeq990$ and $\simeq590\,M_\odot\,{\rm yr}^{-1}$ for \yswl\ and \ysej, respectively, and under the constant star-formation history assumed here the stellar masses follow as $\log(M_\star/M_\odot)=9.29$ and $9.37$.

These young stellar populations reproduce the prominent \Nv\ and \Civ\ P-Cygni features reasonably well. The fits fail, however, at \Heii. All models in the grid substantially underpredict the observed strength of the observed broad \Heii\ emission, with ${\rm EW}_0=8.8\pm1.2$\,\AA\ for \yswl\ and $3.7\pm1.1$\,\AA\ for \ysej\ (Sect.~\ref{sec:feature_heii}). Between the two available upper-mass limits, both fits favour $M_{\rm up}=300\,M_\odot$ over $100\,M_\odot$, yet even these models fall well short of the observed \Heii. Extending the IMF above $100\,M_\odot$ therefore does not by itself produce the observed emission.

Broad stellar \Heii\ emission in typical star-forming galaxies has ${\rm EW}_0$ of order $1$--$2$\,\AA\ \citep[e.g.,][]{Shapley2003, Steidel2016} and is generally attributed to the winds of classical Wolf--Rayet (cWR) stars \citep{Schaerer1998, Brinchmann2008}. Individual cWR stars can show strong \Heii\ emission formed in their dense, optically thick winds \citep{Crowther2007,Martins2023}, but they constitute only a small fraction of the massive-star population, while the integrated UV continuum is dominated by the far more numerous OB stars. Their contribution is therefore heavily diluted in galaxy-integrated spectra, and the \texttt{BPASS-only} models, which include these cWR phases, accordingly predict \Heii\ at this modest level \citep[see also][]{Upadhyaya2024}. The much stronger broad \Heii\ in \yswl\ and \ysej\ points to a massive-star population not captured by the \texttt{BPASS-only} models.

\subsubsection{\texttt{BPASS+VMS} models successfully produce the broad \Heii\ emission\label{sec:sps_vms}}

Figs.~\ref{fig:J1450_fit} and \ref{fig:J1429_fit} also show the best-fit \texttt{BPASS+VMS} templates (red curves), with the corresponding parameters summarised in Table~\ref{tab:template_fits}. The best-fit models have $M_{\rm up}=275\,M_\odot$ with an age of $3.2$\,Myr for \yswl, and $M_{\rm up}=325\,M_\odot$ with an age of $25$\,Myr for \ysej, with reddening of $E(B-V)=0.09$ and $0.11$ and stellar masses of $\log(M_\star/M_\odot)=9.24$ and $9.88$, respectively. The same scaling yields ${\rm SFR}\simeq540$ and $\simeq300\,M_\odot\,{\rm yr}^{-1}$ (Table~\ref{tab:template_fits}). These rates exceed the ${\rm SFR_{UV}}\simeq135\,M_\odot\,{\rm yr}^{-1}$ of Sect.~\ref{sec:euvlg} by factors of $\approx2$--$4$ for two reasons. The template-fitting SFRs should be regarded as total star-formation rate, whereas ${\rm SFR_{UV}}$ is the unobscured part. In addition, at the young ages found here the UV continuum has not yet built up to the equilibrium level assumed by a fixed $\kappa_{\rm UV}$, which could bias ${\rm SFR_{UV}}$ low \citep{Schaerer2025}. The notably older age and higher stellar mass of \ysej\ relative to its \texttt{BPASS-only} fit are discussed below. These templates produce substantially stronger \Heii\ emission, while matching the \Nv\ and \Civ\ P-Cygni profiles as well as the \texttt{BPASS-only} fits.

\begin{figure}
	\centering
	\includegraphics[width=\linewidth]{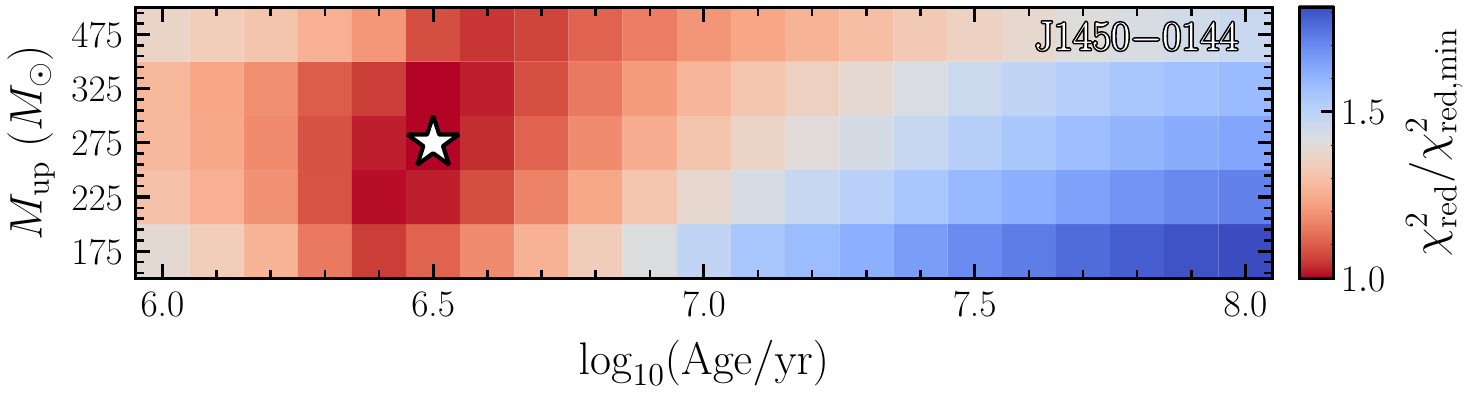}\par\smallskip
	\includegraphics[width=\linewidth]{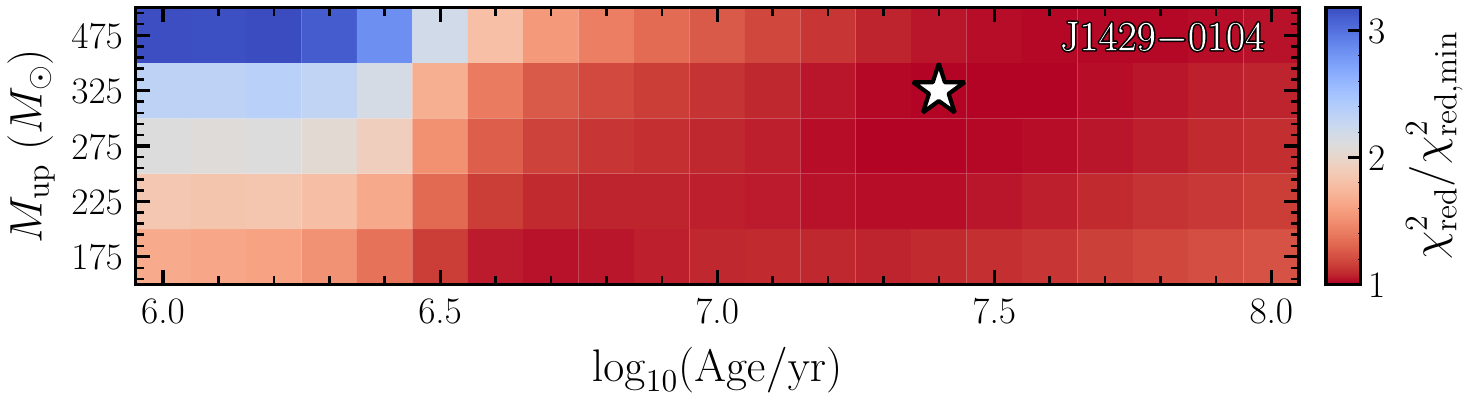}
	\caption{Goodness of fit for the BPASS+VMS template fitting of \yswl\ (top) and \ysej\ (bottom). The colour scale shows the reduced $\chi^2$, normalised to its minimum over the grid, as a function of the age of the constant star-formation model and the IMF upper-mass limit $M_{\rm up}$, at fixed $Z = 0.006$. The $\chi^2$ is evaluated over the \Nv, \Civ, and \Heii\ windows defined in Sect.~\ref{sec:sps}, and white stars mark the best-fit models. For \yswl\ the fit prefers a young population with $\log(t/{\rm yr}) = 6.5$ and $M_{\rm up} = 275\,\msun$. For \ysej\ the surface is shallow at $\log(t/{\rm yr}) \gtrsim 6.5$, with a formal minimum at $\log(t/{\rm yr}) = 7.4$ and $M_{\rm up} = 325\,\msun$, so the age is only weakly constrained, a younger population at $\log(t/{\rm yr}) \lesssim 7$ but lower $M_{\rm up}$ is also consistent with the data.}
	\label{fig:chi2_maps}
\end{figure}

\begin{figure*}
	\includegraphics[width=\linewidth]{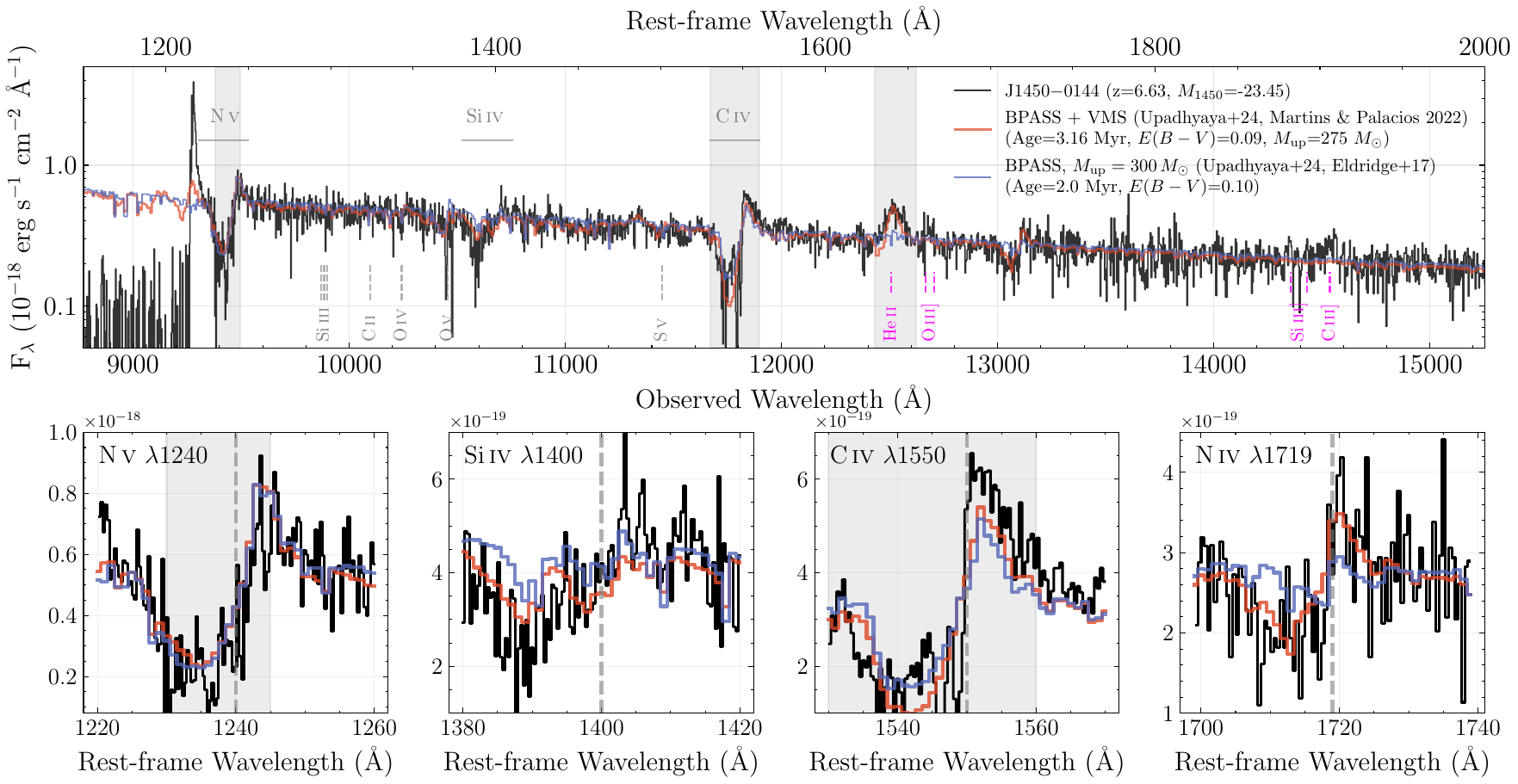}
    \caption{Stellar population fitting for \yswl. The upper panel shows the observed JWST/NIRSpec spectrum (black) compared with the best-fit \texttt{BPASS-only} \citep[blue;][]{Eldridge2017, Stanway2018,Upadhyaya2024} and \texttt{BPASS+VMS} \citep[red;][]{Upadhyaya2024} templates, both under a constant star-formation history at fixed $Z=0.006$. The best-fit star-formation durations (age), $E(B-V)$, and $M_{\rm up}$ are given in the legend and Table~\ref{tab:template_fits}. Grey bands mark the three windows entering the continuum-normalised fit (\Nv, \Civ, and \Heii; Sect.~\ref{sec:sps_models}). For display, each template is reddened with its best-fit $E(B-V)$ and scaled to the observed flux. The lower panels zoom in on the wind profiles \Nv\,$\lambda1240$, \Siiv\,$\lambda1400$, \Civ\,$\lambda1550$, and \Niv\,$\lambda1719$, with the fitted pixels again shaded; \Siiv\ and \Niv\,$\lambda1719$ are shown for comparison but are excluded from the fit, and the \Heii\ profiles are compared in Fig.~\ref{fig:vms_lines}.}
    \label{fig:J1450_fit}
\end{figure*}

\begin{figure*}
	\includegraphics[width=\linewidth]{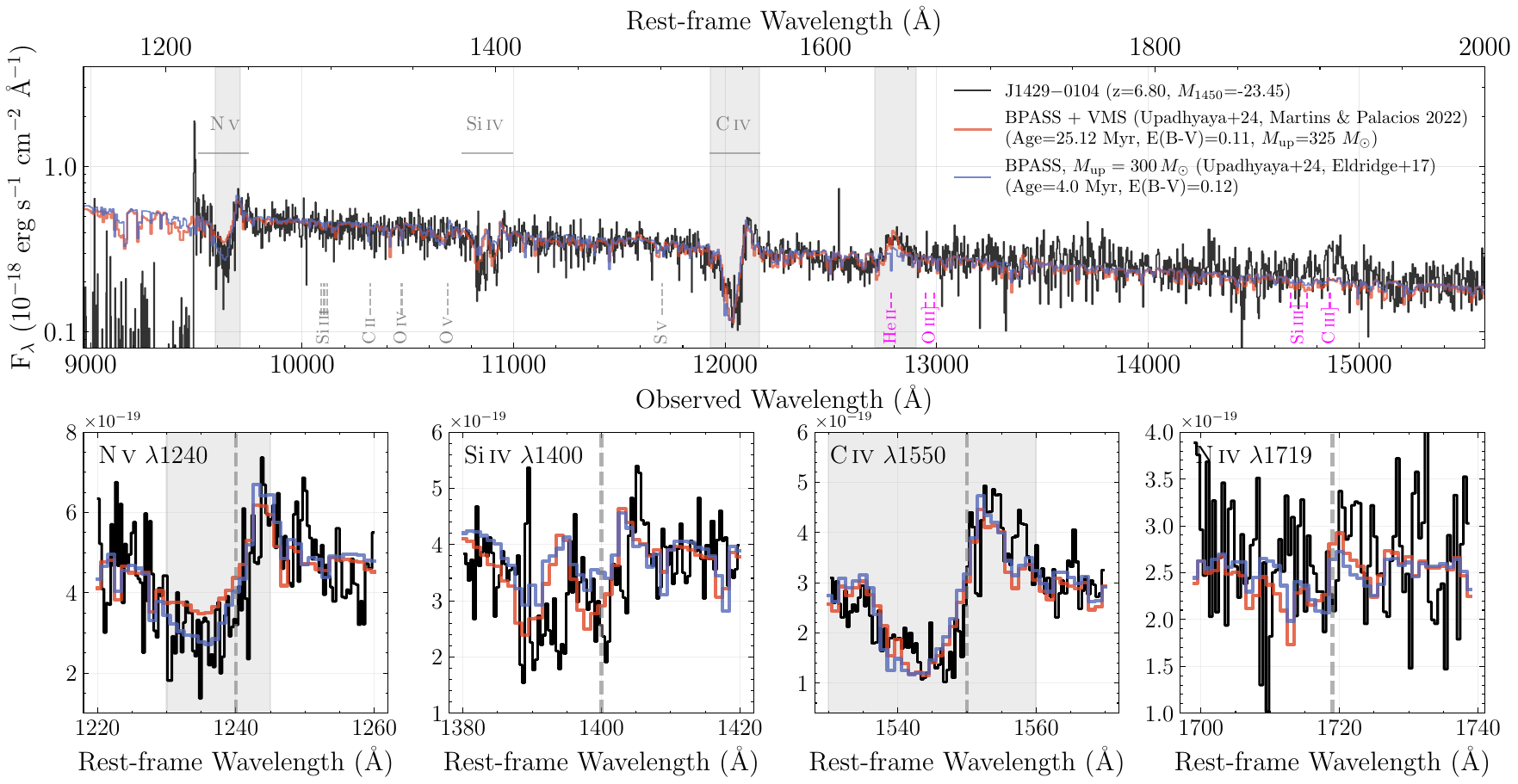}
    \caption{Stellar population fitting for \ysej, same as Fig.~\ref{fig:J1450_fit}.} 
    \label{fig:J1429_fit}
\end{figure*}

As mentioned in Sect.~\ref{sec:sps_bpass}, simply extending the IMF above $M_{\rm up}=100\,M_\odot$ is insufficient to reproduce the observed \Heii\ strength \citep[see also][]{Upadhyaya2024, Marques-Chaves2026}. The strong \Heii\ emerges only when the models include the dedicated VMS atmospheres of \citet{Martins2022}, which is based on the mass-loss prescription of \citet{Grafener2021} calibrated on the VMS observed in R136 in the LMC. The relevant ingredient is therefore not just the presence of stars above $100\,M_\odot$, but the distinct wind physics of very massive stars. At sufficiently high masses and luminosities, the increasing Eddington factor drives dense, optically thick winds \citep{Vink2011, Grafener2011, Bestenlehner2014}, and these stars exhibit Wolf--Rayet-like spectra already on the main sequence, spectroscopically resembling the WNh stars \citep[e.g.,][]{Crowther2010}. Because they are also exceptionally luminous in the UV, even a small population contributes disproportionately to the integrated UV light of a young stellar population. Their wind signatures therefore become prominent in the galaxy-integrated spectrum, producing the strong \Nv\ and \Civ\ P-Cygni profiles together with the broad \Heii\ emission observed in \yswl\ and \ysej.

The best-fit \texttt{BPASS+VMS} template for \yswl\ additionally shows broad \Niv]\,$\lambda1486$ emission and a \Niv\,$\lambda1719$ P-Cygni feature, and both features are present in the data. \citet{Martins2022} identify these as further characteristic signatures of very massive stars, lending additional support to the presence of a VMS population in \yswl.

The best-fit parameters of \ysej\ are noticeably different from those in the \texttt{BPASS-only} fit. The \texttt{BPASS+VMS} model returns an age older by a factor of six and a stellar mass higher by $0.5$\,dex. To assess how well these parameters are actually constrained, Fig.~\ref{fig:chi2_maps} shows the reduced $\chi^2$ over the age--$M_{\rm up}$ grid for both galaxies. For \yswl\ the minimum is well localised at ages of $2$--$4$\,Myr, consistent with the \texttt{BPASS-only} result, because only very young populations with a substantial VMS contribution approach the observed strength of its wind features. For \ysej, in contrast, the $\chi^2$ surface is shallow and the formal best fit carries little significance. Under a constant star-formation history the VMS population reaches a steady state within a few Myr, while the UV continuum from longer-lived stars continues to build, so the wind-line equivalent widths decline slowly with age; raising $M_{\rm up}$ increases them at all ages. The two parameters therefore compensate, and the moderate wind-line strengths of \ysej\ are reproduced comparably well across most of the grid: a young population with a lower $M_{\rm up}$ and an older population with a higher $M_{\rm up}$ are indistinguishable at the current signal-to-noise ratio, and only the combination of ages $\lesssim3$\,Myr and $M_{\rm up}\gtrsim325\,M_\odot$ is excluded. The age and $M_{\rm up}$ of \ysej\ are thus individually only weakly constrained by the template fitting, and its inferred stellar mass, which under CSFH scales with the assumed age, inherits this degeneracy, spanning $\log(M_\star/M_\odot)\simeq 9.5$--$9.9$ across the acceptable region. In Sect.~\ref{sec:discussion_mup}, we attempt to constrain the $M_{\rm up}$ with \Heii-based diagnostics.

The $E(B-V)$ values from the \texttt{BPASS+VMS} fits are consistent with those from \texttt{BPASS-only}, as expected since they are constrained mainly by the continuum slopes $\beta$ (Table~\ref{tab:source_properties}). These values should be treated as upper limits on the stellar attenuation: we do not include a separate nebular-continuum component, which is not negligible at the young ages found here and is especially relevant for galaxies hosting VMS \citep{Schaerer2025}. The dust properties of both sources remain uncertain. In Sect.~\ref{sec:alma} we report a dust-continuum detection in \ysej\ whose peak is offset by $\sim5$\,kpc from the rest-UV core, and the effective attenuation of the UV-emitting regions in both galaxies will require further investigation.

In summary, the template fitting shows that the rest-frame UV spectra of \yswl\ and \ysej, and in particular their broad \Heii\ emission, are reproduced by the \texttt{BPASS+VMS} models but not by the \texttt{BPASS-only} models, indicating that both galaxies host substantial VMS populations. In Sect.~\ref{sec:vms_discussion} we further characterise the VMS populations more directly, comparing the measured equivalent widths of the wind lines with the model predictions to constrain $M_{\rm up}$ and to estimate the number of VMS in each galaxy. We also briefly discuss the broader implications of the VMS interpretation, for the high-mass end of the IMF, the chemical enrichment, and the eventual fate of these systems.

\subsection{Molecular gas and dust\label{sec:alma}}

\begin{figure}
	\includegraphics[width=\linewidth]{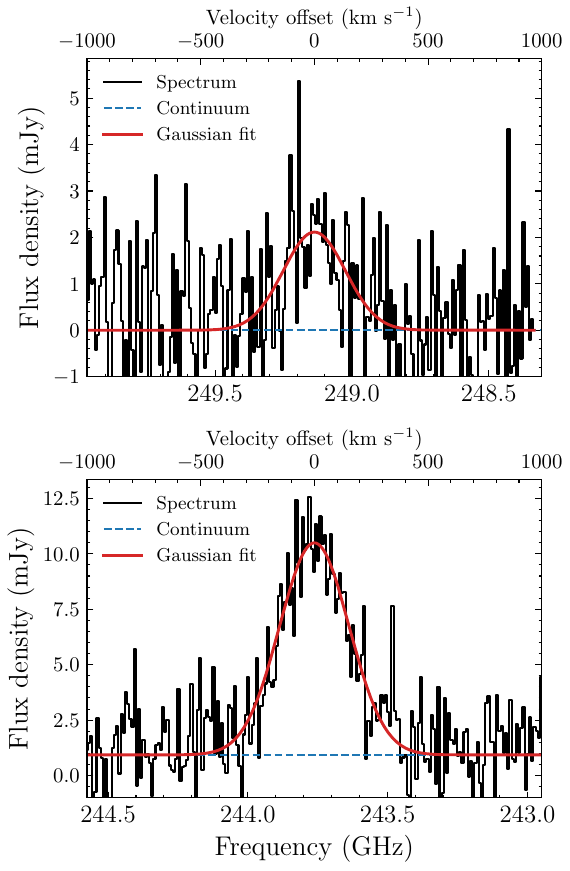}
    \caption{ALMA [\Cii]\,$158\,\mu{\rm m}$ spectra of \yswl\ (top) and \ysej\ (bottom),
    with the best-fit single-Gaussian model. The blue dashed line indicates the fitted continuum level. The upper axis shows the velocity offset relative to
    the systemic [\Cii] redshift of each source.}
    \label{fig:cii}
\end{figure}

\begin{figure*}
	\includegraphics[width=\linewidth]{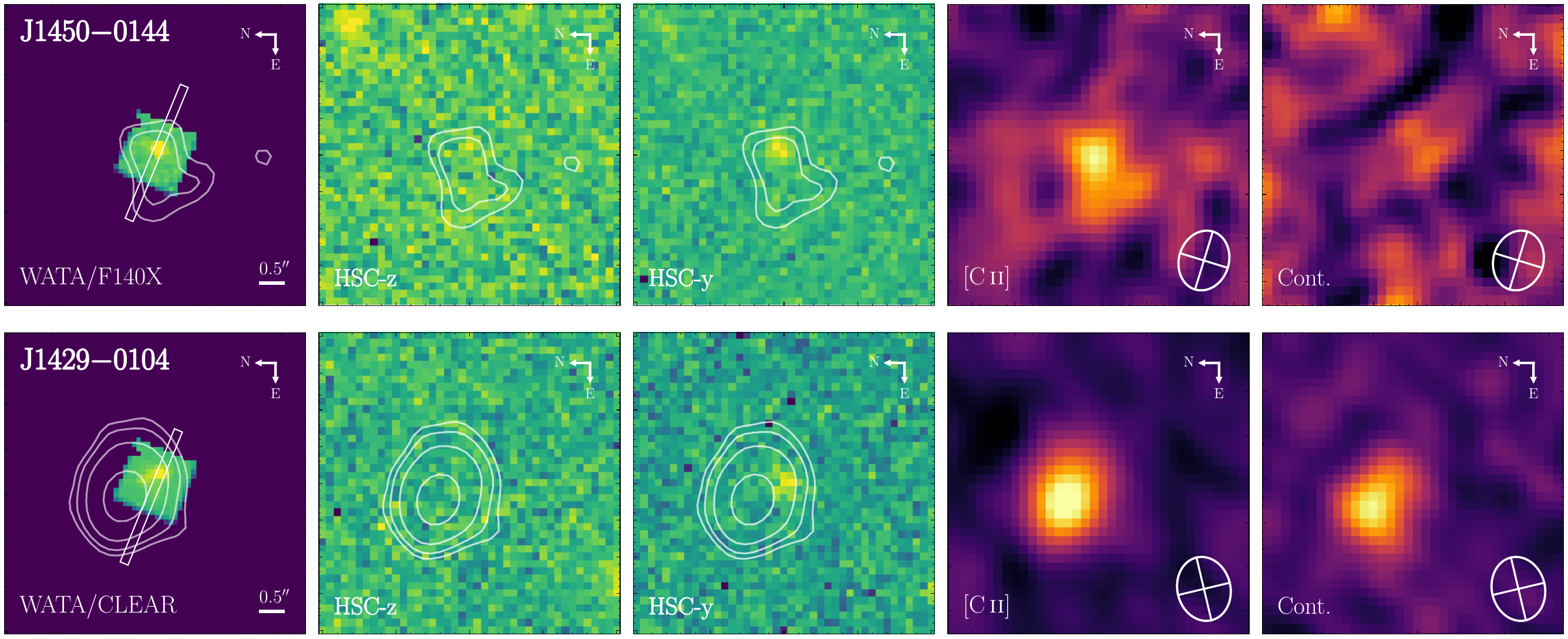}
    \caption{
    Multi-wavelength imaging of \yswl\ (top row) and \ysej\ (bottom row), each panel has a size of $6.9^{\prime\prime}\times6.9^{\prime\prime}$ and is oriented with north up and east left. From left to right: \texttt{JWST}/\texttt{NIRSpec} WATA acquisition image (F140X filter for \yswl, CLEAR for \ysej), Subaru/HSC $z$-band, $y$-band, ALMA band-6 $[\mathrm{C\,\textsc{ii}}]\,158\,\mu\mathrm{m}$ moment-0 map, and ALMA band-6 dust continuum map. White contours show the $[\mathrm{C\,\textsc{ii}}]$ emission at $2$, $3$, $5$, $10\,\sigma$, overlaid on the WATA, HSC $z$, and HSC $y$ panels. The NIRSpec slit position is indicated on the WATA panels. ALMA synthesised beams are shown in the lower right of the ALMA panels. \ysej\ shows a clear $\sim1^{\prime\prime}$ ($\sim5.4\,\mathrm{kpc}$) offset between the rest-frame UV emission and the $[\mathrm{C\,\textsc{ii}}]$/dust-continuum peaks; in \yswl\ the $[\mathrm{C\,\textsc{ii}}]$ emission is approximately co-spatial with the UV centroid, and no continuum is detected at the current sensitivity.
    }
    \label{fig:alma}
\end{figure*}

The ALMA observations provide a view of the cold ISM in these two galaxies, complementing the rest-frame UV information described above. Fig.~\ref{fig:alma} presents matched postage stamps ($6.9^{\prime\prime}\times6.9^{\prime\prime}$) of \yswl\ and \ysej, combining \texttt{JWST}/\texttt{NIRSpec} WATA images, Subaru/HSC ($z$- and $y$-band) imaging, and ALMA band-6 continuum and $[\mathrm{C\,\textsc{ii}}]$ moment-0 maps. The measured $[\mathrm{C\,\textsc{ii}}]$ fluxes and continuum flux densities are listed in Table~\ref{tab:alma}, and the extracted $[\mathrm{C\,\textsc{ii}}]$ spectra are shown in Fig.~\ref{fig:cii}. Details of the continuum and moment-0 map construction, aperture choices, and flux extraction are given in Appendix~\ref{app:alma_measurements}.

Both galaxies show clear detections of $[\mathrm{C\,\textsc{ii}}]$ emission, but with different morphologies: \ysej\ is marginally resolved, while \yswl\ shows an unresolved core with a tentative extended component to the east. \ysej\ also exhibits a significant dust continuum detection, whereas no continuum emission is detected for \yswl\ at the current sensitivity. \ysej\ additionally shows a clear spatial offset between its rest-frame UV emission and the $[\mathrm{C\,\textsc{ii}}]$/dust-continuum peaks, with the UV centroid displaced by $\sim1^{\prime\prime}$ ($\sim5.4\,\mathrm{kpc}$ at its redshift). We discuss the possible origin of this offset in Sect.~\ref{sec:discussion_ir}.

The dust continuum detection in \ysej\ allows us to derive the obscured star-formation rate and dust mass by fitting the observed flux density with a modified blackbody with $T_{\rm dust}=47$\,K, $\beta=1.6$, and $\kappa_{\nu}=0.77\,(\nu/352\,{\rm GHz})^{\beta}$\,cm$^{2}$\,g$^{-1}$ \citep{Beelen2006,Decarli2018}. We obtain $L_{\rm TIR}=(2.6\pm0.4)\times10^{12}\,L_\odot$ and $M_{\rm dust}=(1.2\pm0.2)\times10^8\,M_\odot$, corresponding to an obscured star-formation rate ${\rm SFR_{IR}} = 310 \pm 50\,M_\odot\,{\rm yr}^{-1}$ following \cite{Inami2022}. 
For \yswl, the continuum is not detected at the current sensitivity. We also estimate the molecular gas mass using the $[\mathrm{C\,\textsc{ii}}]$-based calibration of \citet{Zanella2018}, $M_{\rm mol} = \alpha_{\rm [CII]}\, L_{\rm [CII]}$ with $\alpha_{\rm [CII]} = 30\,M_\odot/L_\odot$, yielding $M_{\rm mol} = (1.24\pm0.07)\times10^{11}\,M_\odot$ for \ysej\ and $(0.24\pm0.05)\times10^{11}\,M_\odot$ for \yswl. This calibration was derived from a sample of main-sequence and starbursting galaxies from $z\sim0$ to $z\sim6$ and carries an intrinsic scatter of $\sim0.3$\,dex. Its applicability to EUVLGs is untested, and the much harder and stronger radiation fields of EUVLGs may bias the calibration in either direction. All the measured and derived quantities are listed in Table~\ref{tab:alma}. 
From a two-dimensional Gaussian fit to the [\Cii]\ moment-0 maps, \ysej\ is resolved with a beam-deconvolved major-axis FWHM of $1.27\pm0.32^{\prime\prime}$ ($6.8\pm1.7$\,kpc), whereas \yswl's bright core is consistent being unresolved. Assuming the [\Cii]\ emission traces a rotating disc of diameter $D=1.5\,a_{\rm maj}$, the dynamical mass is $M_{\rm dyn}\,\sin^{2}i=1.16\times10^{5}\,(0.75\,\mathrm{FWHM_{[CII]}}/\mathrm{km\,s^{-1}})^{2}\,(D/\mathrm{kpc})\,\msun$ \citep{Decarli2018}, yielding $M_{\rm dyn}\,\sin^{2}i=(8.6\pm2.3)\times10^{10}\,\msun$ for \ysej\ and, for the unresolved \yswl, an upper limit of $M_{\rm dyn}\,\sin^{2}i\lesssim5\times10^{10}\,\msun$ for any deconvolved size below the synthesised beam. For \ysej\ the dynamical mass is comparable to $M_{\rm mol}$. For \yswl\ the molecular gas mass is consistent with the dynamical limit.
Together, these measurements place \yswl\ and \ysej\ among the most $[\mathrm{C\,\textsc{ii}}]$- and FIR-luminous galaxies known at $z>6.5$. We discuss the broader implications of their IR properties in Sect.~\ref{sec:discussion_ir}.

\begin{table*}
\centering
\caption{ALMA-derived properties of \yswl\ and \ysej. For \yswl, [\Cii] flux 
and luminosity are measured at the peak of the emission given its unresolved 
morphology. For \ysej, values are integrated within an elliptical aperture 
enclosing 90\% of the curve-of-growth flux (Sect.~\ref{sec:alma}). 
$L_{\rm FIR}$ (rest-frame 42.5--122.5\,$\mu$m), $L_{\rm TIR}$ (rest-frame 
8--1000\,$\mu$m), and $M_{\rm dust}$ are derived from the measured continuum 
flux density assuming an optically thin modified blackbody with 
$T_{\rm dust}=47$\,K and $\beta=1.6$, following \citet{Beelen2006} and 
\citet{Decarli2018}. SFR$_{\rm IR}$ is computed as 
$\mathrm{SFR_{IR}}/(M_\odot\,{\rm yr}^{-1}) = 1.2\times10^{-10}\,L_{\rm TIR}/L_\odot$ 
following \citet{Inami2022}. $M_{\rm mol}$ is estimated from $L_{\rm [CII]}$ 
with a conversion factor $\alpha_{\rm [CII]} = 30\,M_{\odot}/L_{\odot}$ 
\citep{Zanella2018}. Quoted uncertainties are statistical only. Several 
derived quantities carry substantial additional systematics, most notably 
the $\sim 0.3$\,dex scatter in the $L_{\rm [CII]}$--$M_{\rm mol}$ relation 
and the unverified applicability of this calibration to UV-luminous sources 
at high redshift.}
\label{tab:alma}
\begin{tabular}{ccccccccc}
\hline
$F_{\rm [CII]}$ & FWHM$_{\rm [CII]}$ & $L_{\rm [CII]}$ & $S_{\nu,\rm cont}$ & $L_{\rm FIR}$ & $L_{\rm TIR}$ & SFR$_{\rm IR}$ & $M_{\rm dust}$ & $M_{\rm mol}$ \\
(Jy km s$^{-1}$) & (km s$^{-1}$) & ($10^{9}\,L_\odot$) & (mJy) & ($10^{12}\,L_\odot$) & ($10^{12}\,L_\odot$) & ($M_\odot\,{\rm yr}^{-1}$) & ($10^{8}\,M_\odot$) & ($10^{11}\,M_\odot$) \\
\hline
\hline
\multicolumn{9}{c}{\yswl} \\
\hline
$0.72 \pm 0.17$ & $322 \pm 56$ & $0.79 \pm 0.18$ & $0.04 \pm 0.05$ & $0.06 \pm 0.09$ & $0.09 \pm 0.13$ & $10.99 \pm 15.14$ & $0.04 \pm 0.06$ & $0.24 \pm 0.05$ \\
\hline
\multicolumn{9}{c}{\ysej} \\
\hline
$3.67 \pm 0.22$ & $360 \pm 16$ & $4.14 \pm 0.24$ & $0.95 \pm 0.14$ & $1.84 \pm 0.27$ & $2.59 \pm 0.39$ & $311.40 \pm 46.29$ & $1.18 \pm 0.17$ & $1.24 \pm 0.07$ \\
\hline
\end{tabular}
\end{table*}

\subsection{Lower limit of the number density of EUVLGs at cosmic dawn\label{sec:density}}

\begin{figure}
	\includegraphics[width=\linewidth]{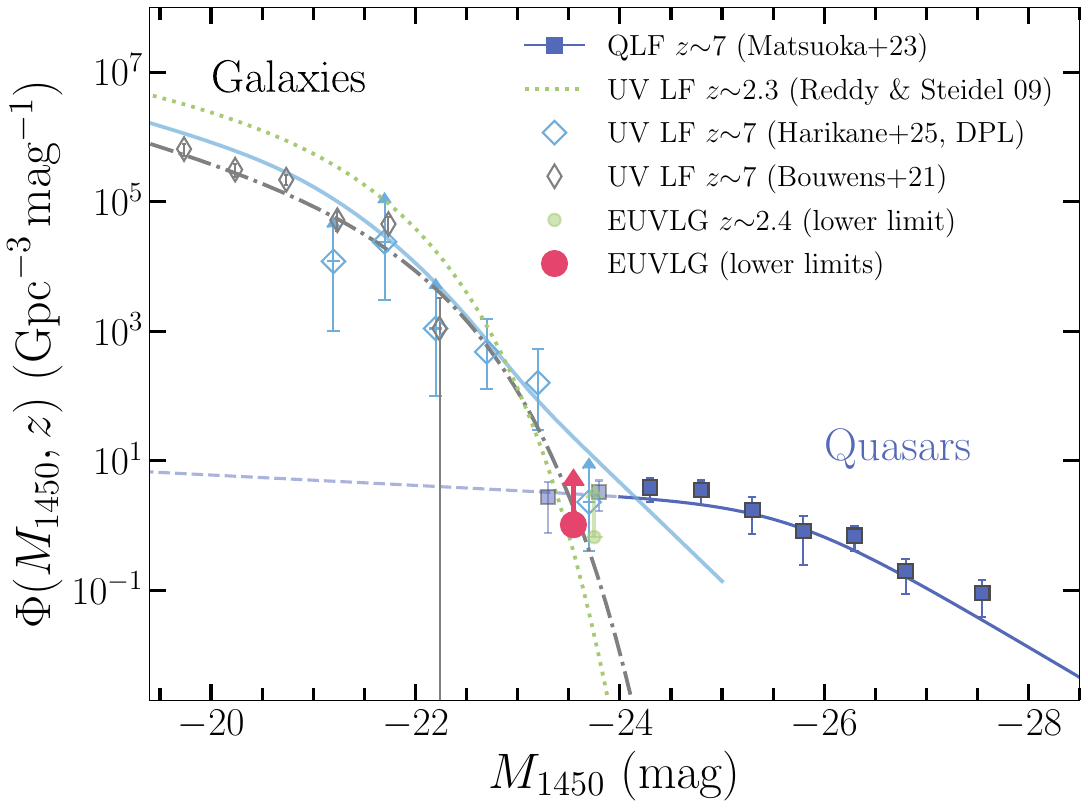}
    \caption{Lower limit on the EUVLG number density at $M_{1450} \approx -23.5$ and $z\sim7$ (red), derived from the SHELLQs selection volume (Sect.~\ref{sec:density}), compared with the $z\sim7$ quasar luminosity function \citep[QLF;][]{Matsuoka2023} and the $z\sim7$ galaxy UV luminosity functions of \citet{Bouwens2021} (Schechter) and \citet{Harikane2025} (DPL). Green points and curves show the cosmic-noon comparison: the $z\sim2.3$ star-forming galaxy UVLF of \citet{Reddy2009} (dotted), and a lower limit on the EUVLG number density at $z\sim2.4$ (circle), obtained by integrating the eBOSS EUVLG sample over $-24.5 < M_{1450} < -23$ at $2 < z < 4$ \citep[][]{Upadhyaya2024,Dessauges-Zavadsky2025}.}
    \label{fig:uvlf}
\end{figure}

The reclassification of \ysej\ and \yswl\ from quasars to star-forming galaxies raises a natural question: how common are such extreme UV-luminous galaxies during the epoch of reionisation?

Because both sources were originally included in the SHELLQs quasar sample, the survey volume and selection function needed to estimate their number density are already in place. \citet{Matsuoka2023} measured the quasar luminosity function at $z \sim 7$ with 35 spectroscopically confirmed objects over the $\sim$670\,deg$^2$ of the HSC-SSP footprint available at the time of that study. With our revised UV magnitudes (Sect.~\ref{sec:euvlg}; Table~\ref{tab:source_properties}), both sources fall at $M_{1450} \approx -23.5$, and we therefore merge the two adjacent bins in the \citet{Matsuoka2023} luminosity function that originally contained them. Of the six sources in these merged bins, two are the EUVLGs presented here. Taking the mean luminosity function of the merged bin and scaling by the EUVLG fraction (2/6), we obtain a lower limit of $\Phi_\mathrm{EUVLG} \gtrsim 0.28$\,Gpc$^{-3}$\,mag$^{-1}$ at $M_{1450} \approx -23.5$ and $z\sim 7$ (Fig.~\ref{fig:uvlf}). This should be regarded as a lower limit because: First, the SHELLQs selection targets quasar-like colours and prominent Ly$\alpha$ emission; EUVLGs with weaker Ly$\alpha$, different colors, or morphologies that fail the point-source criterion would not enter the sample. Second, only these two sources have been reclassified to date; additional SHELLQs objects at comparable luminosities and redshifts may likewise prove to be EUVLGs once NIR spectroscopy becomes available.

In Fig.~\ref{fig:uvlf} we also plot the galaxy UVLF at $z \sim 7$ from \citet{Bouwens2021} (Schechter function) and \citet{Harikane2025} (DPL). At $M_{1450} = -23.5$, these predict galaxy number densities of $\sim 3 \times 10^{-9}$ and $\sim 3 \times 10^{-8}$\,Mpc$^{-3}$\,mag$^{-1}$, respectively. Comparing our EUVLG lower limit to these values implies a fraction of $\sim$1--10\%. 
However, the bright end of the $z\sim 7$ galaxy UVLF remains loosely constrained. In \citet{Harikane2025} only four galaxies with $M_{\rm UV} < -23$ at $z\sim7$, and in their brightest bin, $-24.5 < M_{\rm UV} < -23.5$, the number density is only a lower limit, based on a single data point. Robustly quantifying the EUVLG fraction at $z\sim 7$ therefore awaits larger spectroscopic samples at these luminosities. Nonetheless, the EUVLG phase is likely short-lived if the UV light is boosted by the massive stars, and its appearance in even this small sample raises the possibility that such episodes are common among UV-luminous galaxies during reionisation. At the same time, the faint end of the $z\sim 7$ quasar luminosity function \citep{Matsuoka2023} should be revisited as more candidates receive spectroscopic follow-up. We also note that the GO~9180 target selection is biased towards SHELLQs sources with existing ALMA observations (Sect.~\ref{sec:obs_jwst}). This restricts which sources could have been identified as EUVLGs, but does not affect the lower limit itself.

Finally, the green points and dotted curve in Fig.~\ref{fig:uvlf} place the cosmic-noon EUVLG population in context, alongside the $z\sim 2.3$ star-forming galaxy UV luminosity function of \citet{Reddy2009}. The EUVLG number density is derived from the $\sim 70$ EUVLGs identified at $2 < z < 4$ with $-24.5 < M_{\rm UV} < -23$ over the $\sim 9000$\,deg$^2$ eBOSS/SDSS footprint \citep[][Marques-Chaves et al., in prep.]{Upadhyaya2024, Dessauges-Zavadsky2025}, corresponding to a comoving number density of $\sim 10^{-9}$\,Mpc$^{-3}$, or $\Phi_{\rm EUVLG} \sim 0.7$\,Gpc$^{-3}$\,mag$^{-1}$ when divided by the $1.5$\,mag width of the selection. The cosmic-noon EUVLGs lie well above the extrapolated bright end of the field-galaxy UVLF. However, it is too early to draw conclusions about the EUVLG fraction at either epoch or its evolution with redshift, since the bright end of the UVLF at $M_{\rm UV} < -23$ is poorly constrained at both cosmic noon and cosmic dawn. Both densities are also best regarded as lower limits, and pinning them down will require a systematic search for this population at both epochs.

\section{Discussion}\label{sec:discussion}

\subsection{A combined ultraviolet and far-infrared view}\label{sec:discussion_ir}

\begin{figure*}
	\includegraphics[width=\linewidth]{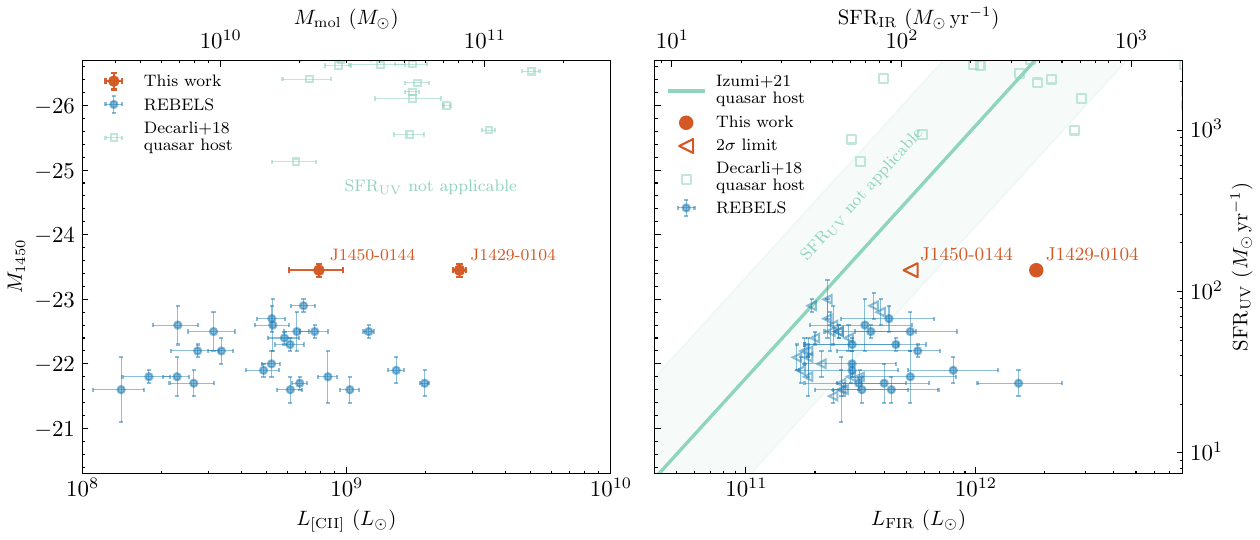}
    \caption{Rest-frame UV luminosity $M_{1450}$ against [\Cii]\,$158\,\mu{\rm m}$ luminosity (left) and far-infrared luminosity (right) for the two galaxies presented in this work (red) and the REBELS sample at $z\sim7$ \citep[blue;][]{Bouwens2022,Inami2022}. \textit{Left:} $L_{\rm [C\,\textsc{ii}]}$ on the lower axis, with $z\gtrsim6$ quasars for comparison \citep[light green;][]{Decarli2018}. The upper axis gives the molecular gas mass $M_{\rm mol}$ inferred from $L_{\rm [C\,\textsc{ii}]}$ via the \citet{Zanella2018} calibration. We caution that this conversion carries substantial systematic uncertainty and is calibrated predominantly on lower-$z$, less extreme star-forming galaxies. \textit{Right:} $L_{\rm FIR}$ (8--1000$\,{\rm \mu m}$) on the lower axis, with the quasar host locus of \citet{Izumi2021} for reference (light green). The upper and right axes give the obscured and unobscured star-formation rates $\mathrm{SFR_{IR}}$ and $\mathrm{SFR_{UV}}$ (see Sect.~\ref{sec:alma}). The $\mathrm{SFR_{UV}}$ does not apply to the quasar comparison samples, whose UV emission is AGN-dominated. \yswl\ is plotted as a $2\sigma$ upper limit on $L_{\rm FIR}$. Our galaxies occupy the extreme high-SFR regime in both unobscured and obscured star formation, appearing as a luminous extension of the REBELS population.}
    \label{fig:muv_lcii_lfir}
\end{figure*}

How do these two galaxies compare to other UV- or [\Cii]-luminous systems at the same epoch? In Fig.~\ref{fig:muv_lcii_lfir}, we compare them with the REBELS sample at $z\sim7$ \citep{Bouwens2022,Inami2022} on the $M_{1450}$--$L_{\rm FIR}$ and $M_{1450}$--$L_{\rm [CII]}$ planes. 
We also show the unobscured star-formation rate $\mathrm{SFR_{UV}}$ converted from $M_{1450}$ on the right axis, using the $\kappa_{\rm UV}$ factor from \citet{Upadhyaya2024} (see Sect.~\ref{sec:euvlg}); the upper axes give the molecular gas mass $M_{\rm mol}$ converted from $L_{\rm [CII]}$ using $\alpha_{\rm [CII]}$ \citep{Zanella2018}, and the obscured star-formation rate $\mathrm{SFR_{IR}}$ converted from $L_{\rm FIR}$ through the $L_{\rm TIR}$ \citep{Inami2022}, both as described in Sect.~\ref{sec:alma}.
We also overplot the $z\sim6$ quasar hosts of \citet{Decarli2018} and \citet{Izumi2021}, as these are among the most [\Cii]-luminous and dust-rich systems known at these redshifts. Both of our galaxies are $\gtrsim1$\,mag brighter in the rest-frame UV than typical REBELS galaxies, yet match the most FIR-luminous REBELS sources in [\Cii] and, for \ysej, in dust continuum. This places them among the most [\Cii]-luminous galaxies known at $z>6.5$, with [\Cii] luminosities approaching those of luminous quasar hosts. \ysej, with $L_{\rm FIR}=1.8\times10^{12}\,L_\odot$, sits firmly in the ultra-luminous infrared regime, and its obscured-to-total star-formation fraction $\mathrm{SFR_{IR}/(SFR_{IR}+SFR_{UV})}\approx 70\%$ is in line with the high obscured fractions seen in the dustiest REBELS sources \citep{Inami2022}. The implied depletion time $\tau_{\rm dep} = M_{\rm mol}/{\rm SFR_{tot}} \sim 300\,\mathrm{Myr}$ is comparable to that of main-sequence galaxies at similar redshifts, suggesting that the present burst can be sustained over a star-formation timescale provided the cold gas reservoir is not disrupted. We note, however, that both $M_{\rm mol}$ and ${\rm SFR_{tot}}$ are subject to substantial systematic uncertainties, and these numbers should therefore be taken as order-of-magnitude estimates.

The picture for \yswl\ is qualitatively different: a clear [\Cii] detection but no dust-continuum signal, and a molecular gas reservoir (inferred from [\Cii]) several times smaller. At face value, \yswl\ is at an earlier stage of the same process, consistent with the younger stellar population age inferred in Sect.~\ref{sec:sps}. But whether the two galaxies represent successive snapshots of the same evolutionary track, or whether \yswl\ has simply converted a smaller fraction of its cold ISM into obscured star formation cannot be settled with the present data. This will be an open question for ALMA follow-up of a larger EUVLG sample during reionisation.

\begin{figure}
	\includegraphics[width=\linewidth]{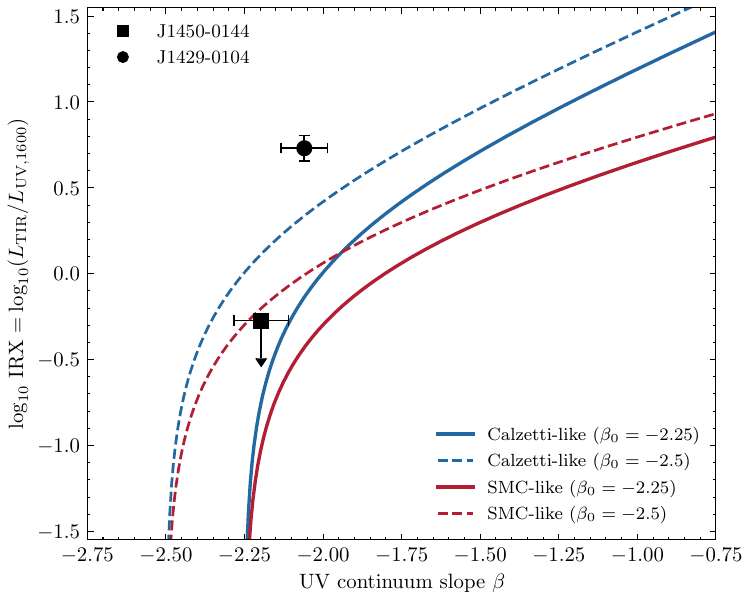}
    \caption{IRX--$\beta$ diagram for \yswl\ and \ysej. The infrared excess ${\rm IRX}=L_{\rm TIR}/L_{\rm UV,1600}$ is computed from $L_{\rm TIR}$ in Table~\ref{tab:alma} and the rest-frame UV luminosity at rest-frame wavelength $\lambda_{\rm rest}=1600\AA$. The UV continuum slopes $\beta$ are measured in Sect.~\ref{sec:feature_continua}. \ysej\ is detected in the dust continuum, while \yswl\ is shown as a $2\sigma$ upper limit on ${\rm IRX}$. Curves show the relations expected for Calzetti-like and SMC-like dust attenuation with $\beta_0=-2.25$ and $-2.5$.}
    \label{fig:irx_beta}
\end{figure}

Fig.~\ref{fig:irx_beta} further illustrates this contrast in the IRX--$\beta$ plane, comparing the infrared excess ${\rm IRX}=L_{\rm TIR}/L_{\rm UV,1600}$ with the observed UV continuum slopes (Sect.~\ref{sec:feature_continua}). \yswl\ is undetected in the dust continuum, and its $2\sigma$ IRX upper limit is consistent with the canonical relations and with little obscured star formation. \ysej, by contrast, lies well above both the Calzetti-like and SMC-like relations even for $\beta_0=-2.5$. A potential explanation is that the UV and FIR emission of \ysej\ arises from spatially distinct regions, consistent with the $\sim5.4\,\mathrm{kpc}$ offset between its rest-frame UV centroid and its [\Cii]/dust-continuum peak (Sect.~\ref{sec:alma}).
Comparable UV--FIR offsets have been reported in a subset of UV-bright galaxies at similar epochs: several REBELS sources show offsets of $\sim0.5$--$1.5^{\prime\prime}$ ($\sim3$--$8$\,kpc) between the rest-UV and dust peaks \citep{Inami2022,Bowler2024}, and roughly $30\%$ of ALPINE main-sequence galaxies at $z\sim4$--$6$ display significant ($>3\sigma$) offsets with a median of $\sim3.5$--$4.5$\,kpc \citep{Killi2024}. Several physical scenarios are typically invoked: an ongoing merger between an unobscured and an obscured component; a single galaxy with an inhomogeneous dust distribution, in which a heavily obscured central region hides the true star-formation peak while the observed UV centroid traces a less-attenuated, off-centre clump \citep{Bowler2024,Ocvirk2025}; or radiation-pressure-driven outflows from the young stellar population that displace dust away from the UV-emitting region \citep{Ferrara2023,Ziparo2023}. With the present data we cannot distinguish between these scenarios. High-resolution imaging in the rest-frame UV/optical and IR, will be critical to determine whether the offset reflects a merger, an obscured central starburst, or feedback-driven dust displacement.

\subsection{Very massive stars at reionisation: constraints and implications\label{sec:vms_discussion}}

\subsubsection{Constraining the upper-mass cutoff $M_{\rm up}$ of the initial mass function with UV wind diagnostics}\label{sec:discussion_mup}

\begin{figure*}
	\includegraphics[width=0.98\linewidth]{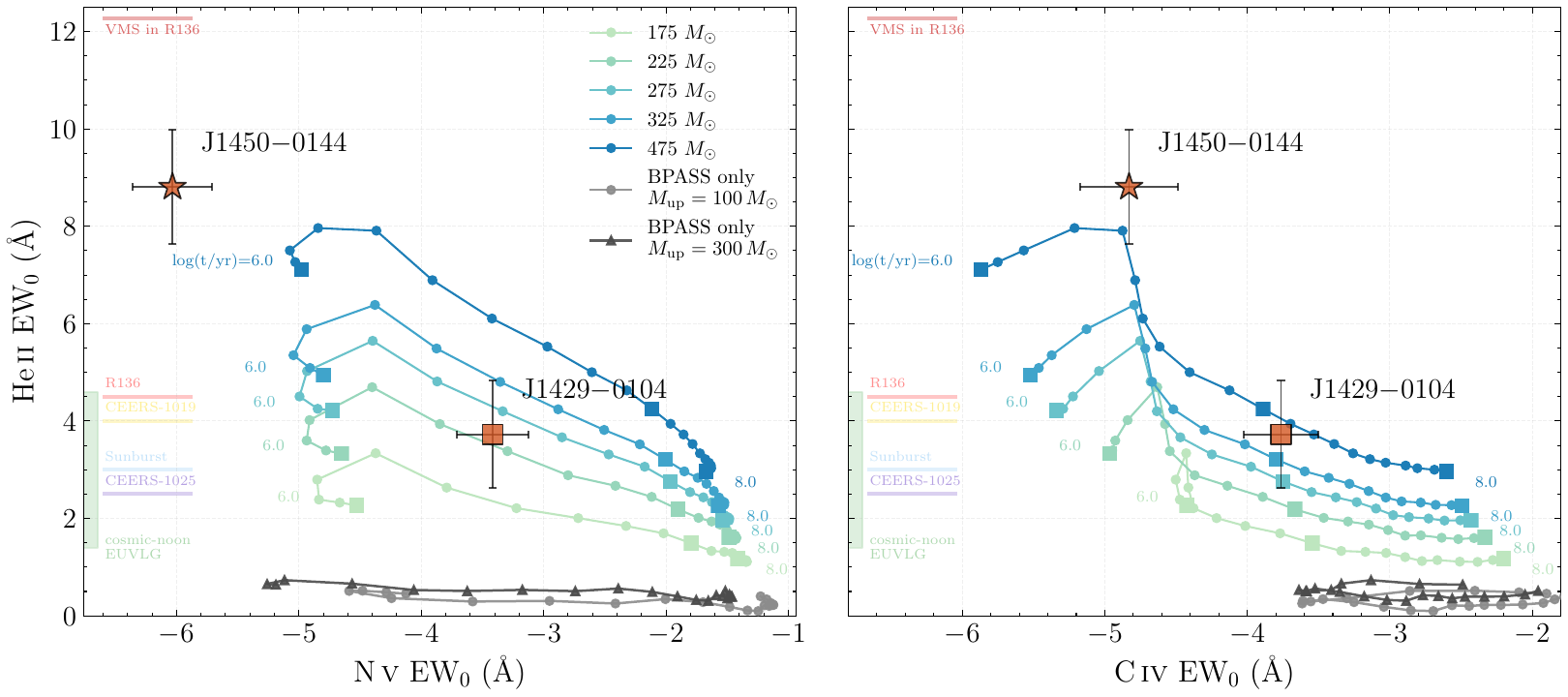}
    \caption{Diagnostic equivalent-width diagrams comparing the rest-frame UV wind features of \yswl\ (star) and \ysej\ (square) with \texttt{BPASS+VMS} population synthesis models from \citet{Upadhyaya2024}, which incorporate VMS spectroscopic models from \citet{Martins2022}. Left: \Heii\,$\lambda 1640$ vs.\ \Nv\,$\lambda 1240$ $\mathrm{EW}_0$. Right: \Heii\,$\lambda 1640$ vs.\ \Civ\,$\lambda 1550$ $\mathrm{EW}_0$. Coloured tracks correspond to models at $Z = 0.006$ with a Salpeter IMF and upper-mass cutoffs $M_{\rm up} = 175$, $225$, $275$, $325$, and $475\,M_\odot$ (light green to blue) with a constant star-formation. Filled circles along each track mark the evolution of the population from $\log(t/{\rm yr}) = 6.0$ to $8.0$ in steps of $\Delta\log(t/{\rm yr}) = 0.1$, with the $\log(t/{\rm yr}) = 6.0$ starting point labelled on each track. For reference, grey tracks show standard BPASS models without the \citet{Martins2022} models for $M_{\rm up} = 100\,M_\odot$ (circles) and $M_{\rm up} = 300\,M_\odot$ (triangles). Coloured bands at the left of each panel mark the \Heii\ $\mathrm{EW}_0$ of literature systems hosting confirmed or candidate VMS, placed along the \Heii\ axis only: The stacked spectrum of seven VMS in R136, the entire R136 cluster (\citealt{Crowther2016}), the $z \simeq 8.7$ galaxies CEERS-1019 and CEERS-1025 (\citealt{Marques-Chaves2026}), the Sunburst cluster (\citealt{Mestric2023, Welch2025}), and the cosmic-noon EUVLG sample, whose vertical band spans the observed range (\citealt{Upadhyaya2024}). Both galaxies fall in the region of the diagram occupied by very young, VMS-rich populations.}
    \label{fig:vms}
\end{figure*}

\begin{figure*}
	\includegraphics[width=0.98\linewidth]{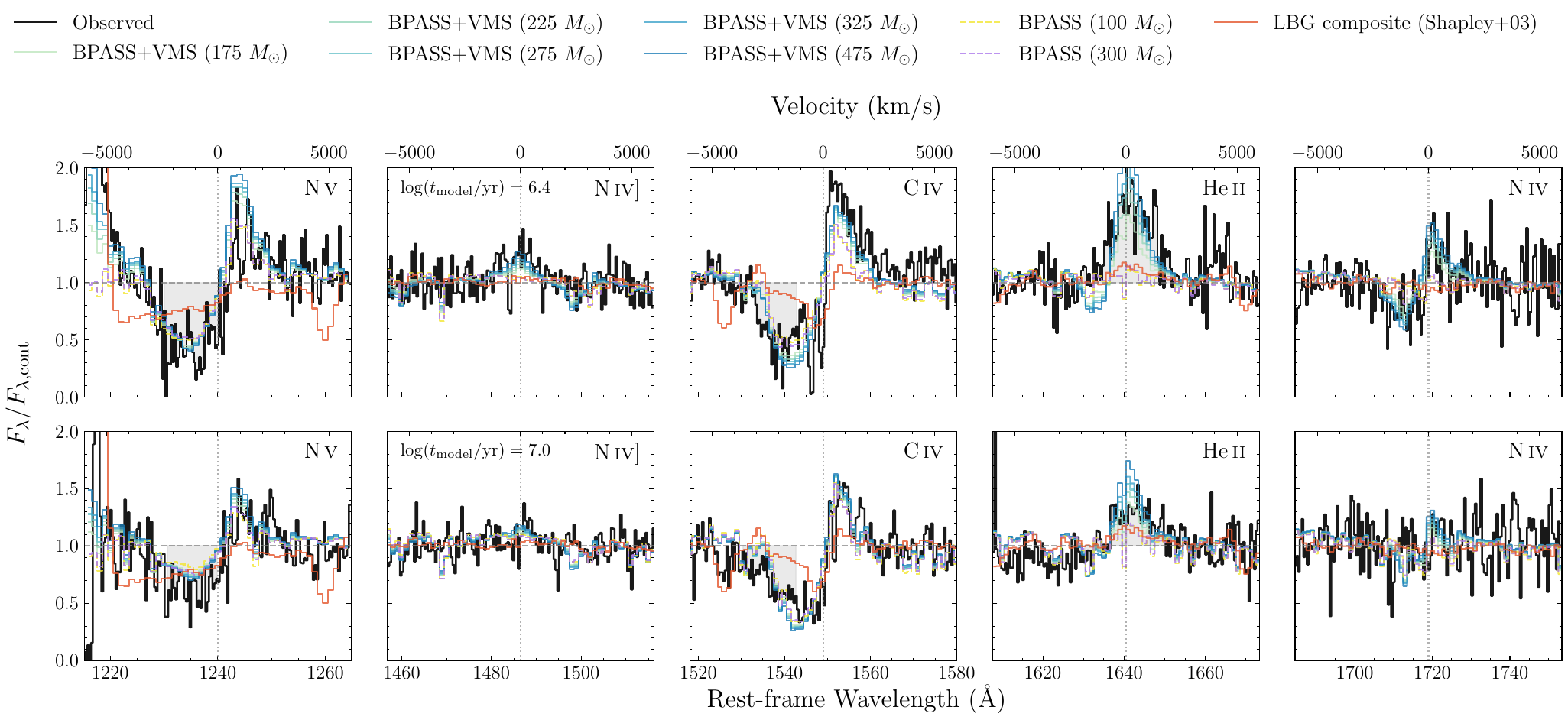}
    \caption{
    Comparison of the rest-frame UV spectra of \yswl\ (top row) and \ysej\ (bottom row) with BPASS population synthesis models including very massive stars, around the five key wind-line and emission diagnostics (left to right): \Nv\,$\lambda 1240$, N\,{\sc iv}]\,$\lambda 1486$, \Civ\,$\lambda 1550$, \Heii\,$\lambda 1640$, and N\,{\sc iv}\,$\lambda 1719$. Black curves show the observed spectra normalised by the local continuum. Coloured curves show BPASS+VMS models from \citet{Upadhyaya2024}, which incorporate the VMS stellar evolution and atmosphere models of \citet{Martins2022}, with IMF upper-mass limits $M_{\rm up} = 175$, $225$, $275$, $325$, and $475\,M_\odot$ (light to dark blue), evaluated at the inferred ages of $\log(t/{\rm yr}) = 6.4$ for \yswl\ and $7.0$ for \ysej\ based on the right panel of Fig.~\ref{fig:vms}. Dashed curves show standard BPASS models without VMS treatment for $M_{\rm up} = 100$ (yellow) and $300\,M_\odot$ (violet). The orange curve is the composite LBG spectrum from \citet{Shapley2003}, shown for reference. Vertical dotted lines mark the systemic wavelengths of each transition. The BPASS+VMS models reproduce the P~Cygni profiles in \Nv\ and \Civ\ and the broad \Heii\,$\lambda 1640$ emission, while standard BPASS and the LBG composite fall well short. \yswl\ also shows clear broad N\,{\sc iv}]\,$\lambda 1486$ and N\,{\sc iv}\,$\lambda 1719$ emission predicted by the VMS models; in \ysej, these features are weaker.
    }
    \label{fig:vms_lines}
\end{figure*}

Sect.~\ref{sec:sps} established that reproducing the broad \Heii\,$\lambda1640$ emission requires models with dedicated very massive star wind physics. To constrain the VMS content more directly, we follow \citet{Upadhyaya2024} and compare the data with the \texttt{BPASS+VMS} grid in the \Heii--\Nv\ and \Heii--\Civ\ equivalent-width planes. 
The three lines trace different parts of the massive-star population. \Nv\ and \Civ\ are resonance lines of high-ionisation states. Even winds of modest density imprint P~Cygni profiles on the photospheric continuum, while maintaining the high ionisation requires the hard radiation field of the hottest stars. They are therefore sensitive tracers of young massive stars in general, not necessarily the very massive ones, and their strengths decline as the population ages. \Heii\ is not a resonance line, so it cannot form by scattering the stellar continuum. Instead it is produced by recombination in the wind itself, with an emissivity that scales as the square of the wind density \citep{Grafener2021, Martins2022}. In the models, only VMS near the Eddington limit sustain winds dense enough to power strong broad \Heii, so its equivalent width serves as a good tracer of the VMS population.
Following \citet{Upadhyaya2024}, the rest-frame equivalent widths are measured by direct integration over fixed windows in both the data and the synthetic spectra (\Heii\ over $1630$--$1655$\,\AA, \Civ\ over $1530$--$1543$\,\AA, \Nv\ over $1230$--$1240$\,\AA), using the spline continuum of Sect.~\ref{sec:specmod} for the data and local sidebands for the models. Adopting local sidebands for the data instead would slightly raise the measured ${\rm EW}_0$ of both sources, moving them further into the VMS regime.

Fig.~\ref{fig:vms} shows the comparison, with \Heii\ ${\rm EW}_0$ plotted against \Nv\ ${\rm EW}_0$ (left) and \Civ\ ${\rm EW}_0$ (right). Coloured tracks correspond to different upper-mass limits, from $M_{\rm up} = 175$ to $475\,M_\odot$, each evolving over $\log(t/{\rm yr}) = 6$--$8$. We also show the two \texttt{BPASS-only} tracks in grey as references. We can see that even extending the IMF to $M_{\rm up} = 300\,M_\odot$, the strength of the \Heii\ emission remains $\lesssim 1$\,\AA\ because their massive stars never develop the optically-thick winds. The measurements for the two galaxies are overplotted as a star (\yswl) and a square (\ysej). Coloured bands at the left of each panel mark the \Heii\ ${\rm EW}_0$ of literature systems (star clusters and galaxies) hosting confirmed or candidate VMS, placed along the \Heii\ axis only: the R136 cluster and the seven VMS within it \citep{Crowther2016}, the $z \simeq 8.7$ galaxies CEERS-1019 and CEERS-1025 \citep{Marques-Chaves2026}, the Sunburst cluster \citep{Mestric2023, Welch2025}, and the cosmic-noon EUVLG sample \citep{Upadhyaya2024}. As a complementary check on the line profiles, Fig.~\ref{fig:vms_lines} overlays the same model grid, evaluated at the closest-matching ages in the \Heii\ versus \Civ\ diagram for the two galaxies ($\log(t/{\rm yr}) = 6.4$ for \yswl\ and $7.0$ for \ysej), on the observed spectra around the five key wind diagnostics (\Nv, \Civ, \Heii\ plus N\,{\sc iv}]\,$\lambda 1486$ and N\,{\sc iv}\,$\lambda1719$). 

We can clearly see that for \yswl, its extreme \Heii\ ${\rm EW}_0$ with a very deep \Nv\ trough place it above the majority of the literature values and beyond the model grid in both panels. Only the composite of the seven VMS in R136 has a stronger but comparable \Heii. \ysej\ lies further along the tracks, suggesting it being a more evolved system than \yswl, and its position in the \Heii--\Nv\ plane is consistent with the $M_{\rm up} = 225\,M_\odot$ track. The relative strengths of \Civ\ and \Nv\ are, however, not fully consistent with the models for either galaxy: at the positions implied by \Heii\ and \Nv, the observed \Civ\ troughs are shallower than predicted, so the \Heii--\Civ\ plane favours somewhat older populations. As seen in Fig.~\ref{fig:vms_lines}, the observed \Civ\ profiles show more complex structure than the models, and emission features superposed on the P-Cygni trough may bias the integrated ${\rm EW}_0$ towards less negative values. Detailed modelling of this profile is deferred to future work.

One minor inconsistency is that the \texttt{BPASS+VMS} models generally predict a P~Cygni profile for \Heii\ as well (see Fig.~\ref{fig:vms_lines}), which is not evident in our spectra (tentative in \ysej), plausibly because the stellar wind is not optically thick to this non-resonant transition. 
We confirmed that repeating the \Heii\ ${\rm EW}_0$ measurements using only the red emission wing of the line for both the data and the models leaves the conclusions unchanged, i.e. both galaxies remain most consistent with $M_{\rm up} \gtrsim 225-475\,M_\odot$.

For \yswl, the diagnostic favours a higher upper-mass limit than the template fitting in Sect.~\ref{sec:sps} ($M_{\rm up} = 275\,M_\odot$). This is somewhat expected as the template fitting need to consider several P~Cygni profiles, which respond only weakly to $M_{\rm up}$, whereas the diagnostic isolates the \Heii\ emission that traces the VMS content most directly. Fig.~\ref{fig:vms_lines} indeed shows that, at these young ages, the models with the largest $M_{\rm up}$ provide a better match to the observed \Heii\ profiles. However, we note that these $M_{\rm up}$ constraints are strongly model dependent. The \citet{Grafener2021} mass-loss prescription underlying the grid is calibrated on the VMS in the LMC, and no empirical constraints on VMS winds exist at other metallicities. \citet{Martins2025} show that the predicted \Heii\ strength at fixed $M_{\rm up}$ changes drastically depending on whether or not VMS mass-loss rates scale with metallicity. In addition, the grid fixes the IMF slope at the Salpeter value. We therefore consider the presence of a substantial VMS population as robust, since no model without appropriate VMS wind physics approaches the observed \Heii\ strengths, while pinning down $M_{\rm up}$ and the detailed shape of the high-mass IMF will require VMS wind models calibrated beyond the LMC and deeper spectroscopy of larger samples.

\subsubsection{The number of very massive stars in these two galaxies}\label{sec:discussion_nvms}

Besides the upper-mass cutoff, another quantity of interest is the total number of VMS that these two galaxies host. For a galaxy with $10^9\,M_\odot$ solar masses formed in a burst with a Salpeter IMF extending to $M_{\rm up} = 300\,M_\odot$, direct integration of the IMF gives $N_{\rm VMS} \approx 2\times10^{5}$ stars above $100\,M_\odot$ at birth. Given the stellar masses of $\log(M_\star/M_\odot) \gtrsim 9$ and ages comparable to the VMS lifetime inferred in Sect.~\ref{sec:sps}, both galaxies are thus expected to host of order $\gtrsim10^{5}$ VMS at birth and a comparable number of living VMS.

We cross-check it based on the UV continuum luminosity. \citet{Martins2022} show that, in a young starburst with a Salpeter IMF extended to $300\,M_\odot$, stars with $M\gtrsim100\,M_\odot$ contribute nearly as much UV light as the rest of the population combined (their Fig.~8), so that VMS account for $f_{\rm VMS}\sim0.5$ of the integrated UV continuum. The depth of the \Civ\ P~Cygni troughs in our spectra supports a comparably large contribution from strong-wind stars. In the integrated spectrum, the trough is a luminosity-weighted average over stars with and without the wind feature. That is, if the absorption in the individual strong-wind stars is saturated, the residual flux at the trough bottom measures the fraction of the continuum contributed by stars without such winds. The observed troughs reach $F_\lambda/F_{\rm cont}\approx0.2$--$0.5$ (Sect.~\ref{sec:feature_pcygni}), implying that $\sim50$--$80$ per cent of the UV continuum arises from stars with dense winds. Since ordinary early-O stars also contribute to the \Civ\ absorption, and incomplete saturation or velocity-dependent overlap of the individual profiles would fill in the trough, this is an upper limit on $f_{\rm VMS}$, broadly consistent with $f_{\rm VMS}\sim0.5$. The number of VMS is then $N_{\rm VMS}\approx f_{\rm VMS}\,L_{\lambda,{\rm UV}}/l_{\lambda,{\rm UV}}^{\rm VMS}$, where $l_{\lambda,{\rm UV}}^{\rm VMS}\sim10^{36}$--$10^{37}\,{\rm erg\,s^{-1}\,\text{\AA}^{-1}}$ for individual VMS with $M\sim150$--$400\,M_\odot$ \citep{Martins2022}. For $M_{\rm UV}=-23.5$ ($L_{\lambda,{\rm UV}}\approx1.5\times10^{42}\,{\rm erg\,s^{-1}\,\text{\AA}^{-1}}$), this yields $N_{\rm VMS}\sim10^{5}$--$10^{6}$, in agreement with the fit-based estimate.

We could imagine that a VMS population of this size could significantly impact the entire galaxy. In the following sections, we discuss what these VMS imply for the high-mass end of the IMF (Sect.~\ref{sec:discussion_imf}), how their winds may enrich the surrounding ISM (Sect.~\ref{sec:discussion_chemical}), and the potential large number of pair-instability supernovae and stellar-mass black holes they will leave behind (Sect.~\ref{sec:discussion_fate}).

\subsubsection{Very massive star formation in EUVLGs and the initial mass function}\label{sec:discussion_imf}

For decades the stellar upper mass limit was thought to be $\lesssim 150\,M_\odot$ \citep[e.g.][]{Figer2005}. This limit has since been revised upward, first by re-analyses of the WNh stars in R136 \citep{Crowther2010,Crowther2016} and the $\sim200\,M_\odot$ stars in the Sunburst LyC cluster at $z=2.37$ \citep{Vanzella2022,Mestric2023}, and more recently by VMS signatures in cosmic-noon EUVLGs \citep{Upadhyaya2024}, in JWST-discovered high-$z$ galaxies \citep{Marques-Chaves2026,Chen2026}, and in the two $z>6$ EUVLGs presented in this work. The local VMS hosts are all young, dense star clusters with central densities $\rho_\star\gtrsim10^{4}\,M_\odot\,{\rm pc}^{-3}$, consistent with theoretical expectations that VMS form in such environments via direct accretion, binary mergers, or runaway collisions \citep[e.g.][]{PortegiesZwart1999,Krumholz2009,Banerjee2012,Vink2015}. The compact sizes, high gas surface densities, and low metallicities common at $z\gtrsim6$ make the epoch of reionisation favourable for VMS formation.

Testing this picture for \yswl\ and \ysej\ requires knowing their internal structure. Reionisation-epoch galaxies observed with JWST are preferentially clumpy in the rest-frame UV, with $\sim150$--$250$\,pc clumps and star-forming complexes identified at $z\simeq6$--$8$ \citep{Chen2023,Zhu2026}. Our current imaging resolves scales of only $\sim0.1^{\prime\prime}$ ($\sim0.5$\,kpc at these redshifts), a factor of $\sim2$--$3$ too coarse to identify such clumps, let alone individual giant molecular clouds. Higher-resolution JWST/NIRCam imaging, or follow-up of strongly lensed analogues, will be needed to link these galaxies to the compact-cluster channel seen locally.

On the IMF side, the detection of a substantial VMS population does not, on its own, require a top-heavy IMF. A standard Salpeter or Kroupa IMF extended to $M_{\rm up}\sim200$--$500\,M_\odot$ already yields $\gtrsim10^{5}$ stars above $100\,M_\odot$ for a galaxy with $\log(M_\star/M_\odot)\sim9$ (Sect.~\ref{sec:discussion_nvms}). The principal observational lever is therefore not the existence of VMS but the upper-mass cutoff $M_{\rm up}$. At the same time, our constraints on $M_{\rm up}$ in Sect.~\ref{sec:discussion_mup} should be regarded as only indicative, since several physical inputs of current VMS models are not yet well understood \citep[e.g.][]{Grafener2021,Martins2022,Sabhahit2023,Schaerer2025}.

Even so, if VMS are widespread at $z\gtrsim6$, they dominate the integrated ionising-photon production, kinetic wind energy, and metal yields per unit stellar mass formed. Standard SFR and stellar-mass calibrations with IMF cutoffs of $100$--$150\,M_\odot$ do not capture these contributions. If compact starbursts of the kind seen here are a common phase among UV-luminous reionisation-epoch galaxies, the high-mass IMF cutoff deserves systematic re-examination in observational calibrations and cosmological simulations.

\subsubsection{Chemical yields and the connection to high-redshift N-emitters}\label{sec:discussion_chemical}

JWST has revealed an unexpected population of high-redshift galaxies with $\log(\mathrm{N/O})$ above the solar value at sub-solar metallicity \citep[e.g.,][]{Bunker2023,Cameron2023,Marques-Chaves2024,Schaerer2024,Ji2024,Topping2024,Berg2025}, roughly an order of magnitude higher than the local N/O--O/H relation \citep{Cataldi2025}. The inferred stellar ages of many of these systems are far shorter than the $\gtrsim30$\,Myr needed for intermediate-mass AGB stars, the standard channel for primary nitrogen enrichment, to return their CNO-processed envelopes to the ISM \citep[e.g.,][]{Cameron2023,Marques-Chaves2024}.

A more rapid alternative is the direct release of CNO-cycle H-burning products through massive-star winds during main-sequence evolution. Proposed polluters operating on Myr timescales include classical Wolf--Rayet stars \citep[e.g.,][]{Isobe2023}, VMS \citep{Vink2023,Shi2026a,Cameron2026}, and hypothetical supermassive stars \citep[SMS; $M\sim10^{3}$--$10^{4}\,M_\odot$;][]{Charbonnel2023}. \citet{Vink2023} argued that VMS are a particularly promising channel. They are observationally established from the local universe to the epoch of reionisation \citep[][and this work]{Marques-Chaves2026}. Their large convective cores and proximity to the Eddington limit enrich their surfaces in N already on the main sequence, while N is being synthesised rather than destroyed. Their dense, optically thick winds then eject this material at terminal velocities well below those of cWR stars, which favours retention within the host star-forming region.

In \yswl\ we detect clear broad \Niv]\,$\lambda1486$ emission together with a prominent \Niv\,$\lambda1719$ P~Cygni feature, while in \ysej\ the \Niv]\ detection is only marginal (Fig.~\ref{fig:vms_lines}). These are likely stellar wind signatures arising in the dense, accelerating outflows of WNh-type VMS \citep{Martins2022}, and they probe the surface and wind composition of the stars rather than the ambient ISM. The \Niii]\,$\lambda1750$ line, the typical nebular signature of JWST-discovered N-emitters, is not detected in our spectra, while \Ciii]\,$\lambda1909$ and \Oiii]\,$\lambda1666$ are detected (Table~\ref{tab:nebular_lines}). The \Ciii] line is broad, however, with FWHM exceeding $1000\,{\rm km\,s}^{-1}$, which suggests a partial stellar-wind contribution. The absence of nebular nitrogen features alongside clear VMS wind signatures could indicate that the enriched material has not yet mixed into the ionised ISM, that the radiation field properties differ, or that the gas-phase abundance pattern genuinely differs from that of canonical N-emitters. We caution that the non-detection of a single line is not an abundance measurement. Either way, EUVLGs with clear stellar VMS signatures are ideal laboratories for testing how VMS yields are transferred into the surrounding ISM. A gas-phase N/O measurement would require deeper limits on nebular \Niv]\ and \Niii]\,$\lambda1750$ relative to \Oiii]\,$\lambda1666$, anchored by rest-frame optical [O\,{\sc iii}] and H$\beta$ to constrain the oxygen abundance and reddening.

\subsubsection{The endpoint of an EUVLG phase}\label{sec:discussion_fate}

VMS lifetimes are short \citep[$\lesssim3$\,Myr; e.g.,][]{Yusof2013,Vink2015,Costa2025}, so at the ages derived in Sect.~\ref{sec:sps} many members of the VMS population must already be reaching the end of their evolution. Their fates depend sensitively on the zero-age main-sequence mass (ZAMS) and on the uncertain wind mass-loss history \citep{Woosley2015,Vink2015b}. At SMC-like metallicity \citep{Yusof2013}, stars of $\sim70$--$100\,M_\odot$ undergo pulsational pair-instability supernovae, which eject the outer layers in a series of thermonuclear pulses but leave $\sim40\,M_\odot$ black holes. Stars of $\sim100$--$290\,M_\odot$ are fully disrupted as pair-instability supernovae (PISNe) and leave no remnant. Above $\sim290\,M_\odot$, the star collapses directly to a black hole with little or no mass ejection. These boundaries are sensitive to mass loss, rotation, and convective overshooting, and vary substantially between stellar-evolution codes \citep{Woosley2015}. For the $N_{\rm VMS}\sim10^{5}$--$10^{6}$ estimated for the two galaxies (Sect.~\ref{sec:discussion_nvms}), the implied PISN rate is of order $10^{-3}$--$10^{-2}\,{\rm yr}^{-1}$ per galaxy, depending on the exact star-formation history.

A potentially more important question is the collective fate of the host clusters themselves. The stars that evade the PISN channel will leave behind a large population of black hole remnants, from the $\sim40\,M_\odot$ PPISN products to the considerably more massive remnants of direct collapse, concentrated in the compact star-forming cores. Whether these remnants can grow significantly by accreting the leftover in-situ gas is unclear. Idealised simulations of compact GMCs by \citet{Shi2026b}, with initial cloud masses up to $\sim10^{9}\,M_\odot$ within $\sim500$\,pc (broadly comparable to the stellar masses and compact morphologies of our two sources), suggest that stellar feedback typically expels the surrounding gas before significant accretion onto the remnants can occur, so individual black holes remain at $\lesssim500\,M_\odot$ within the first $\sim10$\,Myr. Even this bound is reached only in their most extended, lowest-metallicity ($Z=0.01\,Z_\odot$) configuration. At the $Z\sim0.2\,Z_\odot$ inferred for our galaxies, stronger winds and earlier gas expulsion are thought to suppress in-situ growth even further. These results also rest on a limited set of cloud configurations and sub-grid prescriptions, and do not follow the long-term post-starburst evolution. A complementary dynamical channel bypasses this bottleneck. In sufficiently dense young clusters, runaway main-sequence stellar collisions have been proposed to assemble a single supermassive object that collapses into an intermediate-mass black hole (IMBH) seed of $\sim10^{3}\,M_\odot$ \citep[e.g.,][]{PortegiesZwart2004,Gurkan2004,Shi2021}. Repeated tidal disruptions, BH--BH mergers, and hierarchical cluster mergers could in principle grow these seeds further \citep[e.g.,][]{Rantala2024}. Whether any of these pathways can bridge the gap to the $\sim10^{8}$--$10^{9}\,M_\odot$ SMBHs observed in luminous $z\sim6$--$7$ quasars \citep{Fan2023} remains an open question \citep{Inayoshi2020,Volonteri2021}. These high-redshift EUVLGs, which combine large VMS populations, compact star-forming structures, and relatively low metallicities, could nonetheless provide distinctive hints for discriminating between these seeding and growth scenarios.


\section{Conclusions}\label{sec:conclusions}

We have presented JWST/NIRSpec and ALMA Band-6 observations of two SHELLQs sources at $z>6.5$, \yswl\ ($z=6.63$) and \ysej\ ($z=6.80$), originally classified as faint quasars on the basis of their quasar-like UV luminosities ($M_{\rm UV}\approx-23.5$), strong Ly$\alpha$ emission and blue UV continua \citep{Matsuoka2018a,Matsuoka2022}. The new spectroscopy overturns this classification and yields a consistent physical picture of both sources. We summarise our key findings below.

\begin{enumerate}[label=\arabic*.,leftmargin=2.5em,labelsep=0.5em,labelwidth=1.5em,align=left,itemsep=0.5em]

\item \textit{\textbf{Two ``high-$z$ quasars'' reclassified as EUVLGs at the epoch of reionisation.}} The rest-frame UV spectra of both sources exhibit blue continua, prominent P~Cygni profiles in \Nv\,$\lambda1240$, \Siiv\,$\lambda1400$, and \Civ\,$\lambda1550$, broad \Heii\,$\lambda1640$ emission, and narrow nebular lines including \Oii\,$\lambda\lambda3727,3730$ and \Neiii\,$\lambda\lambda3870,3969$. Neither source shows broad Ly$\alpha$ or \Mgii\,$\lambda\lambda2796,2804$ emission. These properties are inconsistent with a type-I quasar interpretation. Instead, both spectra closely match the cosmic-noon EUVLG population \citep{Marques-Chaves2020,Marques-Chaves2021,Marques-Chaves2022,Upadhyaya2024} and resemble the stacked spectrum of the seven most massive stars in the R136 cluster \citep{Crowther2016}. \yswl\ and \ysej\ are therefore the first spectroscopically confirmed EUVLGs at $z>6.5$ with $M_{\rm UV}$ brighter than $-23$ (Sect.~\ref{sec:euvlg}).

\item \textit{\textbf{Young, intensely star-forming stellar populations with very massive stars.}} We fit the continuum-normalised \Nv, \Civ, and \Heii\ wind profiles of both galaxies with \texttt{BPASS-only} and \texttt{BPASS+VMS} models at $Z=0.006$ under a constant star-formation history. The \texttt{BPASS-only} models reproduce the \Nv\ and \Civ\ P~Cygni profiles but underpredict the broad \Heii\ emission (FWHM\,$\simeq1900$ and $2200\,{\rm km\,s}^{-1}$, $\mathrm{EW}_0=8.8\pm1.2$ and $3.7\pm1.1$\,\AA, for \yswl\ and \ysej, respectively), whereas the \texttt{BPASS+VMS} models reproduce all three features, and the dedicated VMS wind physics of \citet{Martins2022} may be the key ingredient. The best-fit \texttt{BPASS+VMS} templates additionally predict broad \Niv]\,$\lambda1486$ emission and a \Niv\,$\lambda1719$ P~Cygni profile, two further characteristic VMS signatures. Both are clearly detected in \yswl, while in \ysej\ the \Niv]\ emission is only marginal. Under the \texttt{BPASS+VMS} models, \yswl\ hosts a tightly constrained, very young population, with a star-formation duration of $2$--$4$\,Myr and $\log(M_\star/M_\odot)\approx9.2$. For \ysej\ the duration and $M_{\rm up}$ are degenerate: only the combination of ages $\lesssim3$\,Myr and $M_{\rm up}\gtrsim325\,M_\odot$ is excluded, and the inferred stellar mass spans $\log(M_\star/M_\odot)\approx9.5$--$9.9$ across the acceptable region. Scaling the best-fit templates to the observed flux implies star-formation rates of ${\rm SFR}\simeq540$ and $\simeq300\,M_\odot\,{\rm yr}^{-1}$ for \yswl\ and \ysej, respectively (Sect.~\ref{sec:sps}).

\item \textit{\textbf{Massive cold gas reservoirs and contrasting obscured star formation.}} ALMA reveals luminous [\Cii]\,$158\,\mu$m emission in both sources ($L_{\rm [CII]}\simeq0.8$ and $4.1\times10^{9}\,L_\odot$ for \yswl\ and \ysej, respectively) and a clear dust-continuum detection in \ysej\ ($L_{\rm TIR}\simeq2.6\times10^{12}\,L_\odot$, $\mathrm{SFR_{IR}}\simeq310\,M_\odot\,{\rm yr}^{-1}$), while \yswl\ remains undetected in the continuum. Both galaxies are $\gtrsim1$\,mag brighter in the rest-frame UV than typical REBELS galaxies yet match the most [\Cii]-luminous REBELS sources, and the implied molecular gas masses of $M_{\rm mol}\sim10^{10}$--$10^{11}\,M_\odot$ correspond to depletion times of a few hundred Myr, sufficient to sustain the present star-formation episodes provided the gas reservoirs are not disrupted. The two galaxies nonetheless differ sharply in obscuration: \ysej\ sits in the ultra-luminous infrared regime with an obscured-to-total star-formation fraction of $\approx70$ per cent and lies well above the Calzetti-like and SMC-like IRX--$\beta$ relations, whereas the $2\sigma$ IRX upper limit of \yswl\ is consistent with little obscured star formation. A potential explanation for the IRX excess of \ysej\ is that its UV and FIR emission arise from spatially distinct regions, consistent with the $\sim1''$ ($\sim5.4$\,kpc) offset between its rest-UV centroid and the [\Cii]/dust peak, whose origin (a merger, an obscured central starburst, or feedback-driven dust displacement) require future high-resolution imaging (Sect.~\ref{sec:alma} and \ref{sec:discussion_ir}).

\item \textit{\textbf{EUVLGs are not vanishingly rare during reionisation.}} Using the SHELLQs selection volume, we derive a lower limit on the EUVLG number density of $\Phi_{\rm EUVLG}\gtrsim0.28\,\mathrm{Gpc}^{-3}\,\mathrm{mag}^{-1}$ at $M_{1450}\approx-23.5$ and $z\sim7$. Compared with the $z\sim7$ galaxy UVLFs, this corresponds to an occupation fraction of $\gtrsim1$--$10$ per cent for the EUVLG phase among UV-luminous galaxies at this magnitude, itself a lower limit given the quasar-oriented SHELLQs selection. Robust quantification awaits larger spectroscopic samples at these luminosities (Sect.~\ref{sec:density}).

\item \textit{\textbf{Constraints on the upper-mass cutoff and the size of the VMS population.}} In the \Heii--\Nv\ and \Heii--\Civ\ equivalent-width planes, \ysej\ is consistent with \texttt{BPASS+VMS} models with $M_{\rm up}\gtrsim225\,M_\odot$, while \yswl\ lies beyond even our most extreme grid, implying $M_{\rm up}\gtrsim475\,M_\odot$. We note that these cutoffs are strongly model dependent, but the presence of a substantial VMS population is robust. The depth of the \Civ\ P~Cygni troughs implies that $\sim50$--$80$ per cent of the UV continuum arises from stars with dense winds, an upper limit on the VMS contribution broadly consistent with the $f_{\rm VMS}\sim0.5$ expected for a young population with a Salpeter IMF extending to $300\,M_\odot$. Both direct IMF integration and the UV luminosity budget imply $N_{\rm VMS}\sim10^{5}$--$10^{6}$ in each galaxy (Sect.~\ref{sec:discussion_mup} and \ref{sec:discussion_nvms}).

\end{enumerate}

Taken together, these results identify \yswl\ and \ysej\ as a new class of laboratory at the very bright end of the UVLF at high redshift: compact, gas-rich, intensely star-forming galaxies that simultaneously host substantial very massive star populations. They sit at the intersection of several open problems --- the IMF in low-metallicity starbursts, the origin of the nitrogen enhancement at high redshift, the escape of ionising radiation from luminous starbursts, and perhaps the formation of supermassive black holes through dense-cluster pathways. None of these connections is resolved by the present data alone, but the existence of even two such systems in a small reclassified sample suggests that many more EUVLGs await identification in the wide-area surveys of the coming decade, including \textit{Euclid} and the \textit{Roman Space Telescope}, which together will begin to chart this population across the reionisation epoch.


\section*{Acknowledgements}

D.~Yang thanks the members of the ENIGMA group at UC Santa Barbara and Leiden University for helpful discussions and valuable feedback. We thank A.~Upadhyaya for sharing the \texttt{BPASS+VMS} models, J.~J.~Eldridge and E.~R.~Stanway for sharing the \texttt{BPASS} models, and P.~Sharda, M.~R.~Krumholz, A.~J.~Cameron, and D.~Schaerer for insightful discussions. 
DY and JFH acknowledge support from the European Research Council (ERC) under the European Union's Horizon 2020 research and innovation program (grant agreement No.~885301). JFH acknowledges support from NSF grant No.~2307180. RAM acknowledges support from the Swiss National Science Foundation (SNSF) through project grant 200020\_207349. HSBA gratefully acknowledges support from Academia Sinica through grant AS-PD-1141-M01-2. J.-T.S. and F.G. acknowledge funding by the Deutsche Forschungsgemeinschaft (DFG, German Research Foundation) - Project number 518006966 - and support by the DFG under Germany’s Excellence Strategy – EXC 2121 „Quantum Universe“ – 390833306.
This work is based on observations made with the NASA/ESA/CSA James Webb Space Telescope, associated with program GO~9180. The data were obtained from the Mikulski Archive for Space Telescopes at the Space Telescope Science Institute, which is operated by the Association of Universities for Research in Astronomy, Inc., under NASA contract NAS~5-03127 for JWST. This paper makes use of the following ALMA data: ADS/JAO.ALMA\#2023.1.00443.S ALMA is a partnership of ESO (representing its member states), NSF (USA) and NINS (Japan), together with NRC (Canada), NSTC and ASIAA (Taiwan), and KASI (Republic of Korea), in cooperation with the Republic of Chile. The Joint ALMA Observatory is operated by ESO, AUI/NRAO and NAOJ.

Claude (Anthropic) and ChatGPT (OpenAI) were used to assist with language editing and revision of the manuscript text and with the development of plotting scripts used to produce the figures. All scientific analysis, results, and conclusions are the authors' own, and the authors take full responsibility for the content of this paper.

\section*{Data Availability}
The JWST/NIRSpec data underlying this article are available in the Mikulski Archive for Space Telescopes (MAST) under program GO 9180, and the ALMA data are available in the ALMA Science Archive under project code 2023.1.00443.S. The reduced spectra, calibrated data cubes, and other data products generated in this work will be shared on reasonable request to the corresponding author.



\bibliographystyle{mnras}
\bibliography{ref} 

@article{Fan2023,
	adsurl = {https://ui.adsabs.harvard.edu/abs/2023ARA&A..61..373F},
	archiveprefix = {arXiv},
	author = {{Fan}, Xiaohui and {Ba{\~n}ados}, Eduardo and {Simcoe}, Robert A.},
	doi = {10.1146/annurev-astro-052920-102455},
	eprint = {2212.06907},
	journal = {\araa},
	month = aug,
	pages = {373-426},
	primaryclass = {astro-ph.GA},
	title = {{Quasars and the Intergalactic Medium at Cosmic Dawn}},
	volume = {61},
	year = 2023}

@ARTICLE{Matsuoka2016,
       author = {{Matsuoka}, Yoshiki and {Onoue}, Masafusa and {Kashikawa}, Nobunari and {Iwasawa}, Kazushi and {Strauss}, Michael A. and {Nagao}, Tohru and {Imanishi}, Masatoshi and {Niida}, Mana and {Toba}, Yoshiki and {Akiyama}, Masayuki and {Asami}, Naoko and {Bosch}, James and {Foucaud}, S{\'e}bastien and {Furusawa}, Hisanori and {Goto}, Tomotsugu and {Gunn}, James E. and {Harikane}, Yuichi and {Ikeda}, Hiroyuki and {Kawaguchi}, Toshihiro and {Kikuta}, Satoshi and {Komiyama}, Yutaka and {Lupton}, Robert H. and {Minezaki}, Takeo and {Miyazaki}, Satoshi and {Morokuma}, Tomoki and {Murayama}, Hitoshi and {Nishizawa}, Atsushi J. and {Ono}, Yoshiaki and {Ouchi}, Masami and {Price}, Paul A. and {Sameshima}, Hiroaki and {Silverman}, John D. and {Sugiyama}, Naoshi and {Tait}, Philip J. and {Takada}, Masahiro and {Takata}, Tadafumi and {Tanaka}, Masayuki and {Tang}, Ji-Jia and {Utsumi}, Yousuke},
        title = "{Subaru High-z Exploration of Low-luminosity Quasars (SHELLQs). I. Discovery of 15 Quasars and Bright Galaxies at 5.7 < z < 6.9}",
      journal = {\apj},
         year = 2016,
        month = sep,
       volume = {828},
       number = {1},
          eid = {26},
        pages = {26},
          doi = {10.3847/0004-637X/828/1/26},
archivePrefix = {arXiv},
       eprint = {1603.02281},
 primaryClass = {astro-ph.GA},
       adsurl = {https://ui.adsabs.harvard.edu/abs/2016ApJ...828...26M}
}

@ARTICLE{Marques-Chaves2022,
       author = {{Marques-Chaves}, R. and {Schaerer}, D. and {{\'A}lvarez-M{\'a}rquez}, J. and {Verhamme}, A. and {Ceverino}, D. and {Chisholm}, J. and {Colina}, L. and {Dessauges-Zavadsky}, M. and {P{\'e}rez-Fournon}, I. and {Saldana-Lopez}, A. and {Upadhyaya}, A. and {Vanzella}, E.},
        title = "{An extreme blue nugget, UV-bright starburst at z = 3.613 with 90 per cent of Lyman continuum photon escape}",
      journal = {\mnras},
         year = 2022,
        month = dec,
       volume = {517},
       number = {2},
        pages = {2972-2989},
          doi = {10.1093/mnras/stac2893},
archivePrefix = {arXiv},
       eprint = {2210.02392},
 primaryClass = {astro-ph.GA},
       adsurl = {https://ui.adsabs.harvard.edu/abs/2022MNRAS.517.2972M}
}

@ARTICLE{Marques-Chaves2021,
       author = {{Marques-Chaves}, R. and {Schaerer}, D. and {{\'A}lvarez-M{\'a}rquez}, J. and {Colina}, L. and {Dessauges-Zavadsky}, M. and {P{\'e}rez-Fournon}, I. and {Saldana-Lopez}, A. and {Verhamme}, A.},
        title = "{The UV-brightest Lyman continuum emitting star-forming galaxy}",
      journal = {\mnras},
         year = 2021,
        month = oct,
       volume = {507},
       number = {1},
        pages = {524-538},
          doi = {10.1093/mnras/stab2187},
archivePrefix = {arXiv},
       eprint = {2107.12313},
 primaryClass = {astro-ph.GA},
       adsurl = {https://ui.adsabs.harvard.edu/abs/2021MNRAS.507..524M}
}

@ARTICLE{Marques-Chaves2020,
       author = {{Marques-Chaves}, R. and {{\'A}lvarez-M{\'a}rquez}, J. and {Colina}, L. and {P{\'e}rez-Fournon}, I. and {Schaerer}, D. and {Dalla Vecchia}, C. and {Hashimoto}, T. and {Jim{\'e}nez-{\'A}ngel}, C. and {Shu}, Y.},
        title = "{The discovery of the most UV-Ly {\ensuremath{\alpha}} luminous star-forming galaxy: a young, dust- and metal-poor starburst with QSO-like luminosities}",
      journal = {\mnras},
         year = 2020,
        month = dec,
       volume = {499},
       number = {1},
        pages = {L105-L110},
          doi = {10.1093/mnrasl/slaa160},
archivePrefix = {arXiv},
       eprint = {2009.02177},
 primaryClass = {astro-ph.GA},
       adsurl = {https://ui.adsabs.harvard.edu/abs/2020MNRAS.499L.105M}
}

@ARTICLE{Upadhyaya2024,
       author = {{Upadhyaya}, A. and {Marques-Chaves}, R. and {Schaerer}, D. and {Martins}, F. and {P{\'e}rez-Fournon}, I. and {Palacios}, A. and {Stanway}, E.~R.},
        title = "{Evidence for very massive stars in extremely UV-bright star-forming galaxies at z {\ensuremath{\sim}} 2.2-3.6}",
      journal = {\aap},
         year = 2024,
        month = jun,
       volume = {686},
          eid = {A185},
        pages = {A185},
          doi = {10.1051/0004-6361/202449184},
archivePrefix = {arXiv},
       eprint = {2401.16165},
 primaryClass = {astro-ph.GA},
       adsurl = {https://ui.adsabs.harvard.edu/abs/2024A&A...686A.185U}
}

@ARTICLE{Crowther2016,
       author = {{Crowther}, Paul A. and {Caballero-Nieves}, S.~M. and {Bostroem}, K.~A. and {Ma{\'\i}z Apell{\'a}niz}, J. and {Schneider}, F.~R.~N. and {Walborn}, N.~R. and {Angus}, C.~R. and {Brott}, I. and {Bonanos}, A. and {de Koter}, A. and {de Mink}, S.~E. and {Evans}, C.~J. and {Gr{\"a}fener}, G. and {Herrero}, A. and {Howarth}, I.~D. and {Langer}, N. and {Lennon}, D.~J. and {Puls}, J. and {Sana}, H. and {Vink}, J.~S.},
        title = "{The R136 star cluster dissected with Hubble Space Telescope/STIS. I. Far-ultraviolet spectroscopic census and the origin of He II {\ensuremath{\lambda}}1640 in young star clusters}",
      journal = {\mnras},
         year = 2016,
        month = may,
       volume = {458},
       number = {1},
        pages = {624-659},
          doi = {10.1093/mnras/stw273},
archivePrefix = {arXiv},
       eprint = {1603.04994},
 primaryClass = {astro-ph.SR},
       adsurl = {https://ui.adsabs.harvard.edu/abs/2016MNRAS.458..624C}
}

@ARTICLE{Roberts-Borsani2024,
       author = {{Roberts-Borsani}, Guido and {Treu}, Tommaso and {Shapley}, Alice and {Fontana}, Adriano and {Pentericci}, Laura and {Castellano}, Marco and {Morishita}, Takahiro and {Bergamini}, Pietro and {Rosati}, Piero},
        title = "{Between the Extremes: A JWST Spectroscopic Benchmark for High-redshift Galaxies Using {\ensuremath{\sim}}500 Confirmed Sources at z {\ensuremath{\geq}} 5}",
      journal = {\apj},
         year = 2024,
        month = dec,
       volume = {976},
       number = {2},
          eid = {193},
        pages = {193},
          doi = {10.3847/1538-4357/ad85d3},
archivePrefix = {arXiv},
       eprint = {2403.07103},
 primaryClass = {astro-ph.GA},
       adsurl = {https://ui.adsabs.harvard.edu/abs/2024ApJ...976..193R}
}

@ARTICLE{Roberts-Borsani2025,
       author = {{Roberts-Borsani}, Guido and {Bagley}, Micaela and {Rojas-Ruiz}, Sof{\'\i}a and {Treu}, Tommaso and {Morishita}, Takahiro and {Finkelstein}, Steven L. and {Trenti}, Michele and {Arrabal Haro}, Pablo and {Ba{\~n}ados}, Eduardo and {Ch{\'a}vez Ortiz}, {\'O}scar A. and {Chworowsky}, Katherine and {Hutchison}, Taylor A. and {Larson}, Rebecca L. and {Leethochawalit}, Nicha and {Leung}, Gene C.~K. and {Mason}, Charlotte and {Somerville}, Rachel S. and {Stiavelli}, Massimo and {Yung}, L.~Y. Aaron and {Kassin}, Susan A. and {Soto}, Christian},
        title = "{The BoRG-JWST Survey: Program Overview and First Confirmations of Luminous Reionization-era Galaxies from Pure-parallel Observations}",
      journal = {\apj},
         year = 2025,
        month = apr,
       volume = {983},
       number = {1},
          eid = {18},
        pages = {18},
          doi = {10.3847/1538-4357/adba60},
archivePrefix = {arXiv},
       eprint = {2407.17551},
 primaryClass = {astro-ph.GA},
       adsurl = {https://ui.adsabs.harvard.edu/abs/2025ApJ...983...18R}
}

@ARTICLE{Matsuoka2018a,
       author = {{Matsuoka}, Yoshiki and {Onoue}, Masafusa and {Kashikawa}, Nobunari and {Iwasawa}, Kazushi and {Strauss}, Michael A. and {Nagao}, Tohru and {Imanishi}, Masatoshi and {Lee}, Chien-Hsiu and {Akiyama}, Masayuki and {Asami}, Naoko and {Bosch}, James and {Foucaud}, S{\'e}bastien and {Furusawa}, Hisanori and {Goto}, Tomotsugu and {Gunn}, James E. and {Harikane}, Yuichi and {Ikeda}, Hiroyuki and {Izumi}, Takuma and {Kawaguchi}, Toshihiro and {Kikuta}, Satoshi and {Kohno}, Kotaro and {Komiyama}, Yutaka and {Lupton}, Robert H. and {Minezaki}, Takeo and {Miyazaki}, Satoshi and {Morokuma}, Tomoki and {Murayama}, Hitoshi and {Niida}, Mana and {Nishizawa}, Atsushi J. and {Oguri}, Masamune and {Ono}, Yoshiaki and {Ouchi}, Masami and {Price}, Paul A. and {Sameshima}, Hiroaki and {Schulze}, Andreas and {Shirakata}, Hikari and {Silverman}, John D. and {Sugiyama}, Naoshi and {Tait}, Philip J. and {Takada}, Masahiro and {Takata}, Tadafumi and {Tanaka}, Masayuki and {Tang}, Ji-Jia and {Toba}, Yoshiki and {Utsumi}, Yousuke and {Wang}, Shiang-Yu},
        title = "{Subaru High-z Exploration of Low-Luminosity Quasars (SHELLQs). II. Discovery of 32 quasars and luminous galaxies at 5.7 < z {\ensuremath{\leq}} 6.8}",
      journal = {PASJ},
         year = 2018,
        month = jan,
       volume = {70},
          eid = {S35},
        pages = {S35},
          doi = {10.1093/pasj/psx046},
archivePrefix = {arXiv},
       eprint = {1704.05854},
 primaryClass = {astro-ph.GA},
       adsurl = {https://ui.adsabs.harvard.edu/abs/2018PASJ...70S..35M}
}

@ARTICLE{Matsuoka2018b,
       author = {{Matsuoka}, Yoshiki and {Iwasawa}, Kazushi and {Onoue}, Masafusa and {Kashikawa}, Nobunari and {Strauss}, Michael A. and {Lee}, Chien-Hsiu and {Imanishi}, Masatoshi and {Nagao}, Tohru and {Akiyama}, Masayuki and {Asami}, Naoko and {Bosch}, James and {Furusawa}, Hisanori and {Goto}, Tomotsugu and {Gunn}, James E. and {Harikane}, Yuichi and {Ikeda}, Hiroyuki and {Izumi}, Takuma and {Kawaguchi}, Toshihiro and {Kato}, Nanako and {Kikuta}, Satoshi and {Kohno}, Kotaro and {Komiyama}, Yutaka and {Lupton}, Robert H. and {Minezaki}, Takeo and {Miyazaki}, Satoshi and {Morokuma}, Tomoki and {Murayama}, Hitoshi and {Niida}, Mana and {Nishizawa}, Atsushi J. and {Oguri}, Masamune and {Ono}, Yoshiaki and {Ouchi}, Masami and {Price}, Paul A. and {Sameshima}, Hiroaki and {Schulze}, Andreas and {Shirakata}, Hikari and {Silverman}, John D. and {Sugiyama}, Naoshi and {Tait}, Philip J. and {Takada}, Masahiro and {Takata}, Tadafumi and {Tanaka}, Masayuki and {Tang}, Ji-Jia and {Toba}, Yoshiki and {Utsumi}, Yousuke and {Wang}, Shiang-Yu and {Yamashita}, Takuji},
        title = "{Subaru High-z Exploration of Low-luminosity Quasars (SHELLQs). IV. Discovery of 41 Quasars and Luminous Galaxies at 5.7 {\ensuremath{\leq}} z {\ensuremath{\leq}} 6.9}",
      journal = {\apjs},
         year = 2018,
        month = jul,
       volume = {237},
       number = {1},
          eid = {5},
        pages = {5},
          doi = {10.3847/1538-4365/aac724},
archivePrefix = {arXiv},
       eprint = {1803.01861},
 primaryClass = {astro-ph.GA},
       adsurl = {https://ui.adsabs.harvard.edu/abs/2018ApJS..237....5M}
}

@ARTICLE{Matsuoka2019,
       author = {{Matsuoka}, Yoshiki and {Iwasawa}, Kazushi and {Onoue}, Masafusa and {Kashikawa}, Nobunari and {Strauss}, Michael A. and {Lee}, Chien-Hsiu and {Imanishi}, Masatoshi and {Nagao}, Tohru and {Akiyama}, Masayuki and {Asami}, Naoko and {Bosch}, James and {Furusawa}, Hisanori and {Goto}, Tomotsugu and {Gunn}, James E. and {Harikane}, Yuichi and {Ikeda}, Hiroyuki and {Izumi}, Takuma and {Kawaguchi}, Toshihiro and {Kato}, Nanako and {Kikuta}, Satoshi and {Kohno}, Kotaro and {Komiyama}, Yutaka and {Koyama}, Shuhei and {Lupton}, Robert H. and {Minezaki}, Takeo and {Miyazaki}, Satoshi and {Murayama}, Hitoshi and {Niida}, Mana and {Nishizawa}, Atsushi J. and {Noboriguchi}, Akatoki and {Oguri}, Masamune and {Ono}, Yoshiaki and {Ouchi}, Masami and {Price}, Paul A. and {Sameshima}, Hiroaki and {Schulze}, Andreas and {Silverman}, John D. and {Sugiyama}, Naoshi and {Tait}, Philip J. and {Takada}, Masahiro and {Takata}, Tadafumi and {Tanaka}, Masayuki and {Tang}, Ji-Jia and {Toba}, Yoshiki and {Utsumi}, Yousuke and {Wang}, Shiang-Yu and {Yamashita}, Takuji},
        title = "{Subaru High-z Exploration of Low-luminosity Quasars (SHELLQs). X. Discovery of 35 Quasars and Luminous Galaxies at 5.7 {\ensuremath{\leq}} z {\ensuremath{\leq}} 7.0}",
      journal = {\apj},
         year = 2019,
        month = oct,
       volume = {883},
       number = {2},
          eid = {183},
        pages = {183},
          doi = {10.3847/1538-4357/ab3c60},
archivePrefix = {arXiv},
       eprint = {1908.07910},
 primaryClass = {astro-ph.GA},
       adsurl = {https://ui.adsabs.harvard.edu/abs/2019ApJ...883..183M}
}

@ARTICLE{Dessauges-Zavadsky2025,
       author = {{Dessauges-Zavadsky}, M. and {Marques-Chaves}, R. and {Schaerer}, D. and {Xiao}, M. -Y. and {Colina}, L. and {Alvarez-Marquez}, J. and {P{\'e}rez-Fournon}, I.},
        title = "{Unveiling dust, molecular gas, and high star-formation efficiency in extremely UV bright star-forming galaxies at z {\ensuremath{\sim}} 2.1{\textendash}3.6}",
      journal = {\aap},
         year = 2025,
        month = jan,
       volume = {693},
          eid = {A17},
        pages = {A17},
          doi = {10.1051/0004-6361/202451832},
archivePrefix = {arXiv},
       eprint = {2410.11121},
 primaryClass = {astro-ph.GA},
       adsurl = {https://ui.adsabs.harvard.edu/abs/2025A&A...693A..17D}
}

@ARTICLE{Marques-Chaves2024,
       author = {{Marques-Chaves}, R. and {Schaerer}, D. and {Vanzella}, E. and {Verhamme}, A. and {Dessauges-Zavadsky}, M. and {Chisholm}, J. and {Leclercq}, F. and {Upadhyaya}, A. and {{\'A}lvarez-M{\'a}rquez}, J. and {Colina}, L. and {Garel}, T. and {Messa}, M.},
        title = "{Witnessing an extreme, highly efficient galaxy formation mode with resolved Lyman-{\ensuremath{\alpha}} and Lyman-continuum emission}",
      journal = {\aap},
         year = 2024,
        month = nov,
       volume = {691},
          eid = {A87},
        pages = {A87},
          doi = {10.1051/0004-6361/202451667},
archivePrefix = {arXiv},
       eprint = {2407.18804},
 primaryClass = {astro-ph.GA},
       adsurl = {https://ui.adsabs.harvard.edu/abs/2024A&A...691A..87M}
}

@ARTICLE{Martins2022,
       author = {{Martins}, F. and {Palacios}, A.},
        title = "{Spectroscopic evolution of very massive stars at Z = 1/2.5 Z$_{{\ensuremath{\odot}}}$}",
      journal = {\aap},
         year = 2022,
        month = mar,
       volume = {659},
          eid = {A163},
        pages = {A163},
          doi = {10.1051/0004-6361/202243048},
archivePrefix = {arXiv},
       eprint = {2202.13703},
 primaryClass = {astro-ph.SR},
       adsurl = {https://ui.adsabs.harvard.edu/abs/2022A&A...659A.163M}
}

@ARTICLE{CASA,
       author = {{CASA Team} and {Bean}, Ben and {Bhatnagar}, Sanjay and {Castro}, Sandra and {Donova\
n Meyer}, Jennifer and {Emonts}, Bjorn and {Garcia}, Enrique and {Garwood}, Robert and {Golap}, Kumar\
 and {Gonzalez Villalba}, Justo and {Harris}, Pamela and {Hayashi}, Yohei and {Hoskins}, Josh and {Hs\
ieh}, Mingyu and {Jagannathan}, Preshanth and {Kawasaki}, Wataru and {Keimpema}, Aard and {Kettenis},\
 Mark and {Lopez}, Jorge and {Marvil}, Joshua and {Masters}, Joseph and {McNichols}, Andrew and {Mehr\
inger}, David and {Miel}, Renaud and {Moellenbrock}, George and {Montesino}, Federico and {Nakazato},\
 Takeshi and {Ott}, Juergen and {Petry}, Dirk and {Pokorny}, Martin and {Raba}, Ryan and {Rau}, Urvas\
hi and {Schiebel}, Darrell and {Schweighart}, Neal and {Sekhar}, Srikrishna and {Shimada}, Kazuhiko a\
nd {Small}, Des and {Steeb}, Jan-Willem and {Sugimoto}, Kanako and {Suoranta}, Ville and {Tsutsumi}, \
Takahiro and {van Bemmel}, Ilse M. and {Verkouter}, Marjolein and {Wells}, Akeem and {Xiong}, Wei and\
 {Szomoru}, Arpad and {Griffith}, Morgan and {Glendenning}, Brian and {Kern}, Jeff},
        title = "{CASA, the Common Astronomy Software Applications for Radio Astronomy}",
      journal = {\pasp},
         year = 2022,
        month = nov,
       volume = {134},
       number = {1041},
          eid = {114501},
        pages = {114501},
          doi = {10.1088/1538-3873/ac9642},
archivePrefix = {arXiv},
       eprint = {2210.02276},
 primaryClass = {astro-ph.IM},
       adsurl = {https://ui.adsabs.harvard.edu/abs/2022PASP..134k4501C}
}

@ARTICLE{Marques-Chaves2025,
       author = {{Marques-Chaves}, R. and {Schaerer}, D. and {Dessauges-Zavadsky}, M. and {{\'A}lvarez-M{\'a}rquez}, J. and {Hashimoto}, T. and {Colina}, L. and {Inoue}, A.~K. and {Blanco-Prieto}, C. and {Nakazato}, Y. and {Costantin}, L. and {Arribas}, S. and {Bakx}, T.~J.~L.~C. and {Ceverino}, D. and {Crespo G{\'o}mez}, A. and {Fudamoto}, Y. and {Hagimoto}, M. and {Hamada}, A. and {Matsuoka}, Y. and {Mawatari}, K. and {Onoue}, M. and {Osone}, W. and {Ren}, Y.~W. and {Sugahara}, Y. and {Terui}, Y. and {Yoshida}, N.},
        title = "{Extremely UV-bright starbursts at the end of cosmic reionization}",
      journal = {arXiv e-prints},
         year = 2025,
        month = oct,
          eid = {arXiv:2510.12411},
        pages = {arXiv:2510.12411},
archivePrefix = {arXiv},
       eprint = {2510.12411},
 primaryClass = {astro-ph.GA},
       adsurl = {https://ui.adsabs.harvard.edu/abs/2025arXiv251012411M}
}

@ARTICLE{Matsuoka2023,
       author = {{Matsuoka}, Yoshiki and {Onoue}, Masafusa and {Iwasawa}, Kazushi and {Strauss}, Michael A. and {Kashikawa}, Nobunari and {Izumi}, Takuma and {Nagao}, Tohru and {Imanishi}, Masatoshi and {Akiyama}, Masayuki and {Silverman}, John D. and {Asami}, Naoko and {Bosch}, James and {Furusawa}, Hisanori and {Goto}, Tomotsugu and {Gunn}, James E. and {Harikane}, Yuichi and {Ikeda}, Hiroyuki and {Inayoshi}, Kohei and {Ishimoto}, Rikako and {Kawaguchi}, Toshihiro and {Kikuta}, Satoshi and {Kohno}, Kotaro and {Komiyama}, Yutaka and {Lee}, Chien-Hsiu and {Lupton}, Robert H. and {Minezaki}, Takeo and {Miyazaki}, Satoshi and {Murayama}, Hitoshi and {Nishizawa}, Atsushi J. and {Oguri}, Masamune and {Ono}, Yoshiaki and {Oogi}, Taira and {Ouchi}, Masami and {Price}, Paul A. and {Sameshima}, Hiroaki and {Sugiyama}, Naoshi and {Tait}, Philip J. and {Takada}, Masahiro and {Takahashi}, Ayumi and {Takata}, Tadafumi and {Tanaka}, Masayuki and {Toba}, Yoshiki and {Wang}, Shiang-Yu and {Yamashita}, Takuji},
        title = "{Quasar Luminosity Function at z = 7}",
      journal = {\apjl},
         year = 2023,
        month = jun,
       volume = {949},
       number = {2},
          eid = {L42},
        pages = {L42},
          doi = {10.3847/2041-8213/acd69f},
archivePrefix = {arXiv},
       eprint = {2305.11225},
 primaryClass = {astro-ph.GA},
       adsurl = {https://ui.adsabs.harvard.edu/abs/2023ApJ...949L..42M}
}

@ARTICLE{Bouwens2021,
       author = {{Bouwens}, R.~J. and {Oesch}, P.~A. and {Stefanon}, M. and {Illingworth}, G. and {Labb{\'e}}, I. and {Reddy}, N. and {Atek}, H. and {Montes}, M. and {Naidu}, R. and {Nanayakkara}, T. and {Nelson}, E. and {Wilkins}, S.},
        title = "{New Determinations of the UV Luminosity Functions from z   9 to 2 Show a Remarkable Consistency with Halo Growth and a Constant Star Formation Efficiency}",
      journal = {\aj},
         year = 2021,
        month = aug,
       volume = {162},
       number = {2},
          eid = {47},
        pages = {47},
          doi = {10.3847/1538-3881/abf83e},
archivePrefix = {arXiv},
       eprint = {2102.07775},
 primaryClass = {astro-ph.GA},
       adsurl = {https://ui.adsabs.harvard.edu/abs/2021AJ....162...47B}
}

@ARTICLE{Appenzeller2005,
       author = {{Appenzeller}, I. and {Stahl}, O. and {Tapken}, C. and {Mehlert}, D. and {Noll}, S.},
        title = "{SDSS J1553+0056: A BALQSO mimicking a Lyman-break galaxy}",
      journal = {\aap},
         year = 2005,
        month = may,
       volume = {435},
       number = {2},
        pages = {465-469},
          doi = {10.1051/0004-6361:20042524},
archivePrefix = {arXiv},
       eprint = {astro-ph/0502207},
 primaryClass = {astro-ph},
       adsurl = {https://ui.adsabs.harvard.edu/abs/2005A&A...435..465A}
}

@ARTICLE{Matsuoka2022,
       author = {{Matsuoka}, Yoshiki and {Iwasawa}, Kazushi and {Onoue}, Masafusa and {Izumi}, Takuma and {Kashikawa}, Nobunari and {Strauss}, Michael A. and {Imanishi}, Masatoshi and {Nagao}, Tohru and {Akiyama}, Masayuki and {Silverman}, John D. and {Asami}, Naoko and {Bosch}, James and {Furusawa}, Hisanori and {Goto}, Tomotsugu and {Gunn}, James E. and {Harikane}, Yuichi and {Ikeda}, Hiroyuki and {Ishimoto}, Rikako and {Kawaguchi}, Toshihiro and {Kato}, Nanako and {Kikuta}, Satoshi and {Kohno}, Kotaro and {Komiyama}, Yutaka and {Lee}, Chien-Hsiu and {Lupton}, Robert H. and {Minezaki}, Takeo and {Miyazaki}, Satoshi and {Murayama}, Hitoshi and {Nishizawa}, Atsushi J. and {Oguri}, Masamune and {Ono}, Yoshiaki and {Ouchi}, Masami and {Price}, Paul A. and {Sameshima}, Hiroaki and {Sugiyama}, Naoshi and {Tait}, Philip J. and {Takada}, Masahiro and {Takahashi}, Ayumi and {Takata}, Tadafumi and {Tanaka}, Masayuki and {Toba}, Yoshiki and {Utsumi}, Yousuke and {Wang}, Shiang-Yu and {Yamashita}, Takuji},
        title = "{Subaru High-z Exploration of Low-luminosity Quasars (SHELLQs). XVI. 69 New Quasars at 5.8 < z < 7.0}",
      journal = {\apjs},
         year = 2022,
        month = mar,
       volume = {259},
       number = {1},
          eid = {18},
        pages = {18},
          doi = {10.3847/1538-4365/ac3d31},
archivePrefix = {arXiv},
       eprint = {2111.12766},
 primaryClass = {astro-ph.GA},
       adsurl = {https://ui.adsabs.harvard.edu/abs/2022ApJS..259...18M}
}

@ARTICLE{Ferrara2023,
       author = {{Ferrara}, Andrea and {Pallottini}, Andrea and {Dayal}, Pratika},
        title = "{On the stunning abundance of super-early, luminous galaxies revealed by JWST}",
      journal = {\mnras},
         year = 2023,
        month = jul,
       volume = {522},
       number = {3},
        pages = {3986-3991},
          doi = {10.1093/mnras/stad1095},
archivePrefix = {arXiv},
       eprint = {2208.00720},
 primaryClass = {astro-ph.GA},
       adsurl = {https://ui.adsabs.harvard.edu/abs/2023MNRAS.522.3986F}
}

@ARTICLE{Dekel2023,
       author = {{Dekel}, Avishai and {Sarkar}, Kartick C. and {Birnboim}, Yuval and {Mandelker}, Nir and {Li}, Zhaozhou},
        title = "{Efficient formation of massive galaxies at cosmic dawn by feedback-free starbursts}",
      journal = {\mnras},
         year = 2023,
        month = aug,
       volume = {523},
       number = {3},
        pages = {3201-3218},
          doi = {10.1093/mnras/stad1557},
archivePrefix = {arXiv},
       eprint = {2303.04827},
 primaryClass = {astro-ph.GA},
       adsurl = {https://ui.adsabs.harvard.edu/abs/2023MNRAS.523.3201D}
}

@INPROCEEDINGS{Cepa2000osiris,
       author = {{Cepa}, Jordi and {Aguiar}, Marta and {Escalera}, Victor G. and {Gonzalez-Serrano}, Ignacio and {Joven-Alvarez}, Enrique and {Peraza}, Lorenzo and {Rasilla}, Jose Luis and {Rodriguez-Ramos}, Luis F. and {Gonzalez}, J. Jesus and {Cobos Duenas}, Francisco J. and {Sanchez}, Beatriz and {Tejada}, Carlos and {Bland-Hawthorn}, Joss and {Militello}, Carmelo and {Rosa}, Fernando},
        title = "{OSIRIS tunable imager and spectrograph}",
    booktitle = {Optical and IR Telescope Instrumentation and Detectors},
         year = 2000,
       editor = {{Iye}, Masanori and {Moorwood}, Alan F.},
       series = {Society of Photo-Optical Instrumentation Engineers (SPIE) Conference Series},
       volume = {4008},
        month = aug,
        pages = {623-631},
          doi = {10.1117/12.395520},
       adsurl = {https://ui.adsabs.harvard.edu/abs/2000SPIE.4008..623C}
}

@ARTICLE{Kashikawa2002FOCAS,
       author = {{Kashikawa}, Nobunari and {Aoki}, Kentaro and {Asai}, Ryo and {Ebizuka}, Noboru and {Inata}, Motoko and {Iye}, Masanori and {Kawabata}, Koji S. and {Kosugi}, George and {Ohyama}, Youichi and {Okita}, Kiichi and {Ozawa}, Tomohiko and {Saito}, Yoshihiko and {Sasaki}, Toshiyuki and {Sekiguchi}, Kazuhiro and {Shimizu}, Yasuhiro and {Taguchi}, Hiroko and {Takata}, Tadafumi and {Yadoumaru}, Yasushi and {Yoshida}, Michitoshi},
        title = "{FOCAS: The Faint Object Camera and Spectrograph for the Subaru Telescope}",
      journal = {\pasj},
         year = 2002,
        month = dec,
       volume = {54},
       number = {6},
        pages = {819-832},
          doi = {10.1093/pasj/54.6.819},
       adsurl = {https://ui.adsabs.harvard.edu/abs/2002PASJ...54..819K}
}

@ARTICLE{Mestric2023,
       author = {{Me{\v{s}}tri{\'c}}, U. and {Vanzella}, E. and {Upadhyaya}, A. and {Martins}, F. and {Marques-Chaves}, R. and {Schaerer}, D. and {Guibert}, J. and {Zanella}, A. and {Grillo}, C. and {Rosati}, P. and {Calura}, F. and {Caminha}, G.~B. and {Bolamperti}, A. and {Meneghetti}, M. and {Bergamini}, P. and {Mercurio}, A. and {Nonino}, M. and {Pascale}, R.},
        title = "{Clues on the presence and segregation of very massive stars in the Sunburst Lyman-continuum cluster at z = 2.37}",
      journal = {\aap},
         year = 2023,
        month = may,
       volume = {673},
          eid = {A50},
        pages = {A50},
          doi = {10.1051/0004-6361/202345895},
archivePrefix = {arXiv},
       eprint = {2301.04672},
 primaryClass = {astro-ph.GA},
       adsurl = {https://ui.adsabs.harvard.edu/abs/2023A&A...673A..50M}
}

@ARTICLE{Crowther2010,
       author = {{Crowther}, Paul A. and {Schnurr}, Olivier and {Hirschi}, Raphael and {Yusof}, Norhasliza and {Parker}, Richard J. and {Goodwin}, Simon P. and {Kassim}, Hasan Abu},
        title = "{The R136 star cluster hosts several stars whose individual masses greatly exceed the accepted 150M$_{solar}$ stellar mass limit}",
      journal = {\mnras},
         year = 2010,
        month = oct,
       volume = {408},
       number = {2},
        pages = {731-751},
          doi = {10.1111/j.1365-2966.2010.17167.x},
archivePrefix = {arXiv},
       eprint = {1007.3284},
 primaryClass = {astro-ph.SR},
       adsurl = {https://ui.adsabs.harvard.edu/abs/2010MNRAS.408..731C}
}

@ARTICLE{Shapley2003,
       author = {{Shapley}, Alice E. and {Steidel}, Charles C. and {Pettini}, Max and {Adelberger}, Kurt L.},
        title = "{Rest-Frame Ultraviolet Spectra of z\raisebox{-0.5ex}\textasciitilde3 Lyman Break Galaxies}",
      journal = {\apj},
         year = 2003,
        month = may,
       volume = {588},
       number = {1},
        pages = {65-89},
          doi = {10.1086/373922},
archivePrefix = {arXiv},
       eprint = {astro-ph/0301230},
 primaryClass = {astro-ph},
       adsurl = {https://ui.adsabs.harvard.edu/abs/2003ApJ...588...65S}
}

@ARTICLE{Schaerer2025,
       author = {{Schaerer}, D. and {Guibert}, J. and {Marques-Chaves}, R. and {Martins}, F.},
        title = "{Observable and ionizing properties of star-forming galaxies with very massive stars and different initial mass functions}",
      journal = {\aap},
         year = 2025,
        month = jan,
       volume = {693},
          eid = {A271},
        pages = {A271},
          doi = {10.1051/0004-6361/202451454},
archivePrefix = {arXiv},
       eprint = {2407.12122},
 primaryClass = {astro-ph.GA},
       adsurl = {https://ui.adsabs.harvard.edu/abs/2025A&A...693A.271S}
}

@ARTICLE{Schaerer1998,
       author = {{Schaerer}, Daniel and {Vacca}, William D.},
        title = "{New Models for Wolf-Rayet and O Star Populations in Young Starbursts}",
      journal = {\apj},
         year = 1998,
        month = apr,
       volume = {497},
       number = {2},
        pages = {618-644},
          doi = {10.1086/305487},
archivePrefix = {arXiv},
       eprint = {astro-ph/9711140},
 primaryClass = {astro-ph},
       adsurl = {https://ui.adsabs.harvard.edu/abs/1998ApJ...497..618S}
}

@ARTICLE{Grafener2021,
       author = {{Gr{\"a}fener}, G{\"o}tz},
        title = "{Physics and evolution of the most massive stars in 30 Doradus. Mass loss, envelope inflation, and a variable upper stellar mass limit}",
      journal = {\aap},
         year = 2021,
        month = mar,
       volume = {647},
          eid = {A13},
        pages = {A13},
          doi = {10.1051/0004-6361/202040037},
archivePrefix = {arXiv},
       eprint = {2101.03837},
 primaryClass = {astro-ph.SR},
       adsurl = {https://ui.adsabs.harvard.edu/abs/2021A&A...647A..13G}
}

@ARTICLE{Vink2001,
       author = {{Vink}, Jorick S. and {de Koter}, A. and {Lamers}, H.~J.~G.~L.~M.},
        title = "{Mass-loss predictions for O and B stars as a function of metallicity}",
      journal = {\aap},
         year = 2001,
        month = apr,
       volume = {369},
        pages = {574-588},
          doi = {10.1051/0004-6361:20010127},
archivePrefix = {arXiv},
       eprint = {astro-ph/0101509},
 primaryClass = {astro-ph},
       adsurl = {https://ui.adsabs.harvard.edu/abs/2001A&A...369..574V}
}

@ARTICLE{Bouwens2022,
       author = {{Bouwens}, R.~J. and {Smit}, R. and {Schouws}, S. and {Stefanon}, M. and {Bowler}, R. and {Endsley}, R. and {Gonzalez}, V. and {Inami}, H. and {Stark}, D. and {Oesch}, P. and {Hodge}, J. and {Aravena}, M. and {da Cunha}, E. and {Dayal}, P. and {de Looze}, I. and {Ferrara}, A. and {Fudamoto}, Y. and {Graziani}, L. and {Li}, C. and {Nanayakkara}, T. and {Pallottini}, A. and {Schneider}, R. and {Sommovigo}, L. and {Topping}, M. and {van der Werf}, P. and {Algera}, H. and {Barrufet}, L. and {Hygate}, A. and {Labb{\'e}}, I. and {Riechers}, D. and {Witstok}, J.},
        title = "{Reionization Era Bright Emission Line Survey: Selection and Characterization of Luminous Interstellar Medium Reservoirs in the z > 6.5 Universe}",
      journal = {\apj},
         year = 2022,
        month = jun,
       volume = {931},
       number = {2},
          eid = {160},
        pages = {160},
          doi = {10.3847/1538-4357/ac5a4a},
archivePrefix = {arXiv},
       eprint = {2106.13719},
 primaryClass = {astro-ph.GA},
       adsurl = {https://ui.adsabs.harvard.edu/abs/2022ApJ...931..160B}
}

@ARTICLE{Inami2022,
       author = {{Inami}, Hanae and {Algera}, Hiddo S.~B. and {Schouws}, Sander and {Sommovigo}, Laura and {Bouwens}, Rychard and {Smit}, Renske and {Stefanon}, Mauro and {Bowler}, Rebecca A.~A. and {Endsley}, Ryan and {Ferrara}, Andrea and {Oesch}, Pascal and {Stark}, Daniel and {Aravena}, Manuel and {Barrufet}, Laia and {da Cunha}, Elisabete and {Dayal}, Pratika and {De Looze}, Ilse and {Fudamoto}, Yoshinobu and {Gonzalez}, Valentino and {Graziani}, Luca and {Hodge}, Jacqueline A. and {Hygate}, Alexander P.~S. and {Nanayakkara}, Themiya and {Pallottini}, Andrea and {Riechers}, Dominik A. and {Schneider}, Raffaella and {Topping}, Michael and {van der Werf}, Paul},
        title = "{The ALMA REBELS Survey: dust continuum detections at z > 6.5}",
      journal = {\mnras},
         year = 2022,
        month = sep,
       volume = {515},
       number = {3},
        pages = {3126-3143},
          doi = {10.1093/mnras/stac1779},
archivePrefix = {arXiv},
       eprint = {2203.15136},
 primaryClass = {astro-ph.GA},
       adsurl = {https://ui.adsabs.harvard.edu/abs/2022MNRAS.515.3126I}
}

@ARTICLE{Izumi2021,
       author = {{Izumi}, Takuma and {Onoue}, Masafusa and {Matsuoka}, Yoshiki and {Strauss}, Michael A. and {Fujimoto}, Seiji and {Umehata}, Hideki and {Imanishi}, Masatoshi and {Kawamuro}, Taiki and {Nagao}, Tohru and {Toba}, Yoshiki and {Kohno}, Kotaro and {Kashikawa}, Nobunari and {Inayoshi}, Kohei and {Kawaguchi}, Toshihiro and {Iwasawa}, Kazushi and {Inoue}, Akio K. and {Goto}, Tomotsugu and {Baba}, Shunsuke and {Schramm}, Malte and {Suh}, Hyewon and {Harikane}, Yuichi and {Ueda}, Yoshihiro and {Silverman}, John D. and {Hashimoto}, Takuya and {Hashimoto}, Yasuhiro and {Ikarashi}, Soh and {Iono}, Daisuke and {Lee}, Chien-Hsiu and {Lee}, Kianhong and {Minezaki}, Takeo and {Nakanishi}, Kouichiro and {Nakano}, Suzuka and {Tamura}, Yoichi and {Tang}, Ji-Jia},
        title = "{Subaru High-z Exploration of Low-luminosity Quasars (SHELLQs). XII. Extended [C II] Structure (Merger or Outflow) in a z = 6.72 Red Quasar}",
      journal = {\apj},
         year = 2021,
        month = feb,
       volume = {908},
       number = {2},
          eid = {235},
        pages = {235},
          doi = {10.3847/1538-4357/abd7ef},
archivePrefix = {arXiv},
       eprint = {2101.01199},
 primaryClass = {astro-ph.GA},
       adsurl = {https://ui.adsabs.harvard.edu/abs/2021ApJ...908..235I}
}

@ARTICLE{Harikane2025,
       author = {{Harikane}, Yuichi and {Inoue}, Akio K. and {Ellis}, Richard S. and {Ouchi}, Masami and {Nakazato}, Yurina and {Yoshida}, Naoki and {Ono}, Yoshiaki and {Sun}, Fengwu and {Sato}, Riku A. and {Ferrami}, Giovanni and {Fujimoto}, Seiji and {Kashikawa}, Nobunari and {McLeod}, Derek J. and {P{\'e}rez-Gonz{\'a}lez}, Pablo G. and {Sawicki}, Marcin and {Sugahara}, Yuma and {Xu}, Yi and {Yamanaka}, Satoshi and {Carnall}, Adam C. and {Cullen}, Fergus and {Dunlop}, James S. and {Egami}, Eiichi and {Grogin}, Norman and {Isobe}, Yuki and {Koekemoer}, Anton M. and {Laporte}, Nicolas and {Lee}, Chien-Hsiu and {Magee}, Dan and {Matsuo}, Hiroshi and {Matsuoka}, Yoshiki and {Mawatari}, Ken and {Nakajima}, Kimihiko and {Nakane}, Minami and {Tamura}, Yoichi and {Umeda}, Hiroya and {Yanagisawa}, Hiroto},
        title = "{JWST, ALMA, and Keck Spectroscopic Constraints on the UV Luminosity Functions at z {\ensuremath{\sim}} 7─14: Clumpiness and Compactness of the Brightest Galaxies in the Early Universe}",
      journal = {\apj},
         year = 2025,
        month = feb,
       volume = {980},
       number = {1},
          eid = {138},
        pages = {138},
          doi = {10.3847/1538-4357/ad9b2c},
archivePrefix = {arXiv},
       eprint = {2406.18352},
 primaryClass = {astro-ph.GA},
       adsurl = {https://ui.adsabs.harvard.edu/abs/2025ApJ...980..138H}
}

@ARTICLE{Mason2023,
       author = {{Mason}, Charlotte A. and {Trenti}, Michele and {Treu}, Tommaso},
        title = "{The brightest galaxies at cosmic dawn}",
      journal = {\mnras},
         year = 2023,
        month = may,
       volume = {521},
       number = {1},
        pages = {497--503},
          doi = {10.1093/mnras/stad035},
archivePrefix = {arXiv},
       eprint = {2207.14808},
 primaryClass = {astro-ph.GA},
       adsurl = {https://ui.adsabs.harvard.edu/abs/2023MNRAS.521..497M}
}

@ARTICLE{Sun2023,
       author = {{Sun}, Guochao and {Faucher-Gigu{\`e}re}, Claude-Andr{\'e} and {Hayward}, Christopher C. and {Shen}, Xuejian and {Wetzel}, Andrew and {Feldmann}, Robert},
        title = "{Bursty star formation naturally explains the abundance of bright galaxies at cosmic dawn}",
      journal = {\apjl},
         year = 2023,
        month = oct,
       volume = {955},
       number = {2},
          eid = {L35},
        pages = {L35},
          doi = {10.3847/2041-8213/acf85a},
archivePrefix = {arXiv},
       eprint = {2307.15305},
 primaryClass = {astro-ph.GA},
       adsurl = {https://ui.adsabs.harvard.edu/abs/2023ApJ...955L..35S}
}

@ARTICLE{Castellano2022,
       author = {{Castellano}, M. and {Fontana}, A. and {Treu}, T. and {Santini}, P. and {Merlin}, E. and {Leethochawalit}, N. and {Roberts-Borsani}, G. and {Trenti}, M. and {Mason}, C.~A. and {Vanzella}, E. and {others}},
        title = "{Early Results from GLASS-JWST. III. Galaxy Candidates at z {\ensuremath{\sim}}9-15}",
      journal = {\apjl},
         year = 2022,
        month = nov,
       volume = {938},
       number = {2},
          eid = {L15},
        pages = {L15},
          doi = {10.3847/2041-8213/ac94d0},
archivePrefix = {arXiv},
       eprint = {2207.09436},
 primaryClass = {astro-ph.GA},
       adsurl = {https://ui.adsabs.harvard.edu/abs/2022ApJ...938L..15C}
}

@ARTICLE{Naidu2022,
       author = {{Naidu}, Rohan P. and {Oesch}, Pascal A. and {van Dokkum}, Pieter and {Nelson}, Erica J. and {Suess}, Katherine A. and {Brammer}, Gabriel and {Whitaker}, Katherine E. and {Illingworth}, Garth and {Bouwens}, Rychard and {Tacchella}, Sandro and {Matthee}, Jorryt},
        title = "{Two Remarkably Luminous Galaxy Candidates at z {\ensuremath{\approx}} 10-12 Revealed by JWST}",
      journal = {\apjl},
         year = 2022,
        month = nov,
       volume = {940},
       number = {1},
          eid = {L14},
        pages = {L14},
          doi = {10.3847/2041-8213/ac9b22},
archivePrefix = {arXiv},
       eprint = {2207.09434},
 primaryClass = {astro-ph.GA},
       adsurl = {https://ui.adsabs.harvard.edu/abs/2022ApJ...940L..14N}
}

@ARTICLE{Yusof2013,
       author = {{Yusof}, Norhasliza and {Hirschi}, Raphael and {Meynet}, Georges and {Crowther}, Paul A. and {Ekstr{\"o}m}, Sylvia and {Frischknecht}, Urs and {Georgy}, Cyril and {Abu Kassim}, Hasan and {Schnurr}, Olivier},
        title = "{Evolution and fate of very massive stars}",
      journal = {\mnras},
         year = 2013,
        month = jul,
       volume = {433},
       number = {2},
        pages = {1114--1132},
          doi = {10.1093/mnras/stt794},
archivePrefix = {arXiv},
       eprint = {1305.2099},
 primaryClass = {astro-ph.SR},
       adsurl = {https://ui.adsabs.harvard.edu/abs/2013MNRAS.433.1114Y}
}

@ARTICLE{Calzetti2000,
       author = {{Calzetti}, Daniela and {Armus}, Lee and {Bohlin}, Ralph C. and {Kinney}, Anne L. and {Koornneef}, Jan and {Storchi-Bergmann}, Thaisa},
        title = "{The Dust Content and Opacity of Actively Star-forming Galaxies}",
      journal = {\apj},
         year = 2000,
        month = apr,
       volume = {533},
       number = {2},
        pages = {682-695},
          doi = {10.1086/308692},
archivePrefix = {arXiv},
       eprint = {astro-ph/9911459},
 primaryClass = {astro-ph},
       adsurl = {https://ui.adsabs.harvard.edu/abs/2000ApJ...533..682C}
}

@ARTICLE{Sanders2025,
       author = {{Sanders}, Ryan L. and {Shapley}, Alice E. and {Topping}, Michael W. and {Reddy}, Naveen A. and {Berg}, Danielle A. and {Khostovan}, Ali Ahmad and {Bouwens}, Rychard J. and {Brammer}, Gabriel and {Carnall}, Adam C. and {Cullen}, Fergus and {Dav{\'e}}, Romeel and {Dunlop}, James S. and {Ellis}, Richard S. and {F{\"o}rster Schreiber}, N.~M. and {Furlanetto}, Steven R. and {Glazebrook}, Karl and {Illingworth}, Garth D. and {Jones}, Tucker and {Kriek}, Mariska and {McLeod}, Derek J. and {McLure}, Ross J. and {Narayanan}, Desika and {Oesch}, Pascal A. and {Pahl}, Anthony J. and {Pettini}, Max and {Schaerer}, Daniel and {Stark}, Daniel P. and {Steidel}, Charles C. and {Tang}, Mengtao and {Clarke}, Leonardo and {Donnan}, Callum T. and {Kehoe}, Emily},
        title = "{The AURORA Survey: High-Redshift Empirical Metallicity Calibrations from Electron Temperature Measurements at z=2-10}",
      journal = {arXiv e-prints},
         year = 2025,
        month = aug,
          eid = {arXiv:2508.10099},
        pages = {arXiv:2508.10099},
          doi = {10.48550/arXiv.2508.10099},
archivePrefix = {arXiv},
       eprint = {2508.10099},
 primaryClass = {astro-ph.GA},
       adsurl = {https://ui.adsabs.harvard.edu/abs/2025arXiv250810099S}
}

@ARTICLE{Asplund2021,
       author = {{Asplund}, M. and {Amarsi}, A.~M. and {Grevesse}, N.},
        title = "{The chemical make-up of the Sun: A 2020 vision}",
      journal = {\aap},
         year = 2021,
        month = sep,
       volume = {653},
          eid = {A141},
        pages = {A141},
          doi = {10.1051/0004-6361/202140445},
archivePrefix = {arXiv},
       eprint = {2105.01661},
 primaryClass = {astro-ph.SR},
       adsurl = {https://ui.adsabs.harvard.edu/abs/2021A&A...653A.141A}
}

@ARTICLE{Zanella2018,
       author = {{Zanella}, A. and {Daddi}, E. and {Magdis}, G. and {Diaz Santos}, T. and {Cormier}, D. and {Liu}, D. and {Cibinel}, A. and {Gobat}, R. and {Dickinson}, M. and {Sargent}, M. and {Popping}, G. and {Madden}, S.~C. and {Bethermin}, M. and {Hughes}, T.~M. and {Valentino}, F. and {Rujopakarn}, W. and {Pannella}, M. and {Bournaud}, F. and {Walter}, F. and {Wang}, T. and {Elbaz}, D. and {Coogan}, R.~T.},
        title = "{The [C II] emission as a molecular gas mass tracer in galaxies at low and high redshifts}",
      journal = {\mnras},
         year = 2018,
        month = dec,
       volume = {481},
       number = {2},
        pages = {1976-1999},
          doi = {10.1093/mnras/sty2394},
archivePrefix = {arXiv},
       eprint = {1808.10331},
 primaryClass = {astro-ph.GA},
       adsurl = {https://ui.adsabs.harvard.edu/abs/2018MNRAS.481.1976Z}
}

@ARTICLE{Beelen2006,
       author = {{Beelen}, Alexandre and {Cox}, Pierre and {Benford}, Dominic J. and {Dowell}, C. Darren and {Kov{\'a}cs}, Attila and {Bertoldi}, Frank and {Omont}, Alain and {Carilli}, Chris L.},
        title = "{350 {\ensuremath{\mu}}m Dust Emission from High-Redshift Quasars}",
      journal = {\apj},
         year = 2006,
        month = may,
       volume = {642},
       number = {2},
        pages = {694-701},
          doi = {10.1086/500636},
archivePrefix = {arXiv},
       eprint = {astro-ph/0603121},
 primaryClass = {astro-ph},
       adsurl = {https://ui.adsabs.harvard.edu/abs/2006ApJ...642..694B}
}

@ARTICLE{DeLooze2014,
       author = {{De Looze}, Ilse and {Cormier}, Diane and {Lebouteiller}, Vianney and {Madden}, Suzanne and {Baes}, Maarten and {Bendo}, George J. and {Boquien}, M{\'e}d{\'e}ric and {Boselli}, Alessandro and {Clements}, David L. and {Cortese}, Luca and {Cooray}, Asantha and {Galametz}, Maud and {Galliano}, Fr{\'e}d{\'e}ric and {Graci{\'a}-Carpio}, Javier and {Isaak}, Kate and {Karczewski}, Oskar {\L}. and {Parkin}, Tara J. and {Pellegrini}, Eric W. and {R{\'e}my-Ruyer}, Aur{\'e}lie and {Spinoglio}, Luigi and {Smith}, Matthew W.~L. and {Sturm}, Eckhard},
        title = "{The applicability of far-infrared fine-structure lines as star formation rate tracers over wide ranges of metallicities and galaxy types}",
      journal = {\aap},
         year = 2014,
        month = aug,
       volume = {568},
          eid = {A62},
        pages = {A62},
          doi = {10.1051/0004-6361/201322489},
archivePrefix = {arXiv},
       eprint = {1402.4075},
 primaryClass = {astro-ph.GA},
       adsurl = {https://ui.adsabs.harvard.edu/abs/2014A&A...568A..62D}
}

@ARTICLE{Decarli2018,
       author = {{Decarli}, Roberto and {Walter}, Fabian and {Venemans}, Bram P. and {Ba{\~n}ados}, Eduardo and {Bertoldi}, Frank and {Carilli}, Chris and {Fan}, Xiaohui and {Farina}, Emanuele Paolo and {Mazzucchelli}, Chiara and {Riechers}, Dominik and {Rix}, Hans-Walter and {Strauss}, Michael A. and {Wang}, Ran and {Yang}, Yujin},
        title = "{An ALMA [C II] Survey of 27 Quasars at z > 5.94}",
      journal = {\apj},
         year = 2018,
        month = feb,
       volume = {854},
       number = {2},
          eid = {97},
        pages = {97},
          doi = {10.3847/1538-4357/aaa5aa},
archivePrefix = {arXiv},
       eprint = {1801.02641},
 primaryClass = {astro-ph.GA},
       adsurl = {https://ui.adsabs.harvard.edu/abs/2018ApJ...854...97D}
}

@ARTICLE{Berg2025,
       author = {{Berg}, Danielle A. and {Naidu}, Rohan P. and {Chisholm}, John and {Atek}, Hakim and {Fujimoto}, Seiji and {Kokorev}, Vasily and {Furtak}, Lukas J. and {Kobayashi}, Chiaki and {Schaerer}, Daniel and {Adamo}, Angela and {Fei}, Qinyue and {Korber}, Damien and {Matthee}, Jorryt and {Marques-Chaves}, Rui and {Martinez}, Zorayda and {Mcquinn}, Kristen B.~W. and {Mu{\~n}oz}, Julian B. and {Oesch}, Pascal A. and {Stark}, Daniel P. and {Stephenson}, Mabel G. and {Hsiao}, Tiger Yu-Yang},
        title = "{A Fleeting GLIMPSE of N/O Enrichment at Cosmic Dawn: Evidence for Wolf Rayet N Stars in a z = 6.1 Galaxy}",
      journal = {arXiv e-prints},
         year = 2025,
        month = nov,
          eid = {arXiv:2511.13591},
        pages = {arXiv:2511.13591},
          doi = {10.48550/arXiv.2511.13591},
archivePrefix = {arXiv},
       eprint = {2511.13591},
 primaryClass = {astro-ph.GA},
       adsurl = {https://ui.adsabs.harvard.edu/abs/2025arXiv251113591B}
}

@ARTICLE{Schaerer2024,
       author = {{Schaerer}, D. and {Marques-Chaves}, R. and {Xiao}, M. and {Korber}, D.},
        title = "{Discovery of a new N-emitter in the epoch of reionization}",
      journal = {\aap},
         year = 2024,
        month = jul,
       volume = {687},
          eid = {L11},
        pages = {L11},
          doi = {10.1051/0004-6361/202450721},
archivePrefix = {arXiv},
       eprint = {2406.08408},
 primaryClass = {astro-ph.GA},
       adsurl = {https://ui.adsabs.harvard.edu/abs/2024A&A...687L..11S}
}

@ARTICLE{Ji2024,
       author = {{Ji}, Xihan and {{\"U}bler}, Hannah and {Maiolino}, Roberto and {D'Eugenio}, Francesco and {Arribas}, Santiago and {Bunker}, Andrew J. and {Charlot}, St{\'e}phane and {Perna}, Michele and {Rodr{\'\i}guez Del Pino}, Bruno and {B{\"o}ker}, Torsten and {Cresci}, Giovanni and {Curti}, Mirko and {Kumari}, Nimisha and {Lamperti}, Isabella},
        title = "{GA-NIFS: an extremely nitrogen-loud and chemically stratified galaxy at z   5.55}",
      journal = {\mnras},
         year = 2024,
        month = nov,
       volume = {535},
       number = {1},
        pages = {881-908},
          doi = {10.1093/mnras/stae2375},
archivePrefix = {arXiv},
       eprint = {2404.04148},
 primaryClass = {astro-ph.GA},
       adsurl = {https://ui.adsabs.harvard.edu/abs/2024MNRAS.535..881J}
}

@ARTICLE{Topping2024,
       author = {{Topping}, Michael W. and {Stark}, Daniel P. and {Senchyna}, Peter and {Plat}, Adele and {Zitrin}, Adi and {Endsley}, Ryan and {Charlot}, St{\'e}phane and {Furtak}, Lukas J. and {Maseda}, Michael V. and {Smit}, Renske and {Mainali}, Ramesh and {Chevallard}, Jacopo and {Molyneux}, Stephen and {Rigby}, Jane R.},
        title = "{Metal-poor star formation at z > 6 with JWST: new insight into hard radiation fields and nitrogen enrichment on 20 pc scales}",
      journal = {\mnras},
         year = 2024,
        month = apr,
       volume = {529},
       number = {4},
        pages = {3301-3322},
          doi = {10.1093/mnras/stae682},
archivePrefix = {arXiv},
       eprint = {2401.08764},
 primaryClass = {astro-ph.GA},
       adsurl = {https://ui.adsabs.harvard.edu/abs/2024MNRAS.529.3301T}
}

@ARTICLE{Bunker2023,
       author = {{Bunker}, Andrew J. and {Saxena}, Aayush and {Cameron}, Alex J. and {Willott}, Chris J. and {Curtis-Lake}, Emma and {Jakobsen}, Peter and {Carniani}, Stefano and {Smit}, Renske and {Maiolino}, Roberto and {Witstok}, Joris and {Curti}, Mirko and {D'Eugenio}, Francesco and {Jones}, Gareth C. and {Ferruit}, Pierre and {Arribas}, Santiago and {Charlot}, Stephane and {Chevallard}, Jacopo and {Giardino}, Giovanna and {de Graaff}, Anna and {Looser}, Tobias J. and {L{\"u}tzgendorf}, Nora and {Maseda}, Michael V. and {Rawle}, Tim and {Rix}, Hans-Walter and {Del Pino}, Bruno Rodr{\'\i}guez and {Alberts}, Stacey and {Egami}, Eiichi and {Eisenstein}, Daniel J. and {Endsley}, Ryan and {Hainline}, Kevin and {Hausen}, Ryan and {Johnson}, Benjamin D. and {Rieke}, George and {Rieke}, Marcia and {Robertson}, Brant E. and {Shivaei}, Irene and {Stark}, Daniel P. and {Sun}, Fengwu and {Tacchella}, Sandro and {Tang}, Mengtao and {Williams}, Christina C. and {Willmer}, Christopher N.~A. and {Baker}, William M. and {Baum}, Stefi and {Bhatawdekar}, Rachana and {Bowler}, Rebecca and {Boyett}, Kristan and {Chen}, Zuyi and {Circosta}, Chiara and {Helton}, Jakob M. and {Ji}, Zhiyuan and {Kumari}, Nimisha and {Lyu}, Jianwei and {Nelson}, Erica and {Parlanti}, Eleonora and {Perna}, Michele and {Sandles}, Lester and {Scholtz}, Jan and {Suess}, Katherine A. and {Topping}, Michael W. and {{\"U}bler}, Hannah and {Wallace}, Imaan E.~B. and {Whitler}, Lily},
        title = "{JADES NIRSpec Spectroscopy of GN-z11: Lyman-{\ensuremath{\alpha}} emission and possible enhanced nitrogen abundance in a z = 10.60 luminous galaxy}",
      journal = {\aap},
         year = 2023,
        month = sep,
       volume = {677},
          eid = {A88},
        pages = {A88},
          doi = {10.1051/0004-6361/202346159},
archivePrefix = {arXiv},
       eprint = {2302.07256},
 primaryClass = {astro-ph.GA},
       adsurl = {https://ui.adsabs.harvard.edu/abs/2023A&A...677A..88B}
}

@ARTICLE{Cameron2023,
       author = {{Cameron}, Alex J. and {Katz}, Harley and {Rey}, Martin P. and {Saxena}, Aayush},
        title = "{Nitrogen enhancements 440 Myr after the big bang: supersolar N/O, a tidal disruption event, or a dense stellar cluster in GN-z11?}",
      journal = {\mnras},
         year = 2023,
        month = aug,
       volume = {523},
       number = {3},
        pages = {3516-3525},
          doi = {10.1093/mnras/stad1579},
archivePrefix = {arXiv},
       eprint = {2302.10142},
 primaryClass = {astro-ph.GA},
       adsurl = {https://ui.adsabs.harvard.edu/abs/2023MNRAS.523.3516C}
}

@ARTICLE{Cameron2026,
       author = {{Cameron}, Alex J. and {Carreira}, Courtney and {Simmonds}, Charlotte and {Bunker}, Andrew J. and {Saxena}, Aayush and {Carniani}, Stefano and {Charlot}, St{\'e}phane and {Chevallard}, Jacopo and {Curtis-Lake}, Emma and {Hainline}, Kevin and {Hausen}, Ryan and {Ji}, Xihan and {Ji}, Zhiyuan and {Johnson}, Benjamin D. and {Rinaldi}, Pierluigi and {Robertson}, Brant and {Scholtz}, Jan and {Silcock}, Maddie S. and {Tacchella}, Sandro and {Trussler}, James A.~A. and {{\"U}bler}, Hannah and {Williams}, Christina C. and {Willmer}, Christopher N.~A. and {Willott}, Chris and {Witstok}, Joris},
        title = "{JADES: Evolution of nitrogen abundances in star-forming galaxies from z \raisebox{-0.5ex}\textasciitilde 1.5-7}",
      journal = {arXiv e-prints},
         year = 2026,
        month = jan,
          eid = {arXiv:2601.15964},
        pages = {arXiv:2601.15964},
          doi = {10.48550/arXiv.2601.15964},
archivePrefix = {arXiv},
       eprint = {2601.15964},
 primaryClass = {astro-ph.GA},
       adsurl = {https://ui.adsabs.harvard.edu/abs/2026arXiv260115964C}
}

@ARTICLE{Vink2023,
       author = {{Vink}, Jorick S.},
        title = "{Very massive stars and nitrogen-emitting galaxies}",
      journal = {\aap},
         year = 2023,
        month = nov,
       volume = {679},
          eid = {L9},
        pages = {L9},
          doi = {10.1051/0004-6361/202347827},
archivePrefix = {arXiv},
       eprint = {2310.10725},
 primaryClass = {astro-ph.GA},
       adsurl = {https://ui.adsabs.harvard.edu/abs/2023A&A...679L...9V}
}

@ARTICLE{Shi2026a,
       author = {{Shi}, Yanlong and {Dai}, Liang and {Murray}, Norman and {Ye}, Claire S. and {Matzner}, Christopher D. and {Pascale}, Massimo},
        title = "{Very Massive Stars and High N/O: A Tale of the Nitrogen-enriched Super Star Cluster in the Sunburst Arc}",
      journal = {\apj},
         year = 2026,
        month = feb,
       volume = {997},
       number = {2},
          eid = {309},
        pages = {309},
          doi = {10.3847/1538-4357/ae3150},
archivePrefix = {arXiv},
       eprint = {2510.15823},
 primaryClass = {astro-ph.GA},
       adsurl = {https://ui.adsabs.harvard.edu/abs/2026ApJ...997..309S}
}

@ARTICLE{Shi2026b,
       author = {{Shi}, Yanlong and {Murray}, Norman},
        title = "{The in-situ growth of stellar-mass ``light'' seed black holes in nuclear star clusters}",
      journal = {arXiv e-prints},
         year = 2026,
        month = mar,
          eid = {arXiv:2603.10581},
        pages = {arXiv:2603.10581},
          doi = {10.48550/arXiv.2603.10581},
archivePrefix = {arXiv},
       eprint = {2603.10581},
 primaryClass = {astro-ph.GA},
       adsurl = {https://ui.adsabs.harvard.edu/abs/2026arXiv260310581S}
}

@ARTICLE{Costa2025,
       author = {{Costa}, G. and {Shepherd}, K.~G. and {Bressan}, A. and {Addari}, F. and {Chen}, Y. and {Fu}, X. and {Volpato}, G. and {Nguyen}, C.~T. and {Girardi}, L. and {Marigo}, P. and {Mazzi}, A. and {Pastorelli}, G. and {Trabucchi}, M. and {Bossini}, D. and {Zaggia}, S.},
        title = "{Evolutionary tracks, ejecta, and ionizing photons from intermediate-mass to very massive stars with PARSEC}",
      journal = {\aap},
         year = 2025,
        month = feb,
       volume = {694},
          eid = {A193},
        pages = {A193},
          doi = {10.1051/0004-6361/202452573},
archivePrefix = {arXiv},
       eprint = {2501.12917},
 primaryClass = {astro-ph.SR},
       adsurl = {https://ui.adsabs.harvard.edu/abs/2025A&A...694A.193C}
}

@INPROCEEDINGS{Woosley2015,
       author = {{Woosley}, Stan. E. and {Heger}, Alexander},
        title = "{The Deaths of Very Massive Stars}",
    booktitle = {Very Massive Stars in the Local Universe},
         year = 2015,
       editor = {{Vink}, Jorick S.},
       series = {Astrophysics and Space Science Library},
       volume = {412},
        month = jan,
        pages = {199},
          doi = {10.1007/978-3-319-09596-7_7},
archivePrefix = {arXiv},
       eprint = {1406.5657},
 primaryClass = {astro-ph.SR},
       adsurl = {https://ui.adsabs.harvard.edu/abs/2015ASSL..412..199W}
}

@ARTICLE{Vink2015,
       author = {{Vink}, Jorick S. and {Heger}, Alexander and {Krumholz}, Mark R. and {Puls}, Joachim and {Banerjee}, S. and {Castro}, N. and {Chen}, K.-J. and {Chen{\`e}}, A.-N. and {Crowther}, P.~A. and {Daminelli}, A. and {Gr{\"a}fener}, G. and {Groh}, J.~H. and {Hamann}, W.-R. and {Heap}, S. and {Herrero}, A. and {Kaper}, L. and {Najarro}, F. and {Oskinova}, L.~M. and {Roman-Lopes}, A. and {Rosen}, A. and {Sander}, A. and {Shirazi}, M. and {Sugawara}, Y. and {Tramper}, F. and {Vanbeveren}, D. and {Voss}, R. and {Wofford}, A. and {Zhang}, Y.},
        title = "{Very Massive Stars in the local Universe}",
      journal = {Highlights of Astronomy},
         year = 2015,
        month = mar,
       volume = {16},
        pages = {51-79},
          doi = {10.1017/S1743921314004657},
archivePrefix = {arXiv},
       eprint = {1302.2021},
 primaryClass = {astro-ph.SR},
       adsurl = {https://ui.adsabs.harvard.edu/abs/2015HiA....16...51V}
}

@INPROCEEDINGS{Vink2015b,
       author = {{Vink}, Jorick S.},
        title = "{Mass-Loss Rates of Very Massive Stars}",
    booktitle = {Very Massive Stars in the Local Universe},
         year = 2015,
       editor = {{Vink}, Jorick S.},
       series = {Astrophysics and Space Science Library},
       volume = {412},
        month = jan,
        pages = {77},
          doi = {10.1007/978-3-319-09596-7_4},
archivePrefix = {arXiv},
       eprint = {1406.5357},
 primaryClass = {astro-ph.SR},
       adsurl = {https://ui.adsabs.harvard.edu/abs/2015ASSL..412...77V}
}

@ARTICLE{PortegiesZwart2004,
       author = {{Portegies Zwart}, Simon F. and {Baumgardt}, Holger and {Hut}, Piet and {Makino}, Junichiro and {McMillan}, Stephen L.~W.},
        title = "{Formation of massive black holes through runaway collisions in dense young star clusters}",
      journal = {\nat},
         year = 2004,
        month = apr,
       volume = {428},
       number = {6984},
        pages = {724-726},
          doi = {10.1038/nature02448},
archivePrefix = {arXiv},
       eprint = {astro-ph/0402622},
 primaryClass = {astro-ph},
       adsurl = {https://ui.adsabs.harvard.edu/abs/2004Natur.428..724P}
}

@ARTICLE{Gurkan2004,
       author = {{G{\"u}rkan}, M. Atakan and {Freitag}, Marc and {Rasio}, Frederic A.},
        title = "{Formation of Massive Black Holes in Dense Star Clusters. I. Mass Segregation and Core Collapse}",
      journal = {\apj},
         year = 2004,
        month = apr,
       volume = {604},
       number = {2},
        pages = {632-652},
          doi = {10.1086/381968},
archivePrefix = {arXiv},
       eprint = {astro-ph/0308449},
 primaryClass = {astro-ph},
       adsurl = {https://ui.adsabs.harvard.edu/abs/2004ApJ...604..632G}
}

@ARTICLE{Rantala2024,
       author = {{Rantala}, Antti and {Naab}, Thorsten and {Lah{\'e}n}, Natalia},
        title = "{FROST-CLUSTERS - I. Hierarchical star cluster assembly boosts intermediate-mass black hole formation}",
      journal = {\mnras},
         year = 2024,
        month = jul,
       volume = {531},
       number = {3},
        pages = {3770-3799},
          doi = {10.1093/mnras/stae1413},
archivePrefix = {arXiv},
       eprint = {2403.10602},
 primaryClass = {astro-ph.GA},
       adsurl = {https://ui.adsabs.harvard.edu/abs/2024MNRAS.531.3770R}
}

@ARTICLE{Inayoshi2020,
       author = {{Inayoshi}, Kohei and {Visbal}, Eli and {Haiman}, Zolt{\'a}n},
        title = "{The Assembly of the First Massive Black Holes}",
      journal = {\araa},
         year = 2020,
        month = aug,
       volume = {58},
        pages = {27-97},
          doi = {10.1146/annurev-astro-120419-014455},
archivePrefix = {arXiv},
       eprint = {1911.05791},
 primaryClass = {astro-ph.GA},
       adsurl = {https://ui.adsabs.harvard.edu/abs/2020ARA&A..58...27I}
}

@ARTICLE{Volonteri2021,
       author = {{Volonteri}, Marta and {Habouzit}, M{\'e}lanie and {Colpi}, Monica},
        title = "{The origins of massive black holes}",
      journal = {Nature Reviews Physics},
         year = 2021,
        month = sep,
       volume = {3},
       number = {11},
        pages = {732-743},
          doi = {10.1038/s42254-021-00364-9},
archivePrefix = {arXiv},
       eprint = {2110.10175},
 primaryClass = {astro-ph.GA},
       adsurl = {https://ui.adsabs.harvard.edu/abs/2021NatRP...3..732V}
}

@ARTICLE{Shi2021,
       author = {{Shi}, Yanlong and {Grudi{\'c}}, Michael Y. and {Hopkins}, Philip F.},
        title = "{The mass budget for intermediate-mass black holes in dense star clusters}",
      journal = {\mnras},
         year = 2021,
        month = aug,
       volume = {505},
       number = {2},
        pages = {2753-2763},
          doi = {10.1093/mnras/stab1470},
archivePrefix = {arXiv},
       eprint = {2008.12290},
 primaryClass = {astro-ph.GA},
       adsurl = {https://ui.adsabs.harvard.edu/abs/2021MNRAS.505.2753S}
}

@ARTICLE{Cataldi2025,
       author = {{Cataldi}, E. and {Belfiore}, F. and {Curti}, M. and {Moreschini}, B. and {Marconi}, A. and {Maiolino}, R. and {Feltre}, A. and {Ginolfi}, M. and {Mannucci}, F. and {Cresci}, G. and {Ji}, X. and {Amiri}, A. and {Arnaboldi}, M. and {Bertola}, E. and {Bracci}, C. and {Ceci}, M. and {Chakraborty}, A. and {Cullen}, F. and {D'Amato}, Q. and {Kobayashi}, C. and {Lamperti}, I. and {Marconcini}, C. and {Scialpi}, M. and {Ulivi}, L. and {Zanchettin}, M.~V.},
        title = "{Tracing nitrogen enrichment across cosmic time with JWST}",
      journal = {arXiv e-prints},
         year = 2025,
        month = dec,
          eid = {arXiv:2512.07955},
        pages = {arXiv:2512.07955},
          doi = {10.48550/arXiv.2512.07955},
archivePrefix = {arXiv},
       eprint = {2512.07955},
 primaryClass = {astro-ph.GA},
       adsurl = {https://ui.adsabs.harvard.edu/abs/2025arXiv251207955C}
}

@software{Newville2025,
       author = {{Newville}, Matthew and {Otten}, Renee and {Nelson}, Andrew and {Stensitzki}, Till and {Ingargiola}, Antonino and {Allan}, Daniel and {Fox}, Austin and {Carter}, Faustin and {Rawlik}, Michal},
        title = "{LMFIT: Non-Linear Least-Squares Minimization and Curve-Fitting for Python}",
         year = 2025,
        month = jul,
          eid = {10.5281/zenodo.16175987},
          doi = {10.5281/zenodo.16175987},
      version = {1.3.4},
    publisher = {Zenodo},
       adsurl = {https://ui.adsabs.harvard.edu/abs/2025zndo..16175987N}
}

@ARTICLE{Killi2024,
       author = {{Killi}, Meghana and {Ginolfi}, Michele and {Popping}, Gerg{\"o} and {Watson}, Darach and {Zamorani}, Giovanni and {Lemaux}, Brian C. and {Fujimoto}, Seiji and {Faisst}, Andreas and {Bethermin}, Matthieu and {Romano}, Michael and {Fudamoto}, Yoshinobu and {Bardelli}, Sandro and {Boquien}, M{\'e}d{\'e}ric and {Carniani}, Stefano and {Dessauges-Zavadsky}, Miroslava and {Gruppioni}, Carlotta and {Hathi}, Nimish and {Ibar}, Eduardo and {Jones}, Gareth C. and {Koekemoer}, Anton M. and {Langan}, Ivanna and {M{\'e}ndez-Hern{\'a}ndez}, Hugo and {Sugahara}, Yuma and {Vallini}, Livia and {Vergani}, Daniela},
        title = "{The ALPINE-ALMA [C II] survey: characterization of spatial offsets in main-sequence galaxies at z   4-6}",
      journal = {\mnras},
         year = 2024,
        month = jul,
       volume = {531},
       number = {3},
        pages = {3222-3241},
          doi = {10.1093/mnras/stae1371},
archivePrefix = {arXiv},
       eprint = {2402.07982},
 primaryClass = {astro-ph.GA},
       adsurl = {https://ui.adsabs.harvard.edu/abs/2024MNRAS.531.3222K}
}

@ARTICLE{Bowler2024,
       author = {{Bowler}, R.~A.~A. and {Inami}, H. and {Sommovigo}, L. and {Smit}, R. and {Algera}, H.~S.~B. and {Aravena}, M. and {Barrufet}, L. and {Bouwens}, R. and {da Cunha}, E. and {Cullen}, F. and {Dayal}, P. and {De Looze}, I. and {Dunlop}, J.~S. and {Fudamoto}, Y. and {Mauerhofer}, V. and {McLure}, R.~J. and {Stefanon}, M. and {Schneider}, R. and {Ferrara}, A. and {Graziani}, L. and {Hodge}, J.~A. and {Nanayakkara}, T. and {Palla}, M. and {Schouws}, S. and {Stark}, D.~P. and {van der Werf}, P.~P.},
        title = "{The ALMA REBELS survey: obscured star formation in massive Lyman-break galaxies at z= 4-8 revealed by the IRX-{\ensuremath{\beta}} and M$_{{\ensuremath{\star}}}$ relations}",
      journal = {\mnras},
         year = 2024,
        month = jan,
       volume = {527},
       number = {3},
        pages = {5808-5828},
          doi = {10.1093/mnras/stad3578},
archivePrefix = {arXiv},
       eprint = {2309.17386},
 primaryClass = {astro-ph.GA},
       adsurl = {https://ui.adsabs.harvard.edu/abs/2024MNRAS.527.5808B}
}

@ARTICLE{Ocvirk2025,
       author = {{Ocvirk}, Pierre and {Lewis}, Joseph S.~W. and {Conaboy}, Luke and {Dubois}, Yohan and {Bethermin}, Matthieu and {Sorce}, Jenny G. and {Aubert}, Dominique and {Shapiro}, Paul R. and {Dawoodbhoy}, Taha and {Lee}, Joohyun and {Teyssier}, Romain and {Yepes}, Gustavo and {Gottl{\"o}ber}, Stefan and {Iliev}, Ilian T. and {Ahn}, Kyungjin and {Park}, Hyunbae and {Palanque}, Mei},
        title = "{Dust-UV offsets in high-redshift galaxies in the Cosmic Dawn III simulation}",
      journal = {\aap},
         year = 2025,
        month = nov,
       volume = {703},
          eid = {A98},
        pages = {A98},
          doi = {10.1051/0004-6361/202452098},
archivePrefix = {arXiv},
       eprint = {2409.05946},
 primaryClass = {astro-ph.GA},
       adsurl = {https://ui.adsabs.harvard.edu/abs/2025A&A...703A..98O}
}

@ARTICLE{Ziparo2023,
       author = {{Ziparo}, Francesco and {Ferrara}, Andrea and {Sommovigo}, Laura and {Kohandel}, Mahsa},
        title = "{Blue monsters. Why are JWST super-early, massive galaxies so blue?}",
      journal = {\mnras},
         year = 2023,
        month = apr,
       volume = {520},
       number = {2},
        pages = {2445-2450},
          doi = {10.1093/mnras/stad125},
archivePrefix = {arXiv},
       eprint = {2209.06840},
 primaryClass = {astro-ph.GA},
       adsurl = {https://ui.adsabs.harvard.edu/abs/2023MNRAS.520.2445Z}
}

@ARTICLE{Marques-Chaves2026,
       author = {{Marques-Chaves}, Rui and {Martins}, Fabrice and {Schaerer}, Daniel and {Dessauges-Zavadsky}, Miroslava and {Palacios}, Ana},
        title = "{Signatures of Very Massive Stars in the Epoch of Reionization}",
      journal = {arXiv e-prints},
         year = 2026,
        month = apr,
          eid = {arXiv:2604.21493},
        pages = {arXiv:2604.21493},
          doi = {10.48550/arXiv.2604.21493},
archivePrefix = {arXiv},
       eprint = {2604.21493},
 primaryClass = {astro-ph.GA},
       adsurl = {https://ui.adsabs.harvard.edu/abs/2026arXiv260421493M}
}

@ARTICLE{PortegiesZwart1999,
       author = {{Portegies Zwart}, S.~F. and {Makino}, J. and {McMillan}, S.~L.~W. and {Hut}, P.},
        title = "{Star cluster ecology. III. Runaway collisions in young compact star clusters}",
      journal = {\aap},
         year = 1999,
        month = aug,
       volume = {348},
        pages = {117-126},
          doi = {10.48550/arXiv.astro-ph/9812006},
archivePrefix = {arXiv},
       eprint = {astro-ph/9812006},
 primaryClass = {astro-ph},
       adsurl = {https://ui.adsabs.harvard.edu/abs/1999A&A...348..117P}
}

@ARTICLE{Krumholz2009,
       author = {{Krumholz}, Mark R. and {Klein}, Richard I. and {McKee}, Christopher F. and {Offner}, Stella S.~R. and {Cunningham}, Andrew J.},
        title = "{The Formation of Massive Star Systems by Accretion}",
      journal = {Science},
         year = 2009,
        month = feb,
       volume = {323},
       number = {5915},
        pages = {754},
          doi = {10.1126/science.1165857},
archivePrefix = {arXiv},
       eprint = {0901.3157},
 primaryClass = {astro-ph.SR},
       adsurl = {https://ui.adsabs.harvard.edu/abs/2009Sci...323..754K}
}

@ARTICLE{Banerjee2012,
       author = {{Banerjee}, Sambaran and {Kroupa}, Pavel and {Oh}, Seungkyung},
        title = "{The emergence of super-canonical stars in R136-type starburst clusters}",
      journal = {\mnras},
         year = 2012,
        month = oct,
       volume = {426},
       number = {2},
        pages = {1416-1426},
          doi = {10.1111/j.1365-2966.2012.21672.x},
archivePrefix = {arXiv},
       eprint = {1208.0826},
 primaryClass = {astro-ph.SR},
       adsurl = {https://ui.adsabs.harvard.edu/abs/2012MNRAS.426.1416B}
}

@ARTICLE{Chen2023,
       author = {{Chen}, Zuyi and {Stark}, Daniel P. and {Endsley}, Ryan and {Topping}, Michael and {Whitler}, Lily and {Charlot}, St{\'e}phane},
        title = "{JWST/NIRCam observations of stars and H II regions in z ≃ 6-8 galaxies: properties of star-forming complexes on 150 pc scales}",
      journal = {\mnras},
         year = 2023,
        month = feb,
       volume = {518},
       number = {4},
        pages = {5607-5619},
          doi = {10.1093/mnras/stac3476},
archivePrefix = {arXiv},
       eprint = {2207.12657},
 primaryClass = {astro-ph.GA},
       adsurl = {https://ui.adsabs.harvard.edu/abs/2023MNRAS.518.5607C}
}

@ARTICLE{Zhu2026,
       author = {{Zhu}, Yongda and {Rieke}, Marcia J. and {Ji}, Zhiyuan and {Bunker}, Andrew J. and {Carreira}, Courtney and {Danhaive}, A. Lola and {Duan}, Qiao and {Egami}, Eiichi and {Eisenstein}, Daniel J. and {Hainline}, Kevin and {Johnson}, Benjamin D. and {Ma}, Zheng and {Pusk{\'a}s}, D{\'a}vid and {Rieke}, George H. and {Rinaldi}, Pierluigi and {Robertson}, Brant and {Tacchella}, Sandro and {{\"U}bler}, Hannah and {Villanueva}, Natalia C. and {Williams}, Christina C. and {Willmer}, Christopher N.~A. and {Wu}, Zihao and {Zhang}, Junyu},
        title = "{Clumps in High-redshift Galaxies: Mass Scaling and Radial Trends from JADES}",
      journal = {\apj},
         year = 2026,
        month = apr,
       volume = {1000},
       number = {2},
          eid = {303},
        pages = {303},
          doi = {10.3847/1538-4357/ae50f1},
archivePrefix = {arXiv},
       eprint = {2601.15965},
 primaryClass = {astro-ph.GA},
       adsurl = {https://ui.adsabs.harvard.edu/abs/2026ApJ..1000..303Z}
}

@ARTICLE{Figer2005,
       author = {{Figer}, Donald F.},
        title = "{An upper limit to the masses of stars}",
      journal = {\nat},
         year = 2005,
        month = mar,
       volume = {434},
       number = {7030},
        pages = {192-194},
          doi = {10.1038/nature03293},
archivePrefix = {arXiv},
       eprint = {astro-ph/0503193},
 primaryClass = {astro-ph},
       adsurl = {https://ui.adsabs.harvard.edu/abs/2005Natur.434..192F}
}

@ARTICLE{Vanzella2022,
       author = {{Vanzella}, E. and {Castellano}, M. and {Bergamini}, P. and {Meneghetti}, M. and {Zanella}, A. and {Calura}, F. and {Caminha}, G.~B. and {Rosati}, P. and {Cupani}, G. and {Me{\v{s}}tri{\'c}}, U. and {Brammer}, G. and {Tozzi}, P. and {Mercurio}, A. and {Grillo}, C. and {Sani}, E. and {Cristiani}, S. and {Nonino}, M. and {Merlin}, E. and {Pignataro}, G.~V.},
        title = "{High star cluster formation efficiency in the strongly lensed Sunburst Lyman-continuum galaxy at z = 2.37}",
      journal = {\aap},
         year = 2022,
        month = mar,
       volume = {659},
          eid = {A2},
        pages = {A2},
          doi = {10.1051/0004-6361/202141590},
archivePrefix = {arXiv},
       eprint = {2106.10280},
 primaryClass = {astro-ph.GA},
       adsurl = {https://ui.adsabs.harvard.edu/abs/2022A&A...659A...2V}
}

@ARTICLE{Sabhahit2023,
       author = {{Sabhahit}, Gautham N. and {Vink}, Jorick S. and {Sander}, Andreas A.~C. and {Higgins}, Erin R.},
        title = "{Very massive stars and pair-instability supernovae: mass-loss framework for low metallicity}",
      journal = {\mnras},
         year = 2023,
        month = sep,
       volume = {524},
       number = {1},
        pages = {1529-1546},
          doi = {10.1093/mnras/stad1888},
archivePrefix = {arXiv},
       eprint = {2306.11785},
 primaryClass = {astro-ph.SR},
       adsurl = {https://ui.adsabs.harvard.edu/abs/2023MNRAS.524.1529S}
}

@ARTICLE{Isobe2023,
       author = {{Isobe}, Yuki and {Ouchi}, Masami and {Tominaga}, Nozomu and {Watanabe}, Kuria and {Nakajima}, Kimihiko and {Umeda}, Hiroya and {Yajima}, Hidenobu and {Harikane}, Yuichi and {Fukushima}, Hajime and {Xu}, Yi and {Ono}, Yoshiaki and {Zhang}, Yechi},
        title = "{JWST Identification of Extremely Low C/N Galaxies with [N/O] {\ensuremath{\gtrsim}} 0.5 at z 6-10 Evidencing the Early CNO-cycle Enrichment and a Connection with Globular Cluster Formation}",
      journal = {\apj},
         year = 2023,
        month = dec,
       volume = {959},
       number = {2},
          eid = {100},
        pages = {100},
          doi = {10.3847/1538-4357/ad09be},
archivePrefix = {arXiv},
       eprint = {2307.00710},
 primaryClass = {astro-ph.GA},
       adsurl = {https://ui.adsabs.harvard.edu/abs/2023ApJ...959..100I}
}

@ARTICLE{Charbonnel2023,
       author = {{Charbonnel}, C. and {Schaerer}, D. and {Prantzos}, N. and {Ram{\'\i}rez-Galeano}, L. and {Fragos}, T. and {Kuruvanthodi}, A. and {Marques-Chaves}, R. and {Gieles}, M.},
        title = "{N-enhancement in GN-z11: First evidence for supermassive stars nucleosynthesis in proto-globular clusters-like conditions at high redshift?}",
      journal = {\aap},
         year = 2023,
        month = may,
       volume = {673},
          eid = {L7},
        pages = {L7},
          doi = {10.1051/0004-6361/202346410},
archivePrefix = {arXiv},
       eprint = {2303.07955},
 primaryClass = {astro-ph.GA},
       adsurl = {https://ui.adsabs.harvard.edu/abs/2023A&A...673L...7C}
}

@ARTICLE{Martins2023,
       author = {{Martins}, F. and {Schaerer}, D. and {Marques-Chaves}, R. and {Upadhyaya}, A.},
        title = "{Inferring the presence of very massive stars in local star-forming regions}",
      journal = {\aap},
         year = 2023,
        month = oct,
       volume = {678},
          eid = {A159},
        pages = {A159},
          doi = {10.1051/0004-6361/202346732},
archivePrefix = {arXiv},
       eprint = {2308.14489},
 primaryClass = {astro-ph.GA},
       adsurl = {https://ui.adsabs.harvard.edu/abs/2023A&A...678A.159M}
}

@ARTICLE{Welch2025,
       author = {{Welch}, Brian and {Rivera-Thorsen}, T. Emil and {Rigby}, Jane R. and {Hutchison}, Taylor A. and {Olivier}, Grace M. and {Berg}, Danielle A. and {Sharon}, Keren and {Dahle}, H{\r{a}}kon and {Owens}, M. Riley and {Bayliss}, Matthew B. and {Khullar}, Gourav and {Chisholm}, John and {Hayes}, Matthew and {Kim}, Keunho J.},
        title = "{The Sunburst Arc with JWST. III. An Abundance of Direct Chemical Abundances}",
      journal = {\apj},
         year = 2025,
        month = feb,
       volume = {980},
       number = {1},
          eid = {33},
        pages = {33},
          doi = {10.3847/1538-4357/ada76c},
archivePrefix = {arXiv},
       eprint = {2405.06631},
 primaryClass = {astro-ph.GA},
       adsurl = {https://ui.adsabs.harvard.edu/abs/2025ApJ...980...33W}
}

@ARTICLE{Senchyna2021,
       author = {{Senchyna}, Peter and {Stark}, Daniel P. and {Charlot}, St{\'e}phane and {Chevallard}, Jacopo and {Bruzual}, Gustavo and {Vidal-Garc{\'\i}a}, Alba},
        title = "{Ultraviolet spectra of extreme nearby star-forming regions: Evidence for an overabundance of very massive stars}",
      journal = {\mnras},
         year = 2021,
        month = jun,
       volume = {503},
       number = {4},
        pages = {6112-6135},
          doi = {10.1093/mnras/stab884},
archivePrefix = {arXiv},
       eprint = {2008.09780},
 primaryClass = {astro-ph.GA},
       adsurl = {https://ui.adsabs.harvard.edu/abs/2021MNRAS.503.6112S}
}

@ARTICLE{Leitherer2018,
       author = {{Leitherer}, Claus and {Byler}, Nell and {Lee}, Janice C. and {Levesque}, Emily M.},
        title = "{Physical Properties of II Zw 40's Super Star Cluster and Nebula: New Insights and Puzzles from UV Spectroscopy}",
      journal = {\apj},
         year = 2018,
        month = sep,
       volume = {865},
       number = {1},
          eid = {55},
        pages = {55},
          doi = {10.3847/1538-4357/aada84},
archivePrefix = {arXiv},
       eprint = {1808.04332},
 primaryClass = {astro-ph.GA},
       adsurl = {https://ui.adsabs.harvard.edu/abs/2018ApJ...865...55L}
}

@ARTICLE{Smith2023,
       author = {{Smith}, Linda J. and {Oey}, M.~S. and {Hernandez}, Svea and {Ryon}, Jenna and {Leitherer}, Claus and {Charlot}, Stephane and {Bruzual}, Gustavo and {Calzetti}, Daniela and {Chu}, You-Hua and {Hayes}, Matthew J. and {James}, Bethan L. and {Jaskot}, Anne E. and {{\"O}stlin}, G{\"o}ran},
        title = "{HST FUV Spectroscopy of Super Star Cluster A in the Green Pea Analog Mrk 71: Revealing the Presence of Very Massive Stars}",
      journal = {\apj},
         year = 2023,
        month = dec,
       volume = {958},
       number = {2},
          eid = {194},
        pages = {194},
          doi = {10.3847/1538-4357/ad00b4},
archivePrefix = {arXiv},
       eprint = {2310.03413},
 primaryClass = {astro-ph.GA},
       adsurl = {https://ui.adsabs.harvard.edu/abs/2023ApJ...958..194S}
}

@ARTICLE{Li2024,
       author = {{Li}, Zhaozhou and {Dekel}, Avishai and {Sarkar}, Kartick C. and {Aung}, Han and {Giavalisco}, Mauro and {Mandelker}, Nir and {Tacchella}, Sandro},
        title = "{Feedback-free starbursts at cosmic dawn: Observable predictions for JWST}",
      journal = {\aap},
         year = 2024,
        month = oct,
       volume = {690},
          eid = {A108},
        pages = {A108},
          doi = {10.1051/0004-6361/202348727},
archivePrefix = {arXiv},
       eprint = {2311.14662},
 primaryClass = {astro-ph.GA},
       adsurl = {https://ui.adsabs.harvard.edu/abs/2024A&A...690A.108L}
}

@ARTICLE{Trinca2024,
       author = {{Trinca}, Alessandro and {Schneider}, Raffaella and {Valiante}, Rosa and {Graziani}, Luca and {Ferrotti}, Arianna and {Omukai}, Kazuyuki and {Chon}, Sunmyon},
        title = "{Exploring the nature of UV-bright z {\ensuremath{\gtrsim}} 10 galaxies detected by JWST: star formation, black hole accretion, or a non-universal IMF?}",
      journal = {\mnras},
         year = 2024,
        month = apr,
       volume = {529},
       number = {4},
        pages = {3563-3581},
          doi = {10.1093/mnras/stae651},
archivePrefix = {arXiv},
       eprint = {2305.04944},
 primaryClass = {astro-ph.GA},
       adsurl = {https://ui.adsabs.harvard.edu/abs/2024MNRAS.529.3563T}
}

@ARTICLE{Cueto2024,
       author = {{Cueto}, Elie R. and {Hutter}, Anne and {Dayal}, Pratika and {Gottl{\"o}ber}, Stefan and {Heintz}, Kasper E. and {Mason}, Charlotte and {Trebitsch}, Maxime and {Yepes}, Gustavo},
        title = "{ASTRAEUS. IX. Impact of an evolving stellar initial mass function on early galaxies and reionisation}",
      journal = {\aap},
         year = 2024,
        month = jun,
       volume = {686},
          eid = {A138},
        pages = {A138},
          doi = {10.1051/0004-6361/202349017},
archivePrefix = {arXiv},
       eprint = {2312.12109},
 primaryClass = {astro-ph.GA},
       adsurl = {https://ui.adsabs.harvard.edu/abs/2024A&A...686A.138C}
}

@ARTICLE{Pallottini2023,
       author = {{Pallottini}, A. and {Ferrara}, A.},
        title = "{Stochastic star formation in early galaxies: Implications for the James Webb Space Telescope}",
      journal = {\aap},
         year = 2023,
        month = sep,
       volume = {677},
          eid = {L4},
        pages = {L4},
          doi = {10.1051/0004-6361/202347384},
archivePrefix = {arXiv},
       eprint = {2307.03219},
 primaryClass = {astro-ph.GA},
       adsurl = {https://ui.adsabs.harvard.edu/abs/2023A&A...677L...4P}
}

@ARTICLE{Shen2023,
       author = {{Shen}, Xuejian and {Vogelsberger}, Mark and {Boylan-Kolchin}, Michael and {Tacchella}, Sandro and {Kannan}, Rahul},
        title = "{The impact of UV variability on the abundance of bright galaxies at z {\ensuremath{\geq}} 9}",
      journal = {\mnras},
         year = 2023,
        month = nov,
       volume = {525},
       number = {3},
        pages = {3254-3261},
          doi = {10.1093/mnras/stad2508},
archivePrefix = {arXiv},
       eprint = {2305.05679},
 primaryClass = {astro-ph.GA},
       adsurl = {https://ui.adsabs.harvard.edu/abs/2023MNRAS.525.3254S}
}

@ARTICLE{Ferrara2025,
       author = {{Ferrara}, A. and {Pallottini}, A. and {Sommovigo}, L.},
        title = "{Blue monsters at z > 10: Where all their dust has gone}",
      journal = {\aap},
         year = 2025,
        month = feb,
       volume = {694},
          eid = {A286},
        pages = {A286},
          doi = {10.1051/0004-6361/202452707},
archivePrefix = {arXiv},
       eprint = {2410.19042},
 primaryClass = {astro-ph.GA},
       adsurl = {https://ui.adsabs.harvard.edu/abs/2025A&A...694A.286F}
}

@ARTICLE{Harikane2023a,
       author = {{Harikane}, Yuichi and {Ouchi}, Masami and {Oguri}, Masamune and {Ono}, Yoshiaki and {Nakajima}, Kimihiko and {Isobe}, Yuki and {Umeda}, Hiroya and {Mawatari}, Ken and {Zhang}, Yechi},
        title = "{A Comprehensive Study of Galaxies at z   9-16 Found in the Early JWST Data: Ultraviolet Luminosity Functions and Cosmic Star Formation History at the Pre-reionization Epoch}",
      journal = {\apjs},
         year = 2023,
        month = mar,
       volume = {265},
       number = {1},
          eid = {5},
        pages = {5},
          doi = {10.3847/1538-4365/acaaa9},
archivePrefix = {arXiv},
       eprint = {2208.01612},
 primaryClass = {astro-ph.GA},
       adsurl = {https://ui.adsabs.harvard.edu/abs/2023ApJS..265....5H}
}

@ARTICLE{Robertson2024,
       author = {{Robertson}, Brant and {Johnson}, Benjamin D. and {Tacchella}, Sandro and {Eisenstein}, Daniel J. and {Hainline}, Kevin and {Arribas}, Santiago and {Baker}, William M. and {Bunker}, Andrew J. and {Carniani}, Stefano and {Cargile}, Phillip A. and {Carreira}, Courtney and {Charlot}, Stephane and {Chevallard}, Jacopo and {Curti}, Mirko and {Curtis-Lake}, Emma and {D'Eugenio}, Francesco and {Egami}, Eiichi and {Hausen}, Ryan and {Helton}, Jakob M. and {Jakobsen}, Peter and {Ji}, Zhiyuan and {Jones}, Gareth C. and {Maiolino}, Roberto and {Maseda}, Michael V. and {Nelson}, Erica and {P{\'e}rez-Gonz{\'a}lez}, Pablo G. and {Pusk{\'a}s}, D{\'a}vid and {Rieke}, Marcia and {Smit}, Renske and {Sun}, Fengwu and {{\"U}bler}, Hannah and {Whitler}, Lily and {Williams}, Christina C. and {Willmer}, Christopher N.~A. and {Willott}, Chris and {Witstok}, Joris},
        title = "{Earliest Galaxies in the JADES Origins Field: Luminosity Function and Cosmic Star Formation Rate Density 300 Myr after the Big Bang}",
      journal = {\apj},
         year = 2024,
        month = jul,
       volume = {970},
       number = {1},
          eid = {31},
        pages = {31},
          doi = {10.3847/1538-4357/ad463d},
archivePrefix = {arXiv},
       eprint = {2312.10033},
 primaryClass = {astro-ph.GA},
       adsurl = {https://ui.adsabs.harvard.edu/abs/2024ApJ...970...31R}
}

@ARTICLE{Finkelstein2023,
       author = {{Finkelstein}, Steven L. and {Bagley}, Micaela B. and {Ferguson}, Henry C. and {Wilkins}, Stephen M. and {Kartaltepe}, Jeyhan S. and {Papovich}, Casey and {Yung}, L.~Y. Aaron and {Arrabal Haro}, Pablo and {Behroozi}, Peter and {Dickinson}, Mark and {Kocevski}, Dale D. and {Koekemoer}, Anton M. and {Larson}, Rebecca L. and {Le Bail}, Aur{\'e}lien and {Morales}, Alexa M. and {P{\'e}rez-Gonz{\'a}lez}, Pablo G. and {Burgarella}, Denis and {Dav{\'e}}, Romeel and {Hirschmann}, Michaela and {Somerville}, Rachel S. and {Wuyts}, Stijn and {Bromm}, Volker and {Casey}, Caitlin M. and {Fontana}, Adriano and {Fujimoto}, Seiji and {Gardner}, Jonathan P. and {Giavalisco}, Mauro and {Grazian}, Andrea and {Grogin}, Norman A. and {Hathi}, Nimish P. and {Hutchison}, Taylor A. and {Jha}, Saurabh W. and {Jogee}, Shardha and {Kewley}, Lisa J. and {Kirkpatrick}, Allison and {Long}, Arianna S. and {Lotz}, Jennifer M. and {Pentericci}, Laura and {Pierel}, Justin D.~R. and {Pirzkal}, Nor and {Ravindranath}, Swara and {Ryan}, Russell E. and {Trump}, Jonathan R. and {Yang}, Guang and {Bhatawdekar}, Rachana and {Bisigello}, Laura and {Buat}, V{\'e}ronique and {Calabr{\`o}}, Antonello and {Castellano}, Marco and {Cleri}, Nikko J. and {Cooper}, M.~C. and {Croton}, Darren and {Daddi}, Emanuele and {Dekel}, Avishai and {Elbaz}, David and {Franco}, Maximilien and {Gawiser}, Eric and {Holwerda}, Benne W. and {Huertas-Company}, Marc and {Jaskot}, Anne E. and {Leung}, Gene C.~K. and {Lucas}, Ray A. and {Mobasher}, Bahram and {Pandya}, Viraj and {Tacchella}, Sandro and {Weiner}, Benjamin J. and {Zavala}, Jorge A.},
        title = "{CEERS Key Paper. I. An Early Look into the First 500 Myr of Galaxy Formation with JWST}",
      journal = {\apjl},
         year = 2023,
        month = mar,
       volume = {946},
       number = {1},
          eid = {L13},
        pages = {L13},
          doi = {10.3847/2041-8213/acade4},
archivePrefix = {arXiv},
       eprint = {2211.05792},
 primaryClass = {astro-ph.GA},
       adsurl = {https://ui.adsabs.harvard.edu/abs/2023ApJ...946L..13F}
}

@ARTICLE{Ono2018,
       author = {{Ono}, Yoshiaki and {Ouchi}, Masami and {Harikane}, Yuichi and {Toshikawa}, Jun and {Rauch}, Michael and {Yuma}, Suraphong and {Sawicki}, Marcin and {Shibuya}, Takatoshi and {Shimasaku}, Kazuhiro and {Oguri}, Masamune and {Willott}, Chris and {Akhlaghi}, Mohammad and {Akiyama}, Masayuki and {Coupon}, Jean and {Kashikawa}, Nobunari and {Komiyama}, Yutaka and {Konno}, Akira and {Lin}, Lihwai and {Matsuoka}, Yoshiki and {Miyazaki}, Satoshi and {Nagao}, Tohru and {Nakajima}, Kimihiko and {Silverman}, John and {Tanaka}, Masayuki and {Taniguchi}, Yoshiaki and {Wang}, Shiang-Yu},
        title = "{Great Optically Luminous Dropout Research Using Subaru HSC (GOLDRUSH). I. UV luminosity functions at z {\ensuremath{\sim}} 4-7 derived with the half-million dropouts on the 100 deg$^{2}$ sky}",
      journal = {\pasj},
         year = 2018,
        month = jan,
       volume = {70},
          eid = {S10},
        pages = {S10},
          doi = {10.1093/pasj/psx103},
archivePrefix = {arXiv},
       eprint = {1704.06004},
 primaryClass = {astro-ph.GA},
       adsurl = {https://ui.adsabs.harvard.edu/abs/2018PASJ...70S..10O}
}

@ARTICLE{Harikane2022a,
       author = {{Harikane}, Yuichi and {Ono}, Yoshiaki and {Ouchi}, Masami and {Liu}, Chengze and {Sawicki}, Marcin and {Shibuya}, Takatoshi and {Behroozi}, Peter S. and {He}, Wanqiu and {Shimasaku}, Kazuhiro and {Arnouts}, Stephane and {Coupon}, Jean and {Fujimoto}, Seiji and {Gwyn}, Stephen and {Huang}, Jiasheng and {Inoue}, Akio K. and {Kashikawa}, Nobunari and {Komiyama}, Yutaka and {Matsuoka}, Yoshiki and {Willott}, Chris J.},
        title = "{GOLDRUSH. IV. Luminosity Functions and Clustering Revealed with  4,000,000 Galaxies at z   2-7: Galaxy-AGN Transition, Star Formation Efficiency, and Implication for Evolution at z > 10}",
      journal = {\apjs},
         year = 2022,
        month = mar,
       volume = {259},
       number = {1},
          eid = {20},
        pages = {20},
          doi = {10.3847/1538-4365/ac3dfc},
archivePrefix = {arXiv},
       eprint = {2108.01090},
 primaryClass = {astro-ph.GA},
       adsurl = {https://ui.adsabs.harvard.edu/abs/2022ApJS..259...20H}
}

@ARTICLE{Bowler2017,
       author = {{Bowler}, R.~A.~A. and {Dunlop}, J.~S. and {McLure}, R.~J. and {McLeod}, D.~J.},
        title = "{Unveiling the nature of bright z ≃ 7 galaxies with the Hubble Space Telescope}",
      journal = {\mnras},
         year = 2017,
        month = apr,
       volume = {466},
       number = {3},
        pages = {3612-3635},
          doi = {10.1093/mnras/stw3296},
archivePrefix = {arXiv},
       eprint = {1605.05325},
 primaryClass = {astro-ph.GA},
       adsurl = {https://ui.adsabs.harvard.edu/abs/2017MNRAS.466.3612B}
}

@ARTICLE{Schouws2023,
       author = {{Schouws}, Sander and {Bouwens}, Rychard and {Smit}, Renske and {Hodge}, Jacqueline and {Stefanon}, Mauro and {Witstok}, Joris and {Hilhorst}, Juli{\"e}tte and {Labb{\'e}}, Ivo and {Algera}, Hiddo and {Boogaard}, Leindert and {Maseda}, Michael and {Oesch}, Pascal and {R{\"o}ttgering}, Huub and {van der Werf}, Paul},
        title = "{ALMA as a Redshift Machine: Using [C II] to Efficiently Confirm Galaxies in the Epoch of Reionization}",
      journal = {\apj},
         year = 2023,
        month = sep,
       volume = {954},
       number = {1},
          eid = {103},
        pages = {103},
          doi = {10.3847/1538-4357/ace10c},
archivePrefix = {arXiv},
       eprint = {2202.04080},
 primaryClass = {astro-ph.GA},
       adsurl = {https://ui.adsabs.harvard.edu/abs/2023ApJ...954..103S}
}

@ARTICLE{Matsuoka2025b,
       author = {{Matsuoka}, Yoshiki and {Iwasawa}, Kazushi and {Onoue}, Masafusa and {Izumi}, Takuma and {Strauss}, Michael A. and {Akiyama}, Masayuki and {Aoki}, Kentaro and {Arita}, Junya and {Ding}, Xuheng and {Imanishi}, Masatoshi and {Kashikawa}, Nobunari and {Kawaguchi}, Toshihiro and {Kikuta}, Satoshi and {Kohno}, Kotaro and {Lee}, Chien-Hsiu and {Nagao}, Tohru and {Phillips}, Camryn L. and {Sawamura}, Mahoshi and {Silverman}, John D. and {Takahashi}, Ayumi and {Toba}, Yoshiki},
        title = "{Subaru High-z Exploration of Low-luminosity Quasars (SHELLQs). XXIV. 54 New Quasars and Candidate Obscured Quasars at 5.71 {\ensuremath{\leq}} z {\ensuremath{\leq}} 7.02}",
      journal = {\apjs},
         year = 2025,
        month = oct,
       volume = {280},
       number = {2},
          eid = {68},
        pages = {68},
          doi = {10.3847/1538-4365/ae0035},
archivePrefix = {arXiv},
       eprint = {2508.21229},
 primaryClass = {astro-ph.GA},
       adsurl = {https://ui.adsabs.harvard.edu/abs/2025ApJS..280...68M}
}

@ARTICLE{Prochaska2020,
       author = {{Prochaska}, J. and {Hennawi}, Joseph and {Westfall}, Kyle and {Cooke}, Ryan and {Wang}, Feige and {Hsyu}, Tiffany and {Davies}, Frederick and {Farina}, Emanuele and {Pelliccia}, Debora},
        title = "{PypeIt: The Python Spectroscopic Data Reduction Pipeline}",
      journal = {The Journal of Open Source Software},
         year = 2020,
        month = dec,
       volume = {5},
       number = {56},
          eid = {2308},
        pages = {2308},
          doi = {10.21105/joss.02308},
archivePrefix = {arXiv},
       eprint = {2005.06505},
 primaryClass = {astro-ph.IM},
       adsurl = {https://ui.adsabs.harvard.edu/abs/2020JOSS....5.2308P}
}

@ARTICLE{Horne1986,
       author = {{Horne}, K.},
        title = "{An optimal extraction algorithm for CCD spectroscopy.}",
      journal = {\pasp},
         year = 1986,
        month = jun,
       volume = {98},
        pages = {609-617},
          doi = {10.1086/131801},
       adsurl = {https://ui.adsabs.harvard.edu/abs/1986PASP...98..609H}
}

@ARTICLE{Crowther2007,
       author = {{Crowther}, Paul A.},
        title = "{Physical Properties of Wolf-Rayet Stars}",
      journal = {\araa},
         year = 2007,
        month = sep,
       volume = {45},
       number = {1},
        pages = {177-219},
          doi = {10.1146/annurev.astro.45.051806.110615},
archivePrefix = {arXiv},
       eprint = {astro-ph/0610356},
 primaryClass = {astro-ph},
       adsurl = {https://ui.adsabs.harvard.edu/abs/2007ARA&A..45..177C}
}

@ARTICLE{Eldridge2017,
       author = {{Eldridge}, J.~J. and {Stanway}, E.~R. and {Xiao}, L. and {McClelland}, L.~A.~S. and {Taylor}, G. and {Ng}, M. and {Greis}, S.~M.~L. and {Bray}, J.~C.},
        title = "{Binary Population and Spectral Synthesis Version 2.1: Construction, Observational Verification, and New Results}",
      journal = {\pasa},
         year = 2017,
        month = nov,
       volume = {34},
          eid = {e058},
        pages = {e058},
          doi = {10.1017/pasa.2017.51},
archivePrefix = {arXiv},
       eprint = {1710.02154},
 primaryClass = {astro-ph.SR},
       adsurl = {https://ui.adsabs.harvard.edu/abs/2017PASA...34...58E}
}

@ARTICLE{Stanway2018,
       author = {{Stanway}, E.~R. and {Eldridge}, J.~J.},
        title = "{Re-evaluating old stellar populations}",
      journal = {\mnras},
         year = 2018,
        month = sep,
       volume = {479},
       number = {1},
        pages = {75-93},
          doi = {10.1093/mnras/sty1353},
archivePrefix = {arXiv},
       eprint = {1805.08784},
 primaryClass = {astro-ph.GA},
       adsurl = {https://ui.adsabs.harvard.edu/abs/2018MNRAS.479...75S}
}

@ARTICLE{Martins2025,
       author = {{Martins}, F. and {Palacios}, A. and {Schaerer}, D. and {Marques-Chaves}, R.},
        title = "{Very massive stars at low metallicity: Evolution, synthetic spectroscopy, and impact on the integrated light of starbursts}",
      journal = {\aap},
         year = 2025,
        month = jun,
       volume = {698},
          eid = {A262},
        pages = {A262},
          doi = {10.1051/0004-6361/202554421},
archivePrefix = {arXiv},
       eprint = {2505.02993},
 primaryClass = {astro-ph.SR},
       adsurl = {https://ui.adsabs.harvard.edu/abs/2025A&A...698A.262M}
}

@ARTICLE{Steidel2016,
       author = {{Steidel}, Charles C. and {Strom}, Allison L. and {Pettini}, Max and {Rudie}, Gwen C. and {Reddy}, Naveen A. and {Trainor}, Ryan F.},
        title = "{Reconciling the Stellar and Nebular Spectra of High-redshift Galaxies}",
      journal = {\apj},
         year = 2016,
        month = aug,
       volume = {826},
       number = {2},
          eid = {159},
        pages = {159},
          doi = {10.3847/0004-637X/826/2/159},
archivePrefix = {arXiv},
       eprint = {1605.07186},
 primaryClass = {astro-ph.GA},
       adsurl = {https://ui.adsabs.harvard.edu/abs/2016ApJ...826..159S}
}

@ARTICLE{Bestenlehner2014,
       author = {{Bestenlehner}, J.~M. and {Gr{\"a}fener}, G. and {Vink}, J.~S. and {Najarro}, F. and {de Koter}, A. and {Sana}, H. and {Evans}, C.~J. and {Crowther}, P.~A. and {H{\'e}nault-Brunet}, V. and {Herrero}, A. and {Langer}, N. and {Schneider}, F.~R.~N. and {Sim{\'o}n-D{\'\i}az}, S. and {Taylor}, W.~D. and {Walborn}, N.~R.},
        title = "{The VLT-FLAMES Tarantula Survey. XVII. Physical and wind properties of massive stars at the top of the main sequence}",
      journal = {\aap},
         year = 2014,
        month = oct,
       volume = {570},
          eid = {A38},
        pages = {A38},
          doi = {10.1051/0004-6361/201423643},
archivePrefix = {arXiv},
       eprint = {1407.1837},
 primaryClass = {astro-ph.SR},
       adsurl = {https://ui.adsabs.harvard.edu/abs/2014A&A...570A..38B}
}

@ARTICLE{Reddy2009,
       author = {{Reddy}, Naveen A. and {Steidel}, Charles C.},
        title = "{A Steep Faint-End Slope of the UV Luminosity Function at z \raisebox{-0.5ex}\textasciitilde 2-3: Implications for the Global Stellar Mass Density and Star Formation in Low-Mass Halos}",
      journal = {\apj},
         year = 2009,
        month = feb,
       volume = {692},
       number = {1},
        pages = {778-803},
          doi = {10.1088/0004-637X/692/1/778},
archivePrefix = {arXiv},
       eprint = {0810.2788},
 primaryClass = {astro-ph},
       adsurl = {https://ui.adsabs.harvard.edu/abs/2009ApJ...692..778R}
}

@ARTICLE{Brinchmann2008,
       author = {{Brinchmann}, Jarle and {Pettini}, Max and {Charlot}, St{\'e}phane},
        title = "{New insights into the stellar content and physical conditions of star-forming galaxies at z = 2-3 from spectral modelling}",
      journal = {\mnras},
         year = 2008,
        month = apr,
       volume = {385},
       number = {2},
        pages = {769-782},
          doi = {10.1111/j.1365-2966.2008.12914.x},
archivePrefix = {arXiv},
       eprint = {0801.1678},
 primaryClass = {astro-ph},
       adsurl = {https://ui.adsabs.harvard.edu/abs/2008MNRAS.385..769B}
}

@ARTICLE{Vink2011,
       author = {{Vink}, Jorick S. and {Muijres}, L.~E. and {Anthonisse}, B. and {de Koter}, A. and {Gr{\"a}fener}, G. and {Langer}, N.},
        title = "{Wind modelling of very massive stars up to 300 solar masses}",
      journal = {\aap},
         year = 2011,
        month = jul,
       volume = {531},
          eid = {A132},
        pages = {A132},
          doi = {10.1051/0004-6361/201116614},
archivePrefix = {arXiv},
       eprint = {1105.0556},
 primaryClass = {astro-ph.SR},
       adsurl = {https://ui.adsabs.harvard.edu/abs/2011A&A...531A.132V}
}

@ARTICLE{Grafener2011,
       author = {{Gr{\"a}fener}, G. and {Vink}, J.~S. and {de Koter}, A. and {Langer}, N.},
        title = "{The Eddington factor as the key to understand the winds of the most massive stars. Evidence for a {\ensuremath{\Gamma}}-dependence of Wolf-Rayet type mass loss}",
      journal = {\aap},
         year = 2011,
        month = nov,
       volume = {535},
          eid = {A56},
        pages = {A56},
          doi = {10.1051/0004-6361/201116701},
archivePrefix = {arXiv},
       eprint = {1106.5361},
 primaryClass = {astro-ph.SR},
       adsurl = {https://ui.adsabs.harvard.edu/abs/2011A&A...535A..56G}
}

@ARTICLE{Erb2010,
   author = {{Erb}, D.~K. and {Pettini}, M. and {Shapley}, A.~E. and {Steidel}, C.~C. and {Law}, D.~R. and {Reddy}, N.~A.},
    title = "{Physical Conditions in a Young, Unreddened, Low-metallicity Galaxy at High Redshift}",
  journal = {\apj},
     year = 2010,
   volume = {719},
    pages = {1168},
      doi = {10.1088/0004-637X/719/2/1168},
   adsurl = {https://ui.adsabs.harvard.edu/abs/2010ApJ...719.1168E}
}

@ARTICLE{Gazagnes2018,
   author = {{Gazagnes}, S. and {Chisholm}, J. and {Schaerer}, D. and {Verhamme}, A. and {Rigby}, J.~R. and {Bayliss}, M.},
    title = "{Neutral gas properties of Lyman continuum emitting galaxies: Column densities and covering fractions from UV absorption lines}",
  journal = {\aap},
     year = 2018,
   volume = {616},
    pages = {A29},
      doi = {10.1051/0004-6361/201832759},
   adsurl = {https://ui.adsabs.harvard.edu/abs/2018A&A...616A..29G}
}

@ARTICLE{Reddy2016,
   author = {{Reddy}, N.~A. and {Steidel}, C.~C. and {Pettini}, M. and {Bogosavljevi{\'c}}, M. and {Shapley}, A.~E.},
    title = "{The Connection between Reddening, Gas Covering Fraction, and the Escape of Ionizing Radiation at High Redshift}",
  journal = {\apj},
     year = 2016,
   volume = {828},
    pages = {108},
      doi = {10.3847/0004-637X/828/2/108},
   eprint = {1606.03452},
   adsurl = {https://ui.adsabs.harvard.edu/abs/2016ApJ...828..108R}
}

@ARTICLE{Saldana-Lopez2022,
   author = {{Saldana-Lopez}, A. and {Schaerer}, D. and {Chisholm}, J. and {Flury}, S.~R. and {Jaskot}, A.~E. and {Worseck}, G. and {Makan}, K. and {Gazagnes}, S. and {Mauerhofer}, V. and {Verhamme}, A. and {Amor{\'\i}n}, R.~O. and {Ferguson}, H.~C. and {Giavalisco}, M. and {Grazian}, A. and {Hayes}, M.~J. and {Heckman}, T.~M. and {Henry}, A. and {Ji}, Z. and {Marques-Chaves}, R. and {McCandliss}, S.~R. and {Oey}, M.~S. and {{\"O}stlin}, G. and {Pentericci}, L. and {Thuan}, T.~X. and {Trebitsch}, M. and others},
    title = "{The Low-redshift Lyman Continuum Survey. Unveiling the ISM properties of low-z Lyman-continuum emitters}",
  journal = {\aap},
     year = 2022,
   volume = {663},
    pages = {A59},
      doi = {10.1051/0004-6361/202141864},
   adsurl = {https://ui.adsabs.harvard.edu/abs/2022A&A...663A..59S}
}

@ARTICLE{Chen2026,
       author = {{Chen}, Zuyi and {Stark}, Daniel P. and {Mason}, Charlotte A. and {Senchyna}, Peter and {Tang}, Mengtao and {C.}, Keerthi Vasan G. and {Whitler}, Lily and {Plat}, Adele and {Gelli}, Viola},
        title = "{SPURS: Massive Stars, Dense Gas, and Ly$α$ Escape in GN-z11 at $z = 10.6$}",
      journal = {arXiv e-prints},
         year = 2026,
        month = aug,
          eid = {arXiv:2608.12699},
        pages = {arXiv:2608.12699},
archivePrefix = {arXiv},
       eprint = {2608.12699},
 primaryClass = {astro-ph.GA},
       adsurl = {https://ui.adsabs.harvard.edu/abs/2026arXiv260812699C}
}

@ARTICLE{Yang2026,
       author = {{Yang}, D. and {Hennawi}, J.~F. and {Guarneri}, F. and {Wolf}, J. and {Belladitta}, S. and {Schindler}, J.-T. and {Hughes}, A.~C.~N. and {Ba{\~n}ados}, E. and {Mortlock}, D.~J. and {Yang}, J. and {Wang}, F. and {Fan}, X. and {Jahnke}, K. and {Stern}, D. and {Willott}, C.~J. and {Barth}, A.~J. and {Rottgering}, H.~J.~A. and {Varadaraj}, R.~G. and {Decarli}, R. and {Eilers}, A.-C. and {Ezziati}, M. and {Fu}, Y. and {Huang}, J. and {Jin}, X. and {Kang}, Y. and {Martinez-Ramirez}, L.~N. and {Matsuoka}, Y. and {Onoue}, M. and {Pello}, R. and {Remigio}, R.~P. and {Tee}, W.~L. and {Venemans}, B. and {Vietri}, G. and {Wang}, B. and {Abbo}, L.~J. and {Atek}, H. and {Bisogni}, S. and {Bosman}, S.~E.~I. and {Bowler}, R.~A.~A. and {Conselice}, C.~J. and {Davies}, F.~B. and {Gutierrez}, C.~M. and {Harikane}, Y. and {Rubinur}, K. and {Lovell}, C.~C. and {Magliocchetti}, M. and {Matthee}, J. and {Ricci}, F. and {Scialpi}, M. and {Scott}, D. and {Spinoglio}, L. and {Tarsitano}, F. and {Toba}, Y. and {Walter}, F. and {Weaver}, J.~R. and {Zamorani}, G. and {Altieri}, B. and {Amara}, A. and {Andreon}, S. and {Aussel}, H. and {Baccigalupi}, C. and {Baldi}, M. and {Balestra}, A. and {Bardelli}, S. and {Battaglia}, P. and {Biviano}, A. and {Branchini}, E. and {Brescia}, M. and {Camera}, S. and {Ca{\~n}as-Herrera}, G. and {Capobianco}, V. and {Carbone}, C. and {Carretero}, J. and {Castellano}, M. and {Castignani}, G. and {Cavuoti}, S. and {Chambers}, K.~C. and {Cimatti}, A. and {Colodro-Conde}, C. and {Congedo}, G. and {Conversi}, L. and {Copin}, Y. and {Courbin}, F. and {Courtois}, H.~M. and {Cropper}, M. and {Cuillandre}, J.-C. and {Degaudenzi}, H. and {De Lucia}, G. and {Dolding}, C. and {Dole}, H. and {Douspis}, M. and {Dubath}, F. and {Dupac}, X. and {Dusini}, S. and {Escoffier}, S. and {Farina}, M. and {Farinelli}, R. and {Ferriol}, S. and {Finelli}, F. and {Fourmanoit}, N. and {Frailis}, M. and {Franceschi}, E. and {Fumana}, M. and {Galeotta}, S. and {George}, K. and {Gillis}, B. and {Giocoli}, C. and {G{\'o}mez-Alvarez}, P. and {Gracia-Carpio}, J. and {Grazian}, A. and {Grupp}, F. and {Guzzo}, L. and {Gwyn}, S. and {Haugan}, S.~V.~H. and {Hoekstra}, H. and {Holmes}, W. and {Hook}, I.~M. and {Hormuth}, F. and {Hornstrup}, A. and {Jhabvala}, M. and {Kermiche}, S. and {Kubik}, B. and {Kuijken}, K. and {K{\"u}mmel}, M. and {Kunz}, M. and {Kurki-Suonio}, H. and {Le Brun}, A.~M.~C. and {Ligori}, S. and {Lilje}, P.~B. and {Lindholm}, V. and {Lloro}, I. and {Mainetti}, G. and {Maino}, D. and {Maiorano}, E. and {Mansutti}, O. and {Marggraf}, O. and {Martinelli}, M. and {Martinet}, N. and {Marulli}, F. and {Massey}, R.~J. and {McCracken}, H.~J. and {Medinaceli}, E. and {Mei}, S. and {Mellier}, Y. and {Meneghetti}, M. and {Merlin}, E. and {Meylan}, G. and {Mohr}, J.~J. and {Mora}, A. and {Moresco}, M. and {Moscardini}, L. and {Munari}, E. and {Nakajima}, R. and {Neissner}, C. and {Nichol}, R.~C. and {Niemi}, S.-M. and {Padilla}, C. and {Paltani}, S. and {Pasian}, F. and {Pedersen}, K. and {Percival}, W.~J. and {Pettorino}, V. and {Pires}, S. and {Polenta}, G. and {Poncet}, M. and {Popa}, L.~A. and {Pozzetti}, L. and {Racca}, G.~D. and {Raison}, F. and {Rebolo}, R. and {Renzi}, A. and {Rhodes}, J. and {Riccio}, G. and {Rix}, H.-W. and {Romelli}, E. and {Roncarelli}, M. and {Rosset}, C. and {Rusholme}, B. and {Saglia}, R. and {Sakr}, Z. and {Sapone}, D. and {Sauvage}, M. and {Schirmer}, M. and {Schneider}, P. and {Schrabback}, T. and {Secroun}, A. and {Seidel}, G. and {Serrano}, S. and {Sihvola}, E. and {Simon}, P. and {Sirignano}, C. and {Sirri}, G. and {Stanco}, L. and {Steinwagner}, J. and {Tallada-Cresp{\'\i}}, P. and {Tereno}, I. and {Tessore}, N. and {Toft}, S. and {Toledo-Moreo}, R. and {Torradeflot}, F.},
        title = "{Euclid: Discovery of 31 new quasars at 6.6 < z < 7.8}",
      journal = {\aap},
         year = 2026,
        month = jul,
       volume = {711},
          eid = {A104},
        pages = {A104},
          doi = {10.1051/0004-6361/202658883},
archivePrefix = {arXiv},
       eprint = {2607.03432},
 primaryClass = {astro-ph.GA},
       adsurl = {https://ui.adsabs.harvard.edu/abs/2026A&A...711A.104Y}
}




\appendix

\section{ALMA measurements}\label{app:alma_measurements}

We measure all ALMA quantities directly from the calibrated data cubes. For each source, we use the systemic redshift measured from the JWST/NIRSpec spectrum to predict the observed [\Cii]\ frequency, $\nu_{\rm ref}=\nu_{\rm rest}/(1+z_{\rm NIRSpec})$. We construct the [\Cii]\ moment-0 map using $|\nu-\nu_{\rm ref}|<0.6\,\mathrm{GHz}$, corresponding to approximately $\pm722$ and $\pm738\,\mathrm{km\,s^{-1}}$ for \yswl\ and \ysej, respectively. This initial window is used to identify the line-emitting region and define the extraction aperture. We subsequently fit the extracted spectrum with a free Gaussian centroid; the resulting $z_{[\mathrm{C\,\textsc{ii}}]}$ defines the velocity zero point in Fig.~\ref{fig:cii}.

\textit{Continuum maps.} Continuum maps are constructed by averaging line-free channels in the ranges $262.4$--$267.2$\,GHz for \yswl\ and $256.5$--$259.5$\,GHz for \ysej. The per-pixel noise $\sigma_{\rm cont}$ is estimated as the median absolute deviation (MAD-scaled) of pixels in a background annulus of $30$--$50$\,pixel radius around the source.

\textit{[\Cii]\ moment-0 maps.} For each spatial pixel we subtract the continuum map from the line-window channels and integrate in velocity to produce a moment-0 map. The line-map noise $\sigma_{\rm m0}$ is computed analogously from the same background annulus.

\textit{Aperture selection.} Centred on the moment-0 peak pixel, we compute an elliptical curve of growth using apertures whose semi-axes scale as $k\times$ the synthesised beam ($k=0.5$--$6.0$). The fiducial aperture is the smallest one enclosing $90\%$ of the maximum cumulative flux, yielding $2.252^{\prime\prime}\times1.856^{\prime\prime}$ for \yswl\ and $1.656^{\prime\prime}\times1.364^{\prime\prime}$ for \ysej.

\textit{Spectra and [\Cii]\ fluxes.} Aperture-extracted spectra are obtained by collapsing the cube with the fiducial-aperture weights, divided by the beam area in pixels. The continuum level is taken as the median of the line-free channels, and the channel noise $\sigma_{\rm chan}$ as the MAD on the same channels. We fit a single Gaussian to the continuum-subtracted spectrum within $\pm1.5$ line-window widths and report $F_{[\mathrm{C\,\textsc{ii}}]}=A\sigma_{v}\sqrt{2\pi}$, with the uncertainty propagated from the fitted $A$ and $\sigma_v$. We additionally extract a peak-pixel spectrum at the moment-0 peak and fit it identically. \yswl\ is consistent with an unresolved point source, so we adopt the peak-pixel flux as our fiducial value; \ysej\ is resolved and we adopt the aperture flux. The two spectra are shown in Fig.~\ref{fig:cii}.


\textit{Continuum flux densities.} The aperture continuum flux is $S_{\nu}=\sum_{i\in\mathrm{ap}} w_i\,C_i / \Omega_{\rm beam}$, where $C_i$ is the continuum map, $w_i$ are the exact-pixel aperture weights, and $\Omega_{\rm beam}$ is the beam area in pixels. Its uncertainty is $\sigma_{\rm cont}\sqrt{N_{\rm beams}}$, with $N_{\rm beams}$ the number of beams subtended by the aperture.

\section{Continuum fitting}\label{app:continuum}

\begin{figure*}
    \centering
    \includegraphics[width=0.9\linewidth]{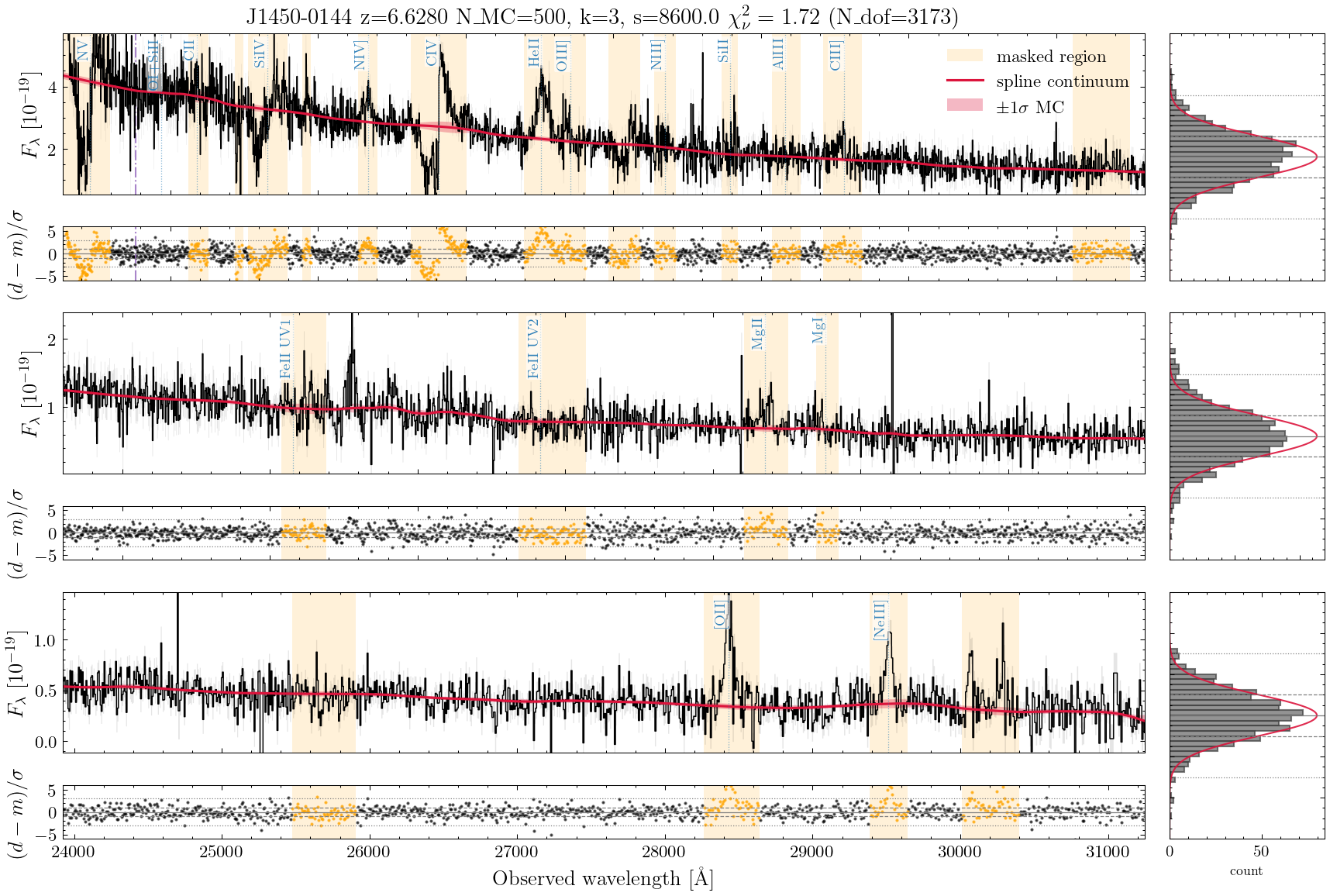}\\[1ex]
    \includegraphics[width=0.9\linewidth]{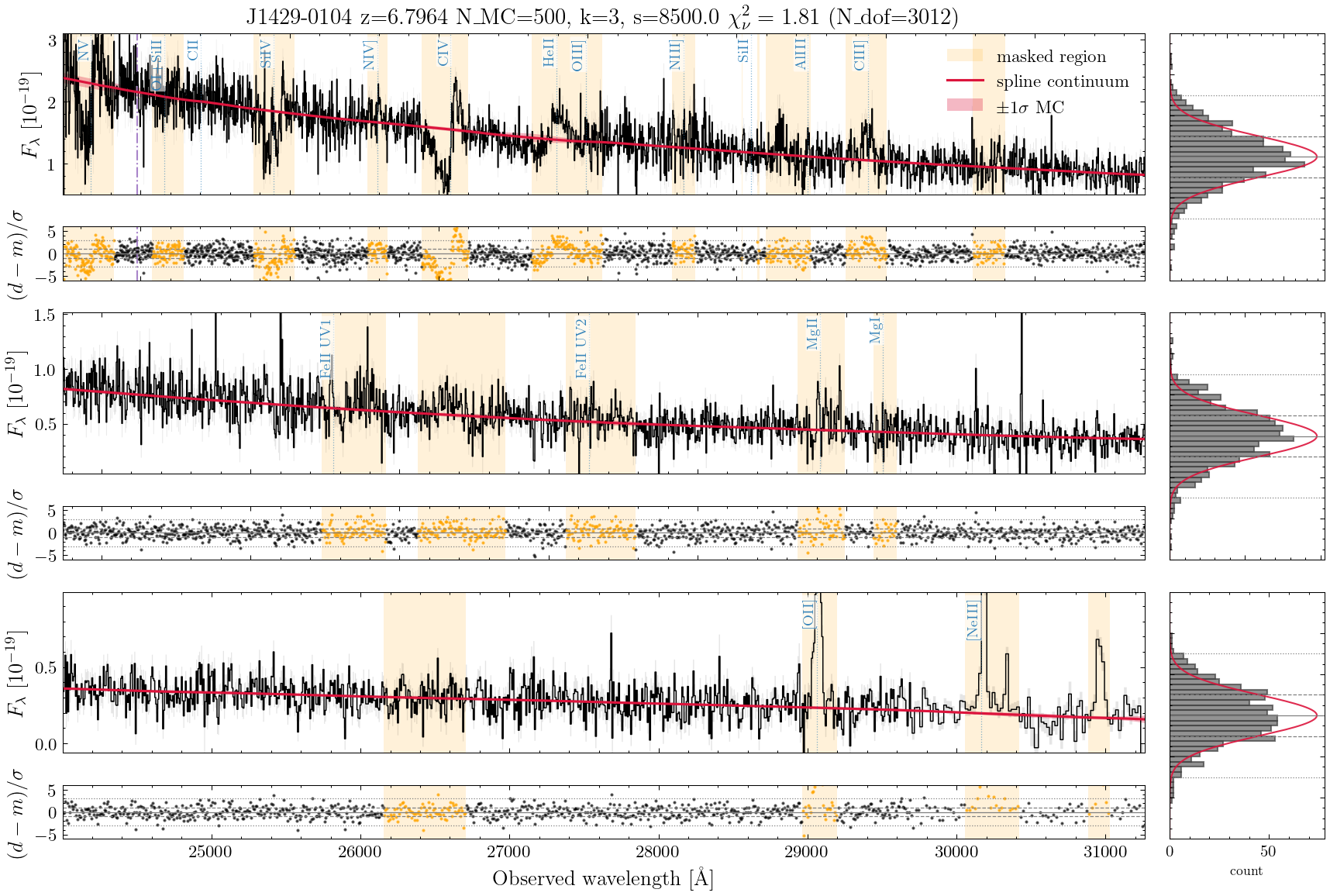}
    \caption{Spline continuum fits of \yswl\ (top) and \ysej\ (bottom). See Sect.~\ref{sec:specmod} for details. Each pair of panels shows the observed spectrum (black) with the median continuum model (red, $\pm1\sigma$ shaded) and the per-pixel residuals $(d-m)/\sigma$ below. Right-hand panels show the residual distributions. Shaded bands mark wavelength windows excluded from the fit.}
    \label{fig:continuum_qa}
\end{figure*}

Fig.~\ref{fig:continuum_qa} shows the spline continuum models of both sources, constructed as described in Sect.~\ref{sec:specmod}, together with the per-pixel residuals and their distributions. These fits define the continuum used for all line and equivalent-width measurements in this work.

\bsp	
\label{lastpage}
\end{document}